\documentclass[12pt]{article}

\usepackage[small,sf]{titlesec}
\usepackage{adjustbox}
\usepackage{amssymb}
\usepackage{amsmath}
\usepackage{amsthm}
\usepackage[round]{natbib}
\usepackage{array}
\usepackage{caption}
\usepackage[labelformat=simple]{subcaption}
	
\usepackage{bm} 
\usepackage{longtable}
\usepackage{mathtools}
\theoremstyle{plain}
\usepackage{tabularx}
\usepackage{float}
\usepackage[toc,title,page]{appendix}
\usepackage[flushleft]{threeparttable}
\usepackage{graphicx}
\graphicspath{ {graphs/} }
  \usepackage{subcaption} 
\usepackage{dcolumn}
\usepackage{booktabs, siunitx}
\usepackage[english]{babel}
\usepackage{hyperref}
\usepackage[dvipsnames]{xcolor}

\newenvironment{tfootnote}[1][\textwidth]{
	\begin{tabular}{m{#1}}  \footnotesize \textit{Notes.}
	}{
	\end{tabular}}

\newtheorem{lemma}{Lemma}
\theoremstyle{definition}
\newtheorem{assumption}{Assumption}
\newtheorem{prediction}{Prediction}
\theoremstyle{plain}
\newtheorem{prop}{Proposition}
\newtheorem{corollary}{Corollary}
\newtheorem{result}{Result}
\theoremstyle{remark}

\newcommand{\dd}{\text{ d}}
\newcommand{\E}[1]{\mathbb{E}\left[#1\right]}
\newcommand{\Ex}[2]{\mathbb{E}_{#1}\left[#2\right]}
\newcommand{\Var}[1]{\text{Var}\left(#1\right)}
\newcommand{\Cov}[1]{\text{Cov}\left(#1\right)}

\newcommand{\sym}[1]{$\overset{^{#1}}{\ }$} 
\newcommand{\sigt}{\sigma_\theta}
\newcommand{\sigs}{\sigma_s}
\newcommand{\vbar}{\overline{v}} 

\ExplSyntaxOn
\cs_set:Npn \expandableinput #1
  { \use:c { @@input } { \file_full_name:n {#1} } }

\AddToHook{env/tabular/begin}
  { \cs_set_eq:NN \input \expandableinput }
\AddToHook{env/tabularx/begin}
  { \cs_set_eq:NN \input \expandableinput }
\ExplSyntaxOff

\usepackage{tikz}
\usetikzlibrary{shapes,arrows,shapes.multipart,shapes.geometric}
\usetikzlibrary{positioning,fit}                       	
\usepackage{pgfplots}
\tikzset{   
        every picture/.style={remember picture,baseline},
        every node/.style={anchor=base,align=center,outer sep=1.5pt},
        every path/.style={thick},
        }
\newcommand\marktopleft[1]{%
    \tikz[overlay,remember picture] 
        \node (marker-#1-a) at (-.3em,.3em) {};%
}
\newcommand\markbottomright[2]{%
    \tikz[overlay,remember picture] 
        \node (marker-#1-b) at (0em,0em) {};%
}
\usetikzlibrary{shapes.callouts}
	\pgfmathdeclarefunction{gauss}{2}{%
	\pgfmathparse{1/(#2*sqrt(2*pi))*exp(-((\x-#1)^2)/(2*#2^2))}%
	}

\usepackage{enumerate}                  	
\usepackage{booktabs}
\usepackage{listings}
\usepackage{mdwlist}                    	
\usepackage[margin=1in]{geometry}      	
\usepackage{graphicx,psfrag}             	
\usepackage{multirow}
    	
\renewcommand{\arraystretch}{1.2}

\usepackage{titlesec}
\renewcommand{\thesection}{\arabic{section}.}
\renewcommand{\thesubsection}{\arabic{section}.\arabic{subsection}}
\titleformat{\section}{\bf\large}{\thesection}{.7em}{}
\titleformat{\subsection}{\large}{\thesubsection}{.7em}{}
\titleformat{\subsubsection}{\rmfamily}{\thesubsubsection}{.7em}{}

\title{Self-selection of Information and Belief Update:
	An Experiment on COVID-19 Vaccine Information Acquisition}
\author{ChienHsun Lin\footnote{Department of Economics, National Taipei University, New Taipei City, Taiwan. Email: \href{mailto:chienhsunlin@gm.ntpu.edu.tw}{chienhsunlin@gm.ntpu.edu.tw}}
	\and Hans H. Tung\footnote{Department of Political Science, National Taiwan University, Taipei, Taiwan. Email: \href{mailto:hans.tung@ntu.edu.tw}{hans.tung@ntu.edu.tw}}}
\date{February 9, 2026 \\ 
	(\href{https://www.dropbox.com/scl/fi/tn6etrugis3q4nh2882s2/vaccine_main.pdf?rlkey=2ec88slkfjxr1nvazjhxau7qr&st=3n5hjzbg&dl=0}{Click here for the latest draft})}

\begin{document}
\maketitle

\begin{abstract}

How does the endogenous selection of information shape belief formation? In observational settings, individuals only consume information they choose, making it impossible to observe how they would respond to information they actively avoid. We address this identification challenge using a randomized experiment on COVID-19 vaccines in Taiwan. After eliciting subjects' preferences over vaccine-specific reports, we randomly assign them to receive either their chosen or unchosen information, orthogonalizing selection from exposure. We find subjects are more likely to select information about vaccines they already perceive as more effective. Conditional on receiving information, belief updating is substantially larger when the information was self-selected, even after controlling for prior-posterior disagreement. These findings highlight endogenous information demand as a central determinant of  persuasion, suggesting that increasing information availability alone may be insufficient when individuals rationally filter out options they perceive as less relevant to their decision-making.

\end{abstract}

\section{Introduction}
Standard economic theory posits that information is instrumental.
People acquire it to update beliefs and optimize decisions (\cite{bohnenblust1949}, \cite{blackwell1953}).
Prior studies have examined how information influences beliefs across various domains, including education, politics, and real estate pricing (\cite{wiswall2015}; \cite{chopra2022}; \cite{fuster2022}),
as well as how beliefs affect behavior, such as social distancing during the COVID-19 pandemic (\cite{allcott2020}), 
and how information directly affects decisions, such as vaccine uptake (\cite{alsan2021}).
However, in the real world, information acquisition is endogenous. 
Individuals actively select which news to read, which experts to consult, and which data to ignore. 
This selection is determined by the ex-ante value the individual assigns to the potential information. 
Because individuals only incur the cost of learning when they believe the information might pivot their decision, 
the act of selection itself reveals their subjective expectations about the information's instrumental value.
Thus, the demand for information depends not only on the uncertainty of beliefs but crucially on the level of those beliefs: 
individuals effectively ``screen'' for information about the options they already perceive as viable. 
Hence, self-selected information is inherently more likely to induce belief revision, as it is more likely to be relevant to their decision-making process.

Nonetheless, the empirical relationship between information selection and belief updating remains unclear due to the endogeneity of information choice. 
In observational settings, we only observe belief updates conditional on the individual choosing to consume the information. 
This creates a missing counterfactual: we rarely observe how an individual would have responded to information they actively avoided. 
Consider an individual deciding whether to adopt a new technology (e.g., a vaccine). 
If she holds a low prior on its quality, she may rationally bypass costly technical reports because she anticipates the signal is unlikely to be pivotal. 
Consequently, we never observe whether a positive signal would have persuaded her had she been forced to see it.

This identification gap is not merely methodological; it can create biased policy implications. 
Without recovering this counterfactual, we cannot distinguish whether belief divergence persists because individuals are immune to evidence or simply because they systematically filter it out. 
If the primary barrier is the endogenous selection of information rather than the interpretation of it, then standard policy interventions that increase the availability of information can be effective, as individuals will continue to filter it out based on their priors.

This paper addresses this identification challenge by designing a controlled field experiment that explicitly decomposes the information acquisition process into its three theoretical components: (i) ex-ante information selection, (ii) belief updating, and (iii) ex-post decision revision. 
By mirroring this sequential structure in an experimental setting, we can observe not just the final outcome, but the transition at each stage of the learning process.

We implement this design in the context of COVID-19 vaccine choice in Taiwan. 
While the setting provides high-stakes, real-world incentives, our experimental innovation lies in the manipulation of information supply conditional on demand. 
After eliciting subjects' preferences for specific vaccine reports, we randomly assign them either to their selected information or to a non-selected alternative. 
This orthogonalization of selection (what they wanted) from reception (what they got) allows us to isolate the causal effect of the signal from the endogeneity of the choice. 
It enables us to recover the counterfactual: how do beliefs evolve when an individual is confronted with evidence they would have otherwise ignored? 

The experiment consists of four main phases.
In Phase 1, subjects report their pre-treatment beliefs about vaccine effectiveness and state their vaccine preferences.
In Phase 2, subjects rank the five COVID-19 vaccine brands based on their willingness to read about each, and then select up to three vaccines for which they wish to receive information.
In Phase 3, we present information about vaccine effectiveness to each subject. The assignment of information is randomized and independent of their selections.
Finally, in Phase 4, we elicit post-treatment beliefs about vaccine effectiveness and post-treatment vaccine preferences.

We organize our main empirical findings along the three stages of the information acquisition process. 
\textit{(i) information selection}: 
 Subjects are more likely to select and read vaccine information when they believe the vaccine is more effective.
\textit{(ii) belief update}: Beliefs respond more strongly to information when subjects receive information they previously selected, after controlling for prior-posterior disagreement.
\textit{(iii) updated decision}: Receiving information about a selected vaccine increases subjects' stated preferences for that vaccine and reduces refusal rates, suggesting effects on the extensive margin of acceptance.
These findings are consistent with the rational information acquisition framework, which predicts that individuals will select information that they expect to be more persuasive and will update their beliefs more in response to information they have chosen to receive.


Since our experimental setting allows for belief updating across multiple vaccines,
we extend the framework to incorporate correlated beliefs.
This allows for the possibility that subjects perceive certain vaccines to be similar---for example, Pfizer and Moderna, which both use mRNA technology---and therefore update their beliefs about one vaccine based on information received about another.
Consistent with this idea, we observe in the data that subjects update their beliefs about a given vaccine even when they do not receive direct information about it.
However, the magnitude of such indirect updates is smaller, suggesting that belief spillovers across vaccines are limited but present.

We contribute to the literature in three main ways.
First, we provide a systematic approach to examine the information acquisition process. 
Our experimental design decomposes the theory into distinct steps, allowing us to diagnose each stage of the information acquisition process.
In the context of our experiment, we find that subjects evaluate information effectively with respect to vaccine effectiveness (i.e., its implications for welfare), but are less sensitive to pure uncertainty reduction.
While other behavioral biases may emerge in different contexts, our approach provides a structured way to identify where and how deviations from theory arise.
This is particularly useful for policy evaluation: by isolating specific points of departure from rational behavior, policymakers can target interventions more precisely.

Second, we extend the scope of rational information acquisition theory to the domain of COVID-19 vaccines, which carries significant public health implications.
We show that the framework is broadly applicable even in complex, real-world environments with multiple alternatives and heterogeneous attributes.
In such settings, the relevance of the theory extends beyond belief accuracy to encompass the behavior induced by those beliefs.

Third, we develop an incentive-compatible method to elicit preferences over information without relying on monetary incentives.
Our design allows subjects to truthfully reveal which vaccine information they are most interested in, and increases the likelihood they receive it---while still preserving exogenous assignment for causal identification.
This methodology is especially useful in domains where financial incentives are infeasible or inappropriate (e.g., political or sensitive health contexts).
Moreover, eliminating extrinsic incentives helps simplify implementation and reduce measurement noise.

\subsection{Literature}


As this paper isolates the stages of information acquisition, it connects three strands of literature: information demand, self-selection and belief updating, and decision-making.
The first strand examines how individuals choose which information to acquire and the factors that influence these choices. \citet{ambuehl2018} directly test whether subjects select more informative signals in a laboratory setting, finding that non-Bayesian belief updating leads subjects to underestimate the marginal value of informativeness. Beyond the lab, a growing literature examines information demand in real-world contexts. \citet{hoffman2016} studies business professionals' willingness to pay for website valuation information, finding systematic over- and underestimation of information's value. \citet{chopra2022} show that fact-checking reduces demand for ideologically aligned news and increases demand for non-aligned news, indicating that demand is shaped not only by instrumental value but also by ideological fit. \citet{mikosch2024} exploit exchange rate uncertainty to show that firms acquire more information than households and increase demand when uncertainty rises, consistent with the rational information acquisition framework.


A natural question emerging from this literature is whether the information people choose is systematically more persuasive.
\cite{fuster2022} show that information demand is heterogeneous and that posteriors shift in response to consumed information; lowering information costs reduces the dispersion in beliefs. However, because their design only presents participants with their most preferred information, the effect of receiving non-selected information remains unclear. \citet{faia2022} conduct two experiments on COVID-19 information, finding that subjects prefer ideologically aligned information and update beliefs more when receiving preferred content. Our study builds on this line of work by extending the setting to multiple correlated alternatives and by measuring the numerical disagreement between prior beliefs and information content. This allows for a direct and richer test of the rational information acquisition framework.

A third strand examines how information affects actual choices. Examples include \citet{wiswall2015} on college major selection, \citet{hoffman2016} on business valuation, and \citet{haaland2021} on racial discrimination beliefs. Within the COVID-19 context, \citet{sadish2021a} study variation in pandemic-related information delivery, and \citet{banerjee2020} examine the role of expert endorsements. Our work contributes to this literature by jointly analyzing information choice, belief updating, and decision change in a realistic, high-stakes public health setting.

The paper proceeds as follows.
In the next section, we describe the experimental design.
In section 3, we provide a theoretical model of information consumption and the belief update,
and we will provide predictions that we may observe in the experiment. 
Section 4 summarizes the data set and the main variables, 
and section 5 presents the main analysis.
Section 6 concludes the paper with a discussion of the findings and future research directions.

\section{Background and Experimental Design}
In real-world settings, information consumption is often shaped by people's interest in the information itself.
As a result, observational data on information acquisition may suffer from selection bias.
For example, when individuals consume only information they are already interested in,
we do not observe how they would respond to information they chose not to acquire.
However, the interests about information can be an unobserved variable that affects their belief updates.
To isolate the effect of information from the influence of preferences,
we design an experimental environment that separately elicits people's information preferences
while ensuring that information exposure is not solely determined by those preferences.

In our experiment, subjects first report their preferences over vaccine-related information.
They are then randomly assigned to one of several treatment arms.
In one treatment, subjects receive the information they selected;
in others, they are randomly exposed to information about vaccines they did not select.
With this design, the subjects are (non-monetarily) incentivized to state their true preferences,
while still allowing for random exposure to less-preferred or unpreferred information.
Consequently, we are able to evaluate how information affects beliefs and decisions,
conditional on both demand and random assignment.
The remainder of this section describes the experimental context,
outlines the design, and provides additional implementation details.

\subsection{Background}
\subsubsection*{\underline{COVID-19 Vaccines}}
Approximately six months after the genetic sequence of the COVID-19 virus was identified,
the first COVID-19 vaccine, \textit{CanSino}, was approved for emergency use by the Chinese government on June 24th, 2020.
Since then, more and more vaccines have been developed to help mitigate the outbreak of the COVID-19 pandemic.
Because full regulatory approval typically requires years of clinical testing,
most of the governments opted to issue \textit{Emergency Use Authorizations} (EUAs) for vaccines that met key safety and efficacy benchmarks---
typically including successful completion of Phase III trials.\footnote{
	In medical research, Phase III clinical trials test a vaccine's efficacy and monitor adverse effects, typically involving 300 to 3,000 participants.
	Phase IV trials, by contrast, assess long-term safety and effectiveness in much larger populations. For more details, see the U.S. Food and Drug Administration (FDA): \url{https://www.fda.gov/patients/drug-development-process/step-3-clinical-research}.
	} 

By November 2021, five COVID-19 vaccines had entered Phase IV trials.
Sinovac (CoronaVac, referred to as \textit{Sinovac} hereafter) was one of the earliest authorized vaccines and is based on an inactivated virus platform.
Oxford-AstraZeneca (Vaxzevria, referred to as \textit{AstraZeneca} hereafter) and Janssen (Jcovden, referred to as \textit{J\&J} hereafter) use viral vector technology to deliver genetic material that prompts an immune response---a platform also widely used in other vaccines.
Pfizer-BioNTech (Comirnaty, referred to as \textit{Pfizer} hereafter) and Moderna (Spikevax, referred to as \textit{Moderna} hereafter) are based on mRNA technology.
These two were the first mRNA vaccines to be deployed at scale in large populations.
However, because mRNA vaccines were relatively novel,
public uncertainty remained regarding their efficacy and potential side effects.
In addition to these internationally developed vaccines,
Taiwan also developed its own domestic options: the MVC COVID-19 vaccine (referred to as \textit{Medigen} hereafter) was the most received domestic vaccine in Taiwan.

\subsubsection*{\underline{The Pandemic and Vaccines in Taiwan}}
The first confirmed COVID-19 case in Taiwan was reported on January 21st, 2020.
However, due to strict border controls and quarantine policies, the case numbers in Taiwan remained low (fewer than $1000$ cases) until March 2021.
The situation changed with the spread of the Alpha variant,
which led to a sharp increase in infections.
By June 2021, Taiwan had reported over 10,000 confirmed cases and more than 500 deaths.

The first COVID-19 vaccine authorized in Taiwan was AstraZeneca,
which received approval and became available in March 2021.
Due to limited supply, vaccines were initially prioritized for healthcare workers and older adults.
Despite early vaccine hesitancy---driven in part by media coverage and public discourse surrounding adverse events---the surge in infections and fatalities in May 2021 significantly increased public willingness to vaccinate.
By the end of June 2021, over two million Taiwanese residents (approximately 10\% of the population) had received at least one dose of a COVID-19 vaccine.

As of November 2021 (the time when this experiment intervention was implemented), four vaccines were available in Taiwan:
AstraZeneca, Pfizer, Moderna, and Medigen.
That same month, Taiwan's Centers for Disease Control (CDC) approved mix-and-match vaccination across different vaccine types,
in part to address supply constraints, particularly for internationally manufactured vaccines.
This policy change created stronger incentives for individuals to seek alternative vaccine combinations.
According to CDC data as of November 22, 2021, 77.1\% of the population had received at least one dose, and 48.9\% had received two doses.
Although J\&J and Sinovac were not granted EUA in Taiwan, some Taiwanese residents traveled abroad to receive these vaccines.
However, the CDC did not recognize these unapproved vaccinations for the purposes of domestic health regulations, such as border control and quarantine policies.

\subsection{Details of the Experiment Design}
\subsubsection*{\underline{Vaccine Performance Information}}

We provided participants with performance information for five COVID-19 vaccines. 
The five vaccines chosen are (in the alphabetical order of the brand names): 
	Johnson \& Johnson, Moderna, Oxford-AstraZeneca, Pfizer, and Sinovac.\footnote{
	The five vaccines were chosen based on two criteria as of September 2021:
	(1) availability of publicly released Phase III trial results; and
	(2) active progression into Phase IV trials.
	Please refer to Appendix \ref{vaccine_detail} for detailed information.}
To enhance the credibility of the information presented to subjects, all performance reports were sourced from peer-reviewed scientific journals and are publicly accessible through the World Health Organization (WHO) website.\footnote{
	Recent studies show that people tend to trust information endorsed by international organizations such as the WHO.
	For example, \citet{sheen2023} conducted a survey experiment in Taiwan and found that subjects were more willing to accept vaccines from China when endorsed by the WHO---especially among those who trusted the WHO more than the origin country of the vaccines.
}

For efficacy and hospitalization prevention rates,
we used the most recent statistics from trials involving the full recommended dosage.
For side effects, we selected the latest data from reports that included both general side effects
and severe adverse events (defined as those exceeding Grade 3 in severity).\footnote{
	Table \ref{tab:vaccine_info_detail} summarizes the information provided.
	}

\subsubsection*{\underline{Subject Recruitment}}

We recruited participants through Facebook advertisements,
targeting Facebook and Instagram users in Taiwan between the ages of 17 and 64.
Upon clicking the advertisement link, individuals were redirected to the online survey platform and invited to complete the survey.
After the subjects finished the survey, 
they would receive a flat payment of a gift card worth \$150 Taiwanese Dollar (approximately \$5 USD).

After excluding participants who reported ages below 17 or above 64,
we obtained a final sample of 1,066 complete responses.
Summary statistics for the sample are provided in Appendix Section \ref{sec:experiment_details}.\footnote{
	Our sample is somewhat younger and more female-skewed than the general Taiwanese population.
	Specifically, 43\% of our participants are under the age of 25 (compared to 19\% nationally),
	and 87\% are under age 45 (compared to 69\% nationally).
	Additionally, 63\% of respondents identify as female,
	compared to approximately 50\% in the population.
	}
Table \ref{tab:subject_detail} reports detailed subject characteristics.
Across all observable covariates, we find no significant predictors of treatment assignment,
suggesting that randomization was balanced.

\subsubsection*{\underline{The Procedure of the Experiment}}

The experiment was implemented using the online survey platform Qualtrics.
The estimated completion time was approximately 20 minutes,
with a median completion time of 15 minutes.
Figure \ref{schedule} illustrates the structure of the experimental flow,
which comprised four main stages:
(1) pre-treatment assessment,
(2) information demand elicitation,
(3) information exposure (treatment), and
(4) post-treatment assessment.
In the pre-treatment assessment stage,
subjects reported their beliefs about the performance of each vaccine
along four dimensions: efficacy, hospitalization prevention rate, adverse event rate, and severe adverse event rate.
In addition, they were asked to state their initial preferences over the five vaccines.

In the information demand elicitation stage,
subjects were informed that they would have an opportunity to read actual scientific reports about COVID-19 vaccines.
Figures \ref{fig:ranking_question} and \ref{fig:selection_question} present sample screenshots of the user interface shown to subjects.
Subjects expressed their demand for vaccine information among five vaccines with the two (strategic method) questions:
\begin{enumerate}[(1)]
\item {\bf [Ranking]} Please rank the listed five vaccines from the highest to the lowest based on how much you want to read the information about the vaccine.
\item {\bf [Selection]} You ranked X the 1st, Y the 2nd, and Z the 3rd. If you have the chance to select, would you like to read the information about your top 3, top 2, top 1, or 0 vaccines?
\end{enumerate}

Subjects were informed that these decisions would increase the probability 
of reading the information about the higher-ranked vaccines.
Thus, it was in their best interest to state their true preferences.

In the information treatment exposure stage, 
the vaccine information was presented to the subjects.
The subjects may receive information about different vaccines based on 
the treatment arms they were assigned,
which will be explained in detail in the next paragraph.
Figure \ref{fig:info_sheet} is an (English translated) sample of the information that subjects observed.
Once on the information page, participants could click buttons to view content for each attribute of the selected vaccines.
After reviewing the information, participants were asked to report:
how much they trusted the information,
which parts were previously unknown to them,
and which parts differed from their prior beliefs.
Finally, in the post-treatment assessment stage,
participants answered the same set of questions as in the pre-treatment stage,
including belief elicitation and vaccine preference measures.

\subsubsection*{\underline{Treatments}}
In the information exposure stage, subjects received different sets of vaccine information based on their assigned treatment arms.
There were three main treatment arms and two supplementary arms.
The main treatment arms varied in whether and how the information was matched to the subject's stated preferences from the earlier elicitation stage.
The three main treatment arms are described below:

\begin{itemize}
\item {\bf Full Compliance (FC).} Subjects received information about the vaccines they ranked highest and explicitly selected,
based on both their responses to the ``Ranking'' and ``Selection'' questions.
\item {\bf Top 3 (T3).} Subjects received information about their top three ranked vaccines,
based solely on their ``Ranking'' responses, regardless of their ``Selection'' choice.
\item {\bf Random Assignment (RA).} Subjects received information about three randomly selected vaccines,
assigned independently of their stated preferences.
\end{itemize}

We provide the following example to illustrate the differences between treatment arms.
Suppose a subject ranks the five vaccines as follows:
(1) Pfizer, (2) Moderna, (3) AstraZeneca, (4) Johnson \& Johnson, and (5) Sinovac. 
She then chooses to \emph{select} only the top two vaccines for information exposure.
If she is assigned the Full Compliance arm, she will receive the information only about Pfizer and Moderna, where both are \emph{selected top 3 vaccines}.\footnote{
	If a subject in the Full Compliance group chooses not to read any of the selected information, they will skip the treatment exposure stage.}
If she is assigned the Top 3 arm, she will receive all three top-ranked vaccine information,
{\it i.e.} Pfizer, Moderna, and AstraZeneca, where AstraZeneca is in top 3 but not acquired by the subject.
If she is in the Random Assignment arm, she will receive information about three random vaccines.\footnote{
	To address potential confusion from not receiving selected information,
	we also included two supplementary treatment arms: T3$^*$ and RA$^*$.
	The subjects in the T3$^*$ treatment only completed the ``Ranking'' question (no ``Selection'')
	and received the information only about their top three ranked vaccines (according to the ``Ranking'' question).
	The subjects in RA$^*$ Treatment did not see either of the ``Ranking'' and ``Selection'' questions,
	and they were assigned information about three randomly selected vaccines.
	}

Note that if a subject ranks the vaccines and selects information truthfully based on her true preference,
she is more likely to receive information about the vaccines she values most.
To illustrate this, consider the example above.
Suppose the subject has the same preference as stated.
Since the subject truly prefers Pfizer and ranks it first, selecting it for information exposure.
In this case, she is guaranteed to receive Pfizer information under either the Full Compliance or Top 3 treatment arms.
In contrast, if she misreports her preferences and either ranks Pfizer lower or does not select it,
she will not receive Pfizer information under those same treatment conditions.
This design structure---combining demand elicitation and randomized assignment---also enables us to isolate the effects of information exposure from preference-driven selection.
Specifically, participants assigned to the Random Assignment arm receive information about three randomly chosen vaccines,
independent of their stated preferences.

We can categorize the vaccines into three groups  based on each subject's elicited information demand:
\begin{enumerate}[(1)]
	\item \emph{Selected Top 3}--the vaccine is ranked top 3 and is selected
	\item \emph{Non-selected Top 3}--the vaccine is ranked top 3 but is not selected
	\item \emph{Not Top 3}--the vaccine is not within the subject's top 3
\end{enumerate}
Within each information demand category, exposure to information is random.\footnote{
	Continuing with the previous example:
	suppose the subject puts Pfizer in the \emph{Selected Top 3} category, 
	the subject will not receive Pfizer information only if she is in the RA treatment arm
	and Pfizer is not one of the three randomly assigned vaccines.
	As whether the subject receives the information about the vaccine is fully determined by the randomness created by the experimenters,
	we can exclude the self-selection of the consumption within the \emph{Selected Top 3} category.
	The same logic applies to the other two categories.
	}
This design allows us to causally identify the effect of receiving vaccine information on belief updating,
separately within each demand category.
We can then compare treatment effects across the three information demand categories to evaluate how prior interest in the vaccine moderates the impact of information exposure.

\section{Theoretic Benchmark}
We apply the environment of COVID-19 vaccine information with the rational information acquisition framework.
The decision maker's (DM's) vaccine information acquisition process can be represented as a three-stage decision problem:\footnote{A similar setup can be found in \cite{fuster2022},
where subjects were asked to predict housing prices based on information sources of varying accuracy.
Participants first chose an information source, then updated their beliefs,
and were ultimately rewarded based on how close their predictions were to the true housing prices.}
\begin{enumerate}
\item The DM decides whether to acquire (potentially costly) information about the vaccine's effectiveness.
\item The DM receives the information and updates the beliefs.
\item The DM decides whether to get vaccinated based on her posterior beliefs.
\end{enumerate}

Throughout the rest of this section, we assume that beliefs about vaccine effectiveness are normally distributed,
and that belief updating follows Bayesian principles upon receiving new information.

\subsection{The Case of Single Vaccine \label{sec:theory_single}}
\subsubsection{Setting}
For simplicity, we start with the case where there is only information for one vaccine available.
The DM holds a prior belief about the vaccine's overall effectiveness,
$\theta \sim N(\mu_\theta,\sigt^2),$
where both $\mu_\theta$ and $\sigt^2$ are known.\footnote{
	This framework can be generalized to $N$ vaccines,
	with normally distributed priors and signals whose variance-covariance matrices are diagonal; that is, there is no correlation across priors or signals for different vaccines.}
Let $\vbar$ denote the reservation value of the outside option.\footnote{
	The reservation value can have multiple interpretations.
	For instance, suppose an agent has decided to receive the vaccine which she believes is the most effective.
	Then the reservation value can be her prior of effectiveness of this vaccine.
	The other vaccine dominates the original one only if the posterior suggests that the new vaccine is (on average) better.}
If the DM does not acquire any information about the vaccine,
she will choose not to take it if the reservation value exceeds her prior mean belief;
otherwise, she will proceed with vaccination.
Thus, the expected value the DM receives without acquiring information, denoted by $v_0$, is given by
$$ v_0 = \max\{ \vbar, \mu_\theta\}.$$

\subsubsection{The Acquisition of the Information}	
The DM can also acquire a signal $s$ about the vaccine at a fixed cost $c$, 
where the signal is given by $s = \theta + \varepsilon$, 
with $\varepsilon\sim N(0,\sigma^2_s)$ and $\sigs^2$ is known to the DM.
We further assume $\theta \perp \varepsilon$.
Therefore, the signal is distributed as $s|\theta\sim N(\theta,\sigs^2)$ and $s \sim N(\mu_\theta, \sigs^2 + \sigt^2)$.

Given a realization of $s$, the Bayesian posterior belief about $\theta$ is
$$\theta|s \sim 
	N\left(\frac{\sigs^2 \mu_\theta + \sigt^2 s}{\sigs^2 + \sigt^2 },
	\frac{\sigs^2\sigt^2}{\sigs^2+\sigt^2}\right).$$

After observing $s$, the DM chooses whether to take the vaccine.
She does so if her posterior expectation exceeds the reservation value $\vbar$, i.e., if $\mathbb{E}[\theta|s]$ exceeds the reservation value $\vbar$.
This reflects the decision rule that the vaccine is only taken when the updated belief about its effectiveness is sufficiently high.
Therefore, after receiving the signal $s$, we can then define the value function as follows:
\begin{equation} \label{eq:value_s}
v(s) = 
\begin{cases}
\mathbb{E}[\theta|s]& \text{ if } \mathbb{E}[\theta|s] \geq \vbar\\
\vbar& \text{ if } \mathbb{E}[\theta|s] <\vbar \end{cases}.
\end{equation} 
To evaluate the expected value of acquiring information,
note that the posterior expectation exceeds $\vbar$ if and only if the signal $s$ exceeds a critical threshold $s^*$:
$$ \mathbb{E}[\theta|s] \geq \vbar \ \Leftrightarrow \ 
s \geq \frac{(\sigt^2+\sigs^2)\vbar-\sigs^2\mu_\theta}{\sigt^2}\equiv s^*,$$
where we define $s^*$ to be the critical value.
That is, the DM will choose to take the vaccine if and only if the realized signal is more optimistic than the critical threshold $s^*$.

Figure \ref{fig:model} demonstrates a characterizing example of belief updating of vaccine effectiveness.
The top panel plots the probability density function of the random variable $\theta$,
which represents the vaccine's true effectiveness.
The gray curve depicts the prior distribution, with $\mu_\theta$ denoting the prior mean.
Let $\vbar$ denote the reservation value.
In this example, since $\mu_\theta<\vbar$, the DM is not taking the vaccine in the absence of any additional information.

Upon receiving a signal $s$, the DM updates her belief to the posterior distribution (shown in red).
The posterior mean is given by $\mathbb{E}[\theta|s]$.
Since $\E{\theta|s}$ implied by the information ($s$) is higher than the reservation value, $\vbar$,
the DM is persuaded to take the vaccine after observing $s$, and the expected value becomes $\mathbb{E}[\theta|s]$.\footnote{
	Note that $\E{\theta|s}= (\sigs^2 \mu_\theta + \sigt^2 s)/(\sigs^2 + \sigt^2)$, which is a linear function in $s$.}
As discussed earlier, the threshold signal $s^*$ is the cutoff that determines whether the DM will be persuaded to take the vaccine.
If $s\leq s^*$, then the DM remains unconvinced and receives utility $\vbar$;
if $s> s^*$, she is persuaded and the utility becomes $\mathbb{E}[\theta|s]$.
The bottom panel of Figure \ref{fig:model} depicts the resulting value function $v(s)$.

Since the DM knows the distribution of $s$ ex ante,
she can compute the expected value of information: $\Ex{s}{v(s)}$, 
which captures the expected utility with information.
Then, the DM's decision to acquire the signal depends on the following criterion:
$$ \left\{ \begin{array}{ll} \text{Acquire the information} & \text{ if } \Ex{s}{v(s)}-v_0-c \geq 0 \\
\text{Refuse the information} & \text{ otherwise } \\ \end{array} \right..$$
Therefore, we can study the behavior of the \textit{net value of information},
($\mathbb{E}_s[v(s)] - v_0 - c$)
to characterize the DM's decision on information acquisition.

\begin{prop} Define the relative accuracy of the prior belief $\gamma\equiv(\sigs^2)/(\sigt^2)$.
Let $V\left(\vbar-\mu_\theta,\gamma,c\right)\equiv\Ex{s}{v(s)}-v_0-c$ be the value of the information.
\label{theory:comp_stat}
	\begin{enumerate}[(a)]
	\item $V(\cdot)$ decreases in $|\vbar-\mu_\theta|$.
	\item $V(\cdot)$ decreases in $\gamma$.
	\item $V(\cdot)$ decreases in $c$.
	\end{enumerate}
\end{prop}
The proof of Proposition \ref{theory:comp_stat} is provided in Appendix \ref{sec:proof_comp_stat}.
As $V(\cdot)$ increases, the agent becomes more likely to acquire the information.
Therefore, Proposition \ref{theory:comp_stat} implies that when the prior belief about the vaccine ($\mu_\theta$) is substantially below the reservation value ($\vbar$), or when the agent holds relatively precise prior beliefs (i.e., a higher $\gamma$), she is less inclined to acquire vaccine information.

Proposition \ref{theory:comp_stat} gives two empirical implications.
First, the DM is more likely to acquire information about vaccines that she believes to be more effective, as indicated by a smaller $|\vbar-\mu_\theta|$.
This explains the confirmatory tendency for individuals to select information about vaccines they already view as more effective.
Second, the DM acquires information about the vaccines that are not her top choice \textit{only if} the DM expects the information suggests that the vaccine is effective enough to persuade her to override her default choice.

If we assume that the prior mean belief about the vaccine is below the reservation value ($\vbar\geq \mu_\theta$), then the first result in Proposition \ref{theory:comp_stat} holds without absolute values.\footnote{Suppose instead that $\vbar< \mu_\theta$. 
	Then it is more natural for the DM to take the vaccine \textit{before} the information acquisition, as they have learned that taking the vaccine is better than \textit{status quo} (which only yields the original reservation value). In that case, the DM may treat her belief about the current vaccine as the new reservation value when comparing alternative vaccines.}
Under the condition that $\mu_\theta \leq \vbar$,
we can further conclude that $V(\cdot)$ is increasing in $\mu_\theta$.
In other words, the more optimistic the DM's prior belief about a vaccine,
the more likely she is to seek additional information about it.

\subsubsection{Change in Beliefs}
Given information $s$, the mean of the posterior belief of $\theta$ is 
$(\sigs^2 \mu_\theta + \sigt^2 s)/(\sigs^2 + \sigt^2)$.
Hence the change in beliefs is 
$$ \delta(s) = \E{\theta|s} - \mu_\theta = \frac{\sigt^2}{\sigs^2+\sigt^2}(s-\mu_\theta)
= \frac{1}{1+\gamma}(s-\mu_\theta).$$
Therefore we have the following comparative statics.
\begin{prop}
The change in beliefs $\delta(s) = \E{\theta|s} - \mu_\theta$  increases in $s-\mu_\theta$.
Furthermore, the absolute change in beliefs $|\delta(s)|$ decreases in $\gamma$.
\end{prop}

Intuitively, when the signal deviates more from the prior mean, or when the signal is relatively more informative,
the DM's posterior belief will exhibit a greater shift. In other words, belief updating is more pronounced when the new information is both strong and reliable.

We can further combine the two predictions to identify the source of endogeneity between information acquisition and persuasion.
Information demand is primarily determined by two factors:
(i) the distance between the outside option value and the mean of the prior belief about vaccine effectiveness, and
(ii) the relative accuracy of the information (signal precision).
Furthermore, persuasion is driven by:
(i) the disagreement between the signal and the prior mean, and
(ii) the relative accuracy of the information.
Since both information acquisition and persuasion are functions of signal accuracy,
we can predict a comparative static relationship between the two:
\begin{corollary}
Given the distance between the mean of the default value and the prior belief  ($|\vbar-\mu_\theta |$), 
the disagreement between the signal and the prior mean $(s - \mu_\theta)$, 
and the cost of the information,
the degree of persuasion, defined as  $\delta(s) = \E{\theta|s} - \mu_\theta $ is increasing in the value of information, $V(\vbar - \mu_\theta ,c,\cdot)$.
\end{corollary}

Corollary 1 provides a benchmark linking selection and persuasion under rational precision-based acquisition.
This phenomenon coincides with the ``confirmatory bias'' addressed in literature,
that people seek information are more likely to be persuaded by the information



\subsection{Multiple Vaccines}
In our experiment, subjects are presented with information and asked to update their beliefs about five vaccines simultaneously.
If the vaccine information is not independent across different vaccine brands,
the information of vaccine brand $A$ \textit{can} help the agent update her belief of another vaccine.
This introduces additional complexity to the information acquisition decision:
the DM must now consider how information about one vaccine may inform her beliefs about others, due to potential correlations across brands.
We leave a detailed discussion of this more general case to Appendix \ref{sec:multi_vaccine},
where beliefs are correlated and cross-vaccine inference is possible.
Nonetheless, the main insights derived from the single-vaccine framework continue to hold under the multi-vaccine setting.

The model provides a benchmark for how information demand and belief updating would be related if individuals selected information solely to reduce uncertainty. 
In the empirical analysis, we use these predictions as reference points to understand how observed behavior aligns with and departs from this benchmark.

\section{Description of the Main Variables}
To align the data more closely with the model's predictions, we summarize the main variables used in our experiment.
The key factors in our theoretical framework are the subjects' beliefs about the vaccines, their information demand, and their preferences for the vaccines.
The following subsections describe how each variable is measured.

\subsection{Beliefs about the Vaccine effectiveness}
Subjects reported their beliefs about vaccine effectiveness, specifically in four areas: efficacy, hospitalization prevention rate, adverse event rate, and severe adverse event rate.\footnote{
	All factors range from 0\% to 100\%.
	The subjects were told how these factors are calculated.
	}
Beliefs were elicited both before and after the treatment phase. 
Each subject provided beliefs about all four effectiveness dimensions for five vaccines, yielding five observations per factor.

Table \ref{tab:summary_main} summarizes the beliefs and changes before and after the information exposure.
The first two rows show the average beliefs on each factor of vaccine effectiveness before and after the treatments.
Though not significant, the average belief in vaccine effectiveness increases after the treatment phase (higher efficacy/hospitalization prevention rate and lower adverse event rates).

Figure \ref{fig:cdf_belief} shows the distribution of each factor's beliefs before and after the treatment.
More than half of the observations report beliefs of at least 70\% efficacy and at least 80\% hospitalization prevention. 
Around 50\% of respondents believe there is no more than a 20\% chance of severe adverse events.
The distributions of beliefs regarding efficacy and hospitalization prevention shift upward, while beliefs about adverse event rates shift downward.
This indicates that our subjects' beliefs generally become more optimistic after the intervention.  

The third and fourth rows of Table \ref{tab:summary_main} report the average and absolute changes in beliefs at the observation level before and after treatment.
On average, the subjects hold more positive beliefs about the vaccines' effectiveness after receiving the information.
The final two rows report the gaps between participants' pre-treatment beliefs and the information provided.
Overall, our subjects tend to undervalue vaccine effectiveness relative to scientific evidence---evidenced by negative gaps for efficacy and hospitalization prevention, and positive gaps for adverse event rates.

\subsection{The Subjects' Demand for the Vaccine Effectiveness Information}
Figure \ref{fig:info_consumption}  illustrates subjects' responses to the two information demand questions
Since vaccines from Pfizer, Moderna, and AstraZeneca were the only available vaccines in Taiwan at the time of the experiment,
information about these vaccines was more frequently selected by the subjects.
Approximately 60\% of participants selected information on the maximum of three vaccines.
There is no significant difference in information demand among the treatments.


\subsection{Preference on Receiving Vaccines}
The subjects also reported their willingness to receive vaccines. 
We used two questions to measure participants' vaccine preferences.
The first asked participants to rate their preference for each vaccine on a 0--100 scale.
The second asked how many weeks they would be willing to wait to receive each vaccine.
Since vaccines were not fully accessible in Taiwan at the time of the study, the waiting-time question served as an alternative measure of relative preference.
Both preference questions were asked before and after the information exposure, immediately following the belief elicitation.

\subsection{Familiarity with the Vaccines and the Receptions}
To control the subjects' knowledge about the vaccines before they received the information,
we asked them to self-assess their familiarity with each vaccine.
Additionally, we included factual quiz questions to objectively assess subjects' vaccine knowledge.\footnote{
	The quiz questions are about (1) the technique platform applied by each vaccine and (2) the recommended number of doses.
	}
In general, the subjects are more familiar with the vaccines available in Taiwan.	
We also find a positive correlation between self-reported familiarity and quiz performance.

The subjects also reported which vaccines they had received.
For each vaccine, we asked the subjects whether they had received or registered for the vaccine (at least one dose).
Summary statistics are presented in Appendix Table \ref{vaccine_bg}.

In our sample, 87\% of the subjects had received at least one vaccine shot.
At the time of the experiment, approximately 70\% of Taiwan's population had received their first vaccine dose, and about 45\% had received a second dose.
Because many individuals were still deciding on their second dose, our information treatment had the potential to influence vaccine choices.
Moreover, subjects were more likely to view information about vaccines they expressed greater interest in, supporting the validity of our preference elicitation method.

\subsection{Information Engagement}
We also collected subjects' interaction data with the information materials.
Specifically, we recorded whether subjects clicked the buttons on the information page (Figure \ref{fig:info_sheet}) and how much time they spent viewing the page.
Table \ref{tab:info_interaction} summarizes subjects' interactions with the information pages.
Overall, subjects were highly engaged with the information provided.
For each vaccine attribute, approximately 80\% of the observations have clicked the buttons and checked the information at least once.
The median time spent on a page is 37.44 seconds.
Moreover, we observe that the subjects are more willing to check the information that they are more interested in, which provides  secondary evidence that justifies our elicitation of the preferences of the information.

\section{Experimental Results}

This section presents the empirical results from our experiment.
Guided by our theoretical framework, we organize the results into three parts: information selection, belief updating, and changes in vaccine preference.
First, we investigate how the subjects' beliefs about the vaccine effectiveness influence their selection of vaccine information.
Second, we analyze how the persuasiveness of the information depends on subjects' information demand.
Finally, we test whether belief updating translates into changes in vaccine preferences and willingness to receive vaccines.

\subsection{Information Selection\label{sec:info_demand}}
We begin by examining the factors that influence subjects' selection of vaccine information.
According to Proposition 1, subjects should be more likely to select information they expect will yield a higher posterior belief.
This leads to the following testable prediction.
\begin{prediction}
Subjects are more likely to select information about vaccines that they believe are more effective.
\end{prediction}

\begin{figure}[htp]
	{\centering
	\begin{subfigure}[b]{\textwidth}
	\centering
	\includegraphics[width = 0.48 \textwidth]{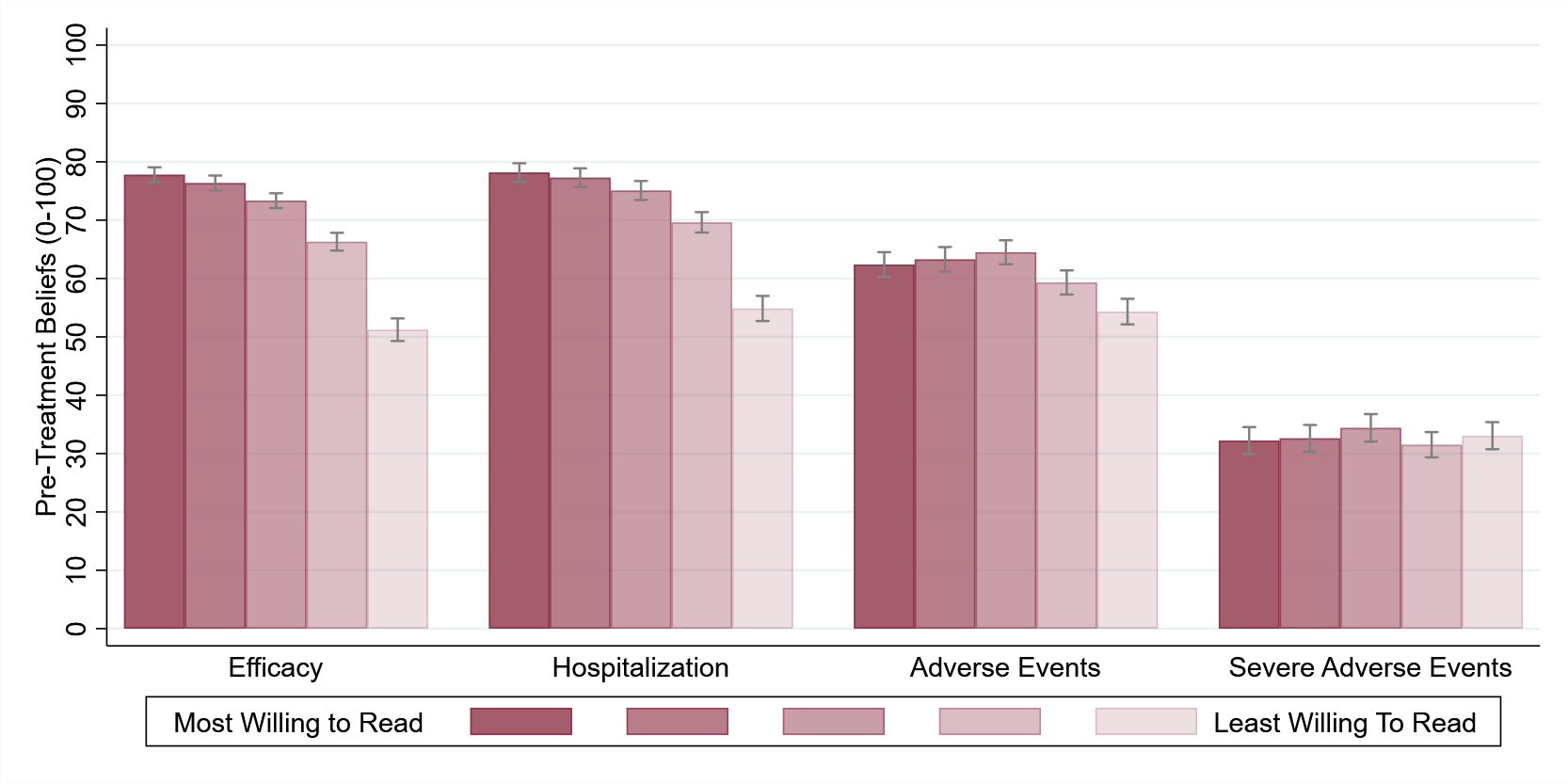}
	\includegraphics[width = 0.48 \textwidth]{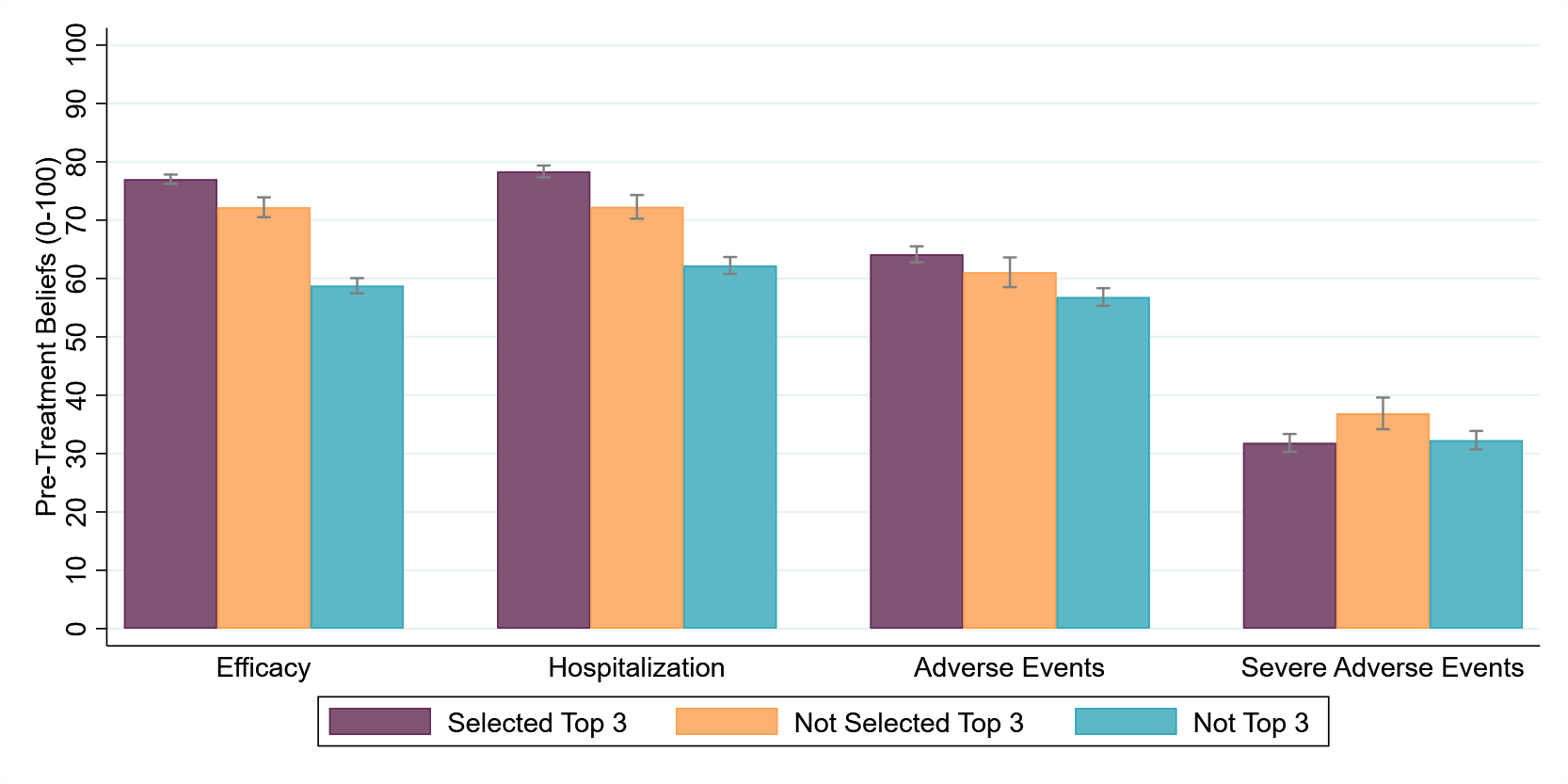}
	\caption{\small Beliefs of effectiveness by willingness to read (left) and information preference (right) \label{subfig:vaccine_pre}}
	\end{subfigure}
	\begin{subfigure}[b]{\textwidth}
	\centering
	\includegraphics[width = 0.48 \textwidth]{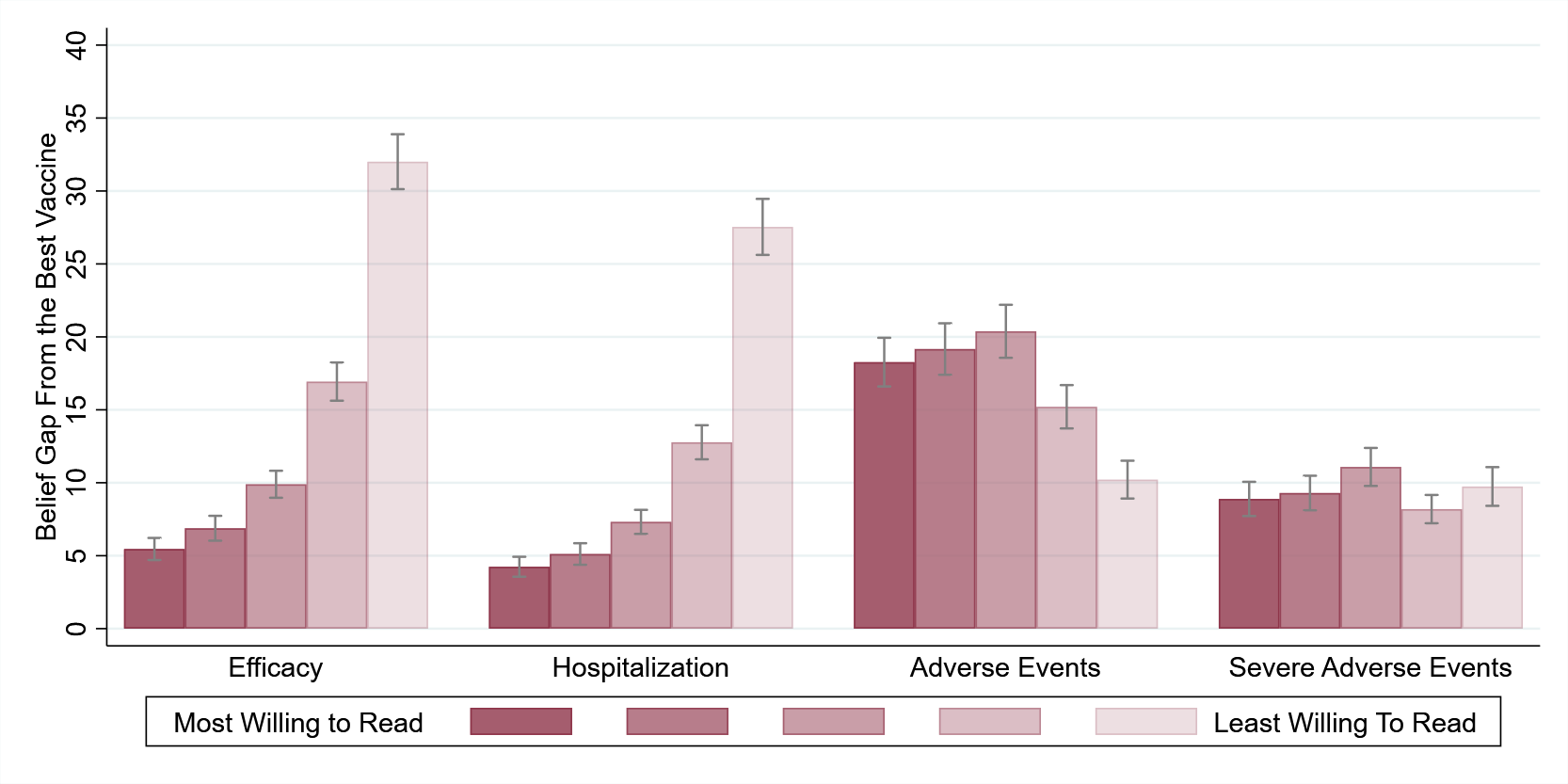}
	\includegraphics[width = 0.48 \textwidth]{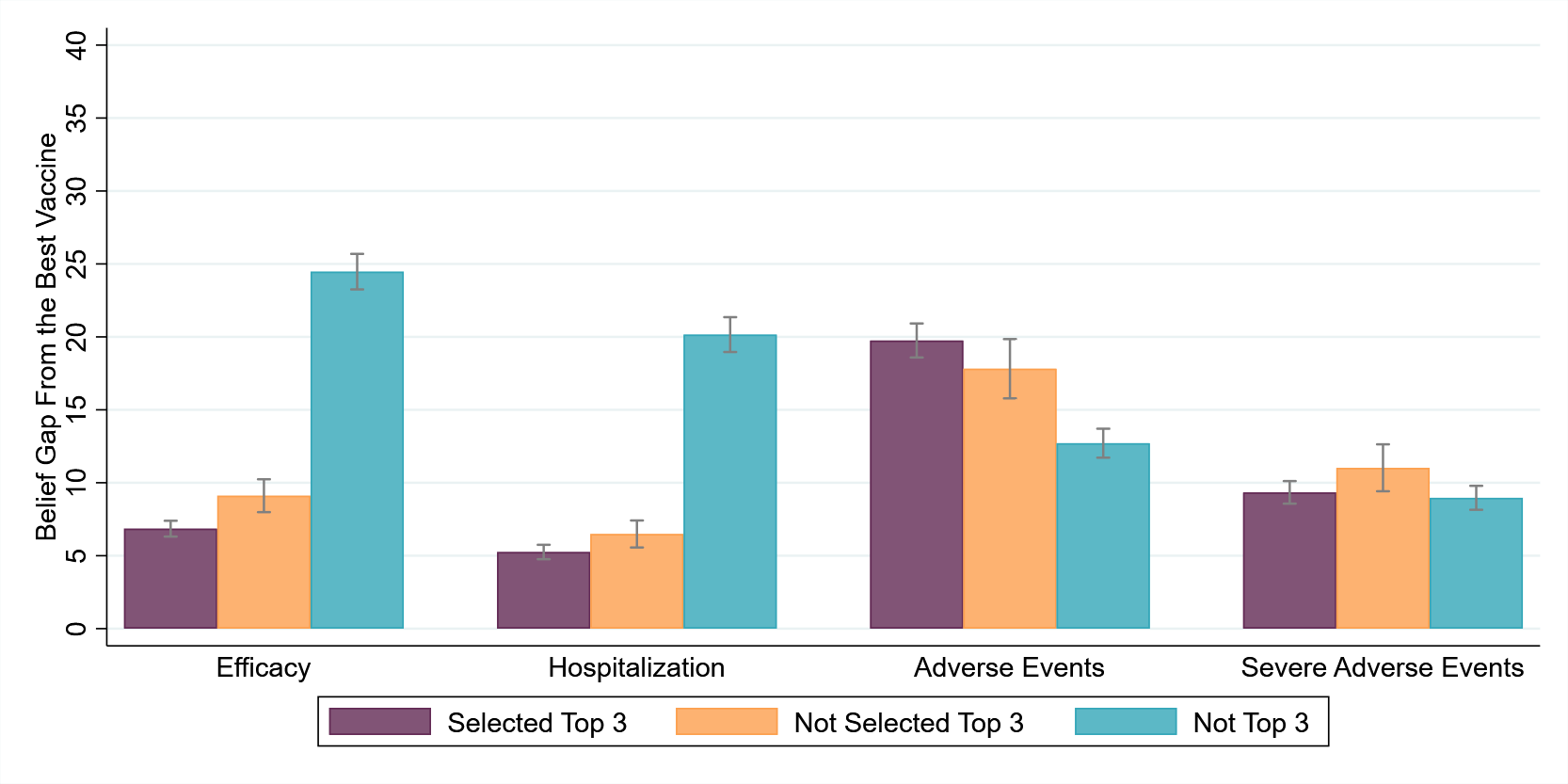}
	\caption{\small Gap between beliefs and default by willingness to read (left) and information preference (right) \label{subfig:from_best}}
	\end{subfigure}
	\begin{subfigure}[b]{\textwidth}
	\centering
	\includegraphics[width = 0.48 \textwidth]{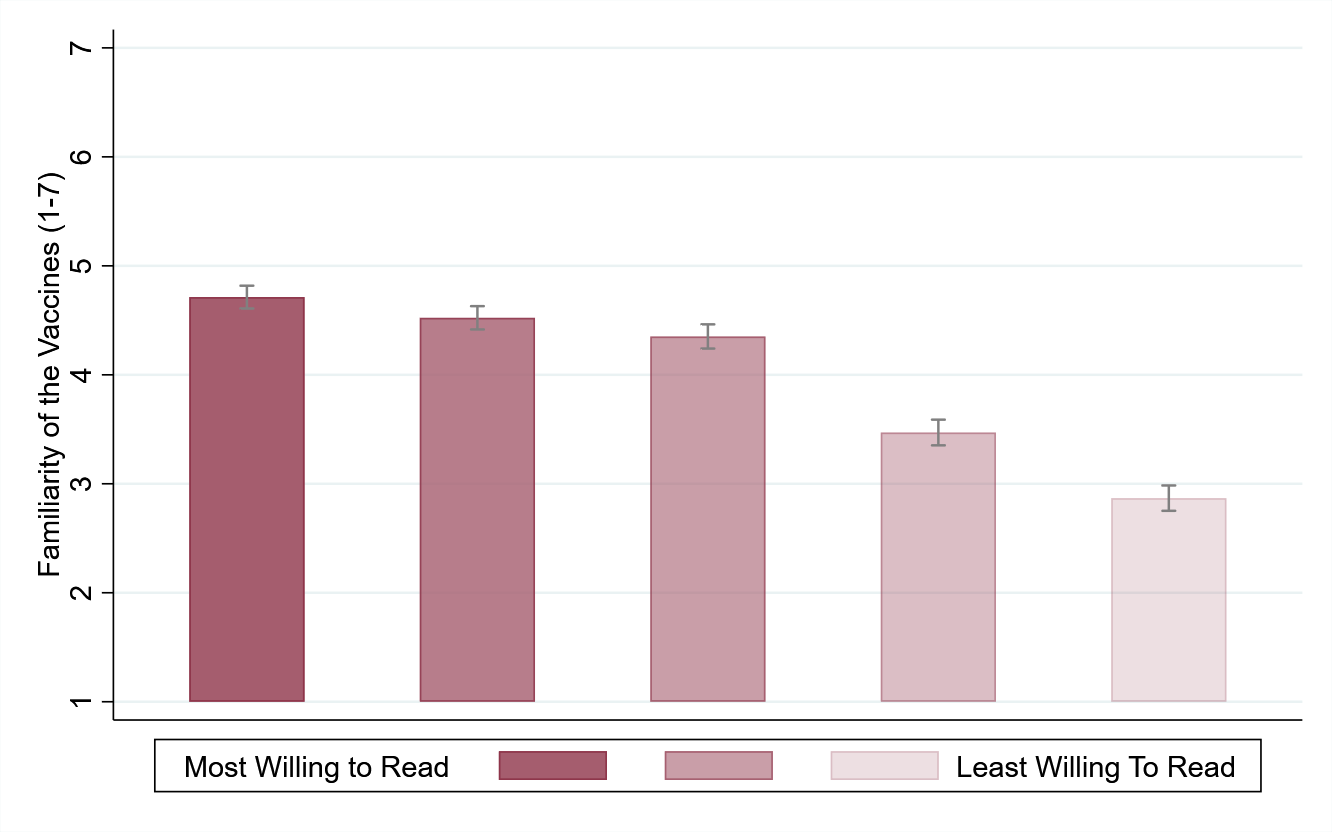}
	\includegraphics[width = 0.48 \textwidth]{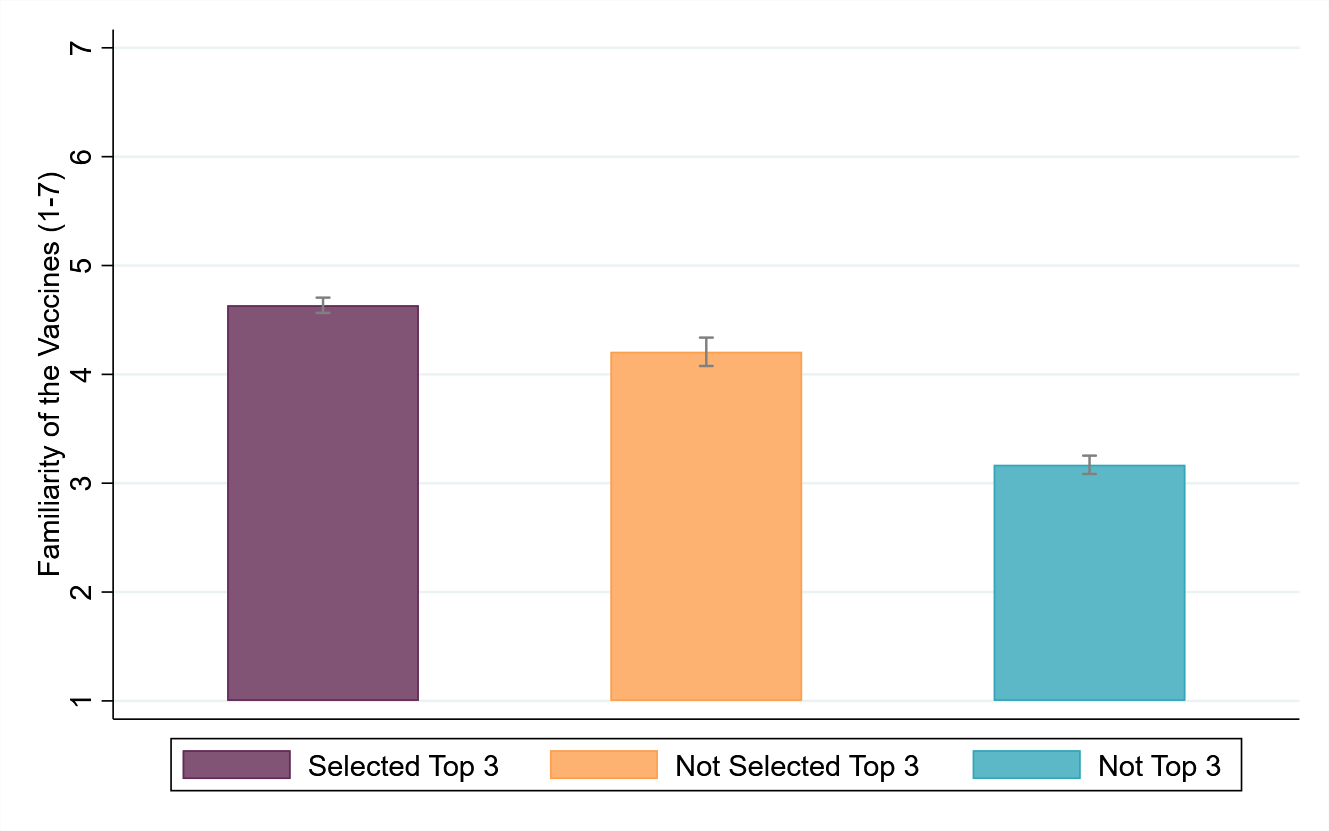}
	\caption{\small Vaccine familiarity by willingness to read (left) and information preference (right) \label{subfig:familiarity}}
	\end{subfigure}
	}
	\caption{Information Preference in Vaccine Effectiveness and Familiarity \label{fig:info_demand_belief}}
	{\begin{tfootnote} 
	The ranks in the willingness to read panels are determined by the question 
		``\textit{Please rank the following vaccines from the highest to the lowest based on how much you want to read the information about the vaccine.}''
	For each of the factors in (b), the distance of beliefs from the best vaccine for some factor is defined as the difference between the belief in that factor of the vaccine and the highest belief in that factor among all vaccines.
	95\% confidence intervals of the means of each bar are included.
	\end{tfootnote}
	}
\end{figure}

Figure \ref{fig:info_demand_belief}(a) displays the stated beliefs for each vaccine across the four effectiveness dimensions.
The left panel sorts the stated beliefs by the willingness to read elicited from the information ranking question.
Vaccines that subjects are more interested in (i.e., more willing to read about) tend to be associated with higher stated beliefs in efficacy and hospitalization prevention. 
The right panel categorizes vaccines based on responses to both information demand questions. 
Vaccines are classified as: (1) `Selected Top 3' if they were selected for reading and ranked in the top 3; (2) `Not Selected Top 3' if ranked in the top 3 but not selected; and (3) `Not Top 3' if not ranked in the top 3.
Selected vaccines also tend to have higher pre-treatment beliefs in efficacy and hospitalization prevention.
These patterns are not observed for adverse event or severe adverse event beliefs.

To evaluate the theory prediction, we use the the highest stated belief among the five vaccines in each of the attributes as the lower bound of the default value in the model.\footnote{ 
	The default value is unobserved under our setting.
	We assume the default value could be higher or equal to the highest stated belief among the five vaccines, but not lower;
	if the unobserved default value is lower than any of the five vaccines' stated beliefs, 
	the agent should take the vaccine whose performance exceeds the default value,
	so the default option should switch to the best vaccine at the time.
	}
Figure \ref{fig:info_demand_belief}(b) depicts the gap between the stated belief of each vaccine and the corresponding default value across the four factors.
The results mirror those in panel (a): vaccines with higher information demand are closer to the highest-performing vaccines in efficacy and hospitalization prevention.
These findings support the rational information acquisition framework: subjects are more interested in information that may improve decision quality---i.e., help them choose more effective vaccines.

Subjects' self-reported familiarity with each vaccine closely mirrors their information demand. 
Because familiarity is formed prior to the experiment, it captures pre-existing engagement with particular vaccines. 
Figure \ref{fig:info_demand_belief}(c) shows that subjects are systematically more familiar with the vaccines they express greater interest in reading about. 
This pattern suggests that information demand in our setting reflects prior engagement  rather than an attempt to reduce uncertainty about unfamiliar options.

To formally test the predictions from Figure \ref{fig:info_demand_belief}, we estimate the following regression models.
\begin{equation}
\text{InfoDemand}_{i,v} = \alpha + \sum_k \beta_k \text{Pre-beliefRank}^k_{i,v} +\lambda \text{Familiarity}_{i,v} + X_i\xi
\label{eq:demand_belief}
\end{equation}
\begin{equation}
\text{InfoDemand}_{i,v} = \alpha + \sum_k \gamma_k (\text{Default}^k_i - \text{Pre-belief}^k_{i,v}) +\lambda \text{Familiarity}_{i,v}+ X_i\xi
\label{eq:demand_gap}
\end{equation}
The dependent variables are the indexes of \textit{information demand}.
We construct two measures based on the information demand questions.
The first is the ranking of vaccine information, ranging from 1 (least interested) to 5 (most interested).
The second is a binary variable indicating whether the subject selected information about a given vaccine.
There are two sets of main explanatory variables;
each includes four variables about the effectiveness factors.
The first set captures pre-treatment rankings, where 1 indicates the lowest believed performance and 5 the highest, for each factor.
The second set measures the gap between each vaccine's belief and the default (i.e., highest stated belief among the five).
We also control for the subject's self-reported familiarity with each vaccine.

Table \ref{tab:info_demand} reports the regression results, which reinforce the patterns observed in Figure \ref{fig:info_demand_belief}.
Columns (1) and (3) show that higher perceived efficacy or hospitalization prevention is associated with higher ranking and a greater likelihood of information selection.
Columns (2) and (4) confirm that vaccines perceived to be further from the `best' option (in efficacy or hospitalization prevention) are less likely to be selected.

Across all specifications, subjects are more likely to select information about vaccines they are more familiar with.
We propose two possible mechanisms to explain the positive correlation between familiarity and information demand.
First, subjects may show less interest in vaccines that are not accessible to them and thus less familiar.
To address this, we restrict the analysis to vaccines available in Taiwan at the time of the experiment (AstraZeneca, Pfizer, and Moderna).
As shown in Table \ref{tab:info_demand_available}, the positive correlation between familiarity and selection persists within this restricted sample.
Second, familiarity may be driven by past vaccination choices, with subjects seeking information to justify previous decisions. 
To examine this, we include both familiarity and prior vaccine receipt in the regression (Table \ref{tab:info_demand_received}). 
While prior receipt is positively associated with information demand, familiarity remains a strong and significant predictor.
Taken together, these results suggest that familiarity captures a broader form of prior engagement with specific vaccines, beyond simple availability or past behavior. 
Information demand in our setting therefore appears to reflect pre-existing cognitive engagement with certain options rather than an attempt to learn about unfamiliar ones.

\begin{result}[Information Selection] 
	People select the information about the vaccines they believe are more effective and are more familiar with.
\end{result}

\subsection{Belief Updates} \label{subsec:belief_update}
In this subsection, we discuss how much the subjects' beliefs are persuaded by the information.
Specifically, we measure the change in each subject's beliefs before and after information exposure.

\begin{figure}[hbp]

{
\centering
	\includegraphics[width=0.45\textwidth]{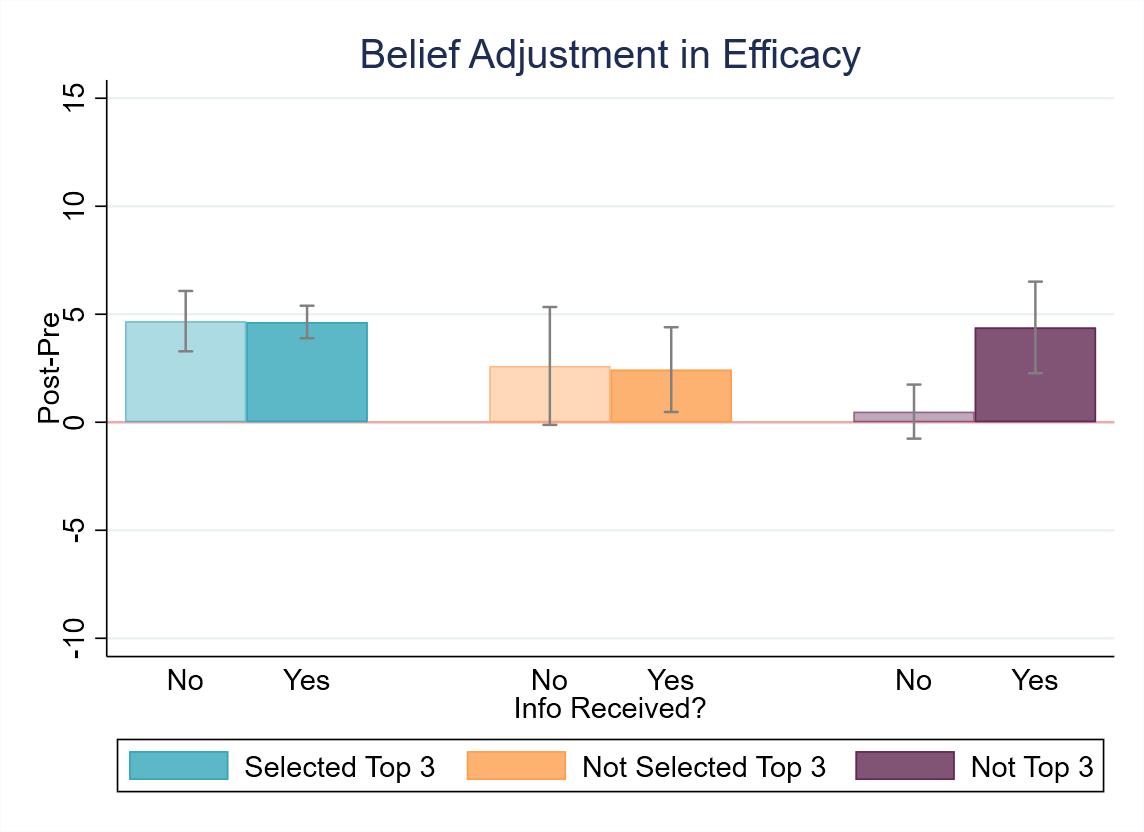}
	\includegraphics[width=0.45\textwidth]{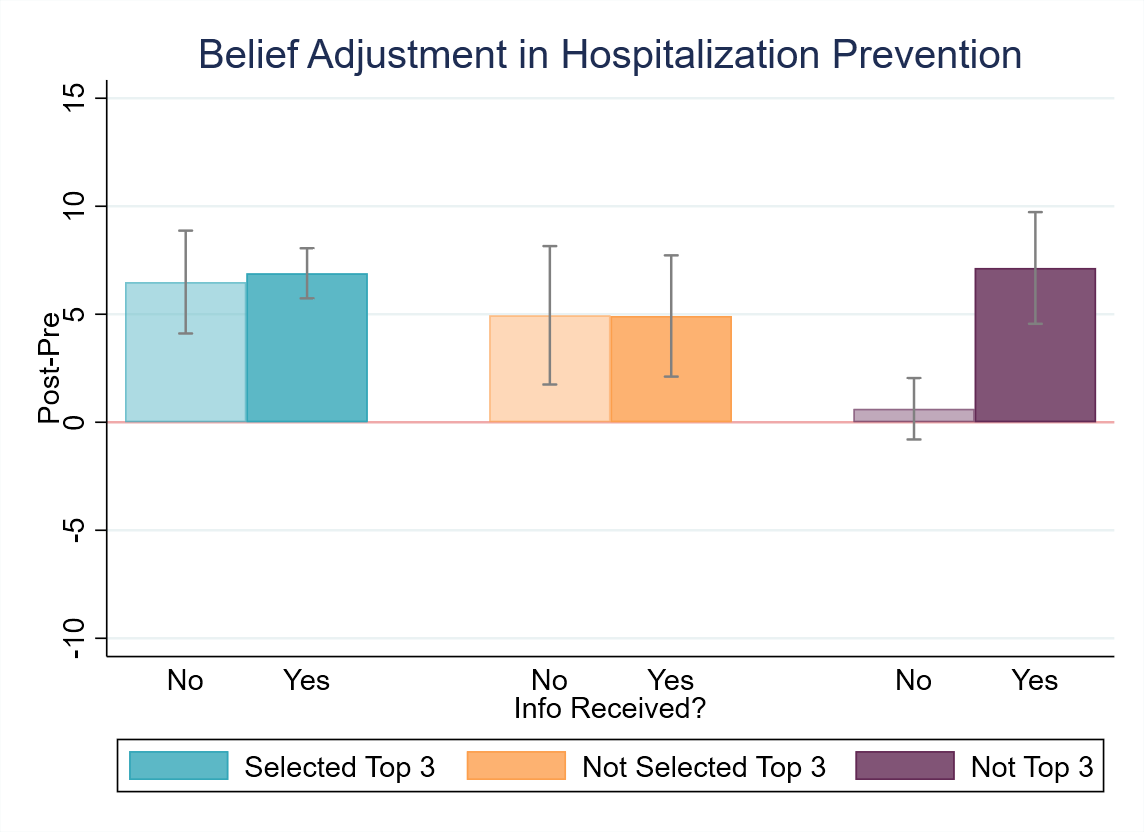} 
	\caption{Changes in Beliefs}
	\label{fig:change_in_beliefs}
	}
\begin{tfootnote}
The mean of the changes in beliefs before and after the information exposure stage are depicted.
The change is defined by Post-treatment belief $-$ Pre-treatment belief.
Within each category, the left/lighter bar is for the observations that the subjects do not receive the vaccine information,
and the right/deeper bar is for the observations that the subjects receive the information.
95\% confidence intervals of the means are plotted for each category.
\end{tfootnote}
\end{figure}

Figure \ref{fig:change_in_beliefs} depicts belief changes before and after the information exposure stage in efficacy and hospitalization prevention rates.
Observations are grouped as in Figure \ref{fig:info_demand_belief} and further subdivided by whether the subject received information about the vaccine.

On average, belief changes are positive for efficacy and hospitalization prevention.\footnote{
	The belief changes for adverse event rates and severe adverse event rates are negative, as shown in Figure \ref{fig:change_in_beliefs_adverse}.
}
In other words, subjects became more optimistic about vaccine effectiveness after reading the information.
Among those who received information (``Yes'' in Figure \ref{fig:change_in_beliefs}), belief adjustments were slightly larger in the \textit{Selected Top 3} group compared to the \textit{Not Selected Top 3} group. However, the difference between \textit{Selected Top 3} and \textit{Not Top 3} was not statistically significant.

Another finding is about the adjustments for the vaccines that the subjects \emph{did not} receive any information from the treatment.
In the absence of new information, beliefs about a given vaccine should remain unchanged, implying an average belief change near zero.
This pattern holds only for the \textit{Not Top 3} category. In contrast, we observe belief updates even when subjects did not receive information about \textit{Top 3} vaccines.
This suggests potential ``information spillovers''---subjects may indirectly update beliefs about one vaccine after receiving information about another.

We now distinguish between two types of belief updates: (1) direct updates from information received about a vaccine, and (2) indirect updates from information about other vaccines, suggesting cross-vaccine belief adjustments

\subsubsection{Direct Update}

As discussed in Section \ref{sec:info_demand}, pre-treatment beliefs may be correlated with information demand.
To estimate the correlation between the belief adjustment and information selection,
we can derive a prediction from Corollary 1.

\begin{prediction} Define \textit{persuasion} as the belief change from pre- to post-treatment
($\E{\theta|s} - \mu_\theta$). Then 
\begin{enumerate}[(i)]
\item persuasion increases with greater disagreement between the signal and the pre-treatment belief ($s-\mu_\theta$);
\item holding constant the distance to the default value 
($|v-\mu_\theta|$) and the signal disagreement ($s-\mu_\theta$), persuasion is greater when the information received was self-selected.
\end{enumerate}
\end{prediction}
Disagreement is measured as the difference between the provided signal and the subject's pre-treatment belief ($s-\mu_\theta$).
The greater the disagreement, the more ``surprising'' the information is to the subject.
The second prediction is adapted from Corollary 1.

To test these predictions, we estimate the following regression model:
\begin{align}
\text{BeliefAdjustment}_{i,v} = \alpha &+ \beta_1 \text{SelectedTop3}_{i,v} + \beta_2 \text{NotSelectedTop3}_{i,v}  \nonumber \\ 
 & + \delta \text{Disagreement}_{i,v}    \nonumber \\
 & + \sum_{k}\gamma_k (\text{Default}_i^k - \text{Pre-belief}^k_{i,v}) \nonumber \\
 & + \lambda \text{Familiarity}_{i,v} + X_{i}\xi \label{eq:reg_persuation}
\end{align}
The default belief gap is defined identically to Equation (\ref{eq:demand_gap}).
If Prediction 2 holds, we expect $\beta_1$ and $\beta_2$  to be positive (indicating greater persuasion for selected vaccines), and $\delta$ to be positive (greater disagreement leads to larger belief changes).

Our analysis focuses on belief updates in efficacy and hospitalization prevention rates.\footnote{
	The information provided for the side effects includes statistics for both the control group and treatment group in the selected scientific reports, where the subjects may not take only one of them to update their beliefs.
	Please see the online appendix for the results with the same specifications.} 
Table \ref{tab:update} summarizes the estimation of equation (\ref{eq:reg_persuation}),
where we include the observations of the vaccines that the subjects receive the information about.
Models (1) and (4) include only the signal disagreement term ($s-\mu_\theta$) and vaccine familiarity.
First, the belief change is positively correlated with signal strength,
which indicates that the subjects are (on average) correctly using the information to update beliefs.
Additionally, familiarity has a positive effect on belief updating, implying that subjects update even \textit{more} when exposed to information about vaccines they are more familiar with.

Building on the findings from Section \ref{sec:info_demand}---that subjects are more willing to read information about vaccines they are more familiar with---we expect greater belief updating when they receive information they selected themselves.
Models (2) and (5) directly test this hypothesis.
We find that subjects tend to update more when they receive and update according to the information they select.
In Models (3) and (6), we add familiarity to fully estimate Equation (\ref{eq:reg_persuation}); the positive effect of selected information persists.
Notably, for vaccines ranked in the top three for willingness to read but not selected, there is no significant belief updating.

We discuss possible reasons for the positive effect of selected information on belief updating.
First, subjects may seek information that justifies their past vaccination choices.
As shown in Table \ref{tab:info_demand_received},
subjects are especially interested in information about vaccines they have already received.
If self-justification drives selection,
vaccine-receipt history should enter positively when added to the regressions in Table \ref{tab:update}.
This test is reported in Table \ref{tab:update_vaccined}.
The coefficients of vaccine-receipt history are positive,
while the familiarity coefficients become statistically insignificant,
supporting the self-justification conjecture.
However, the ``Selected'' dummy remains positive and significant.
Thus, past-vaccine justification cannot fully account for the additional persuasion associated with self-selected information.\footnote{
	We also estimate the specifications with only the observations with the vaccines that the subjects have received before. 
	We do not see the effects in the information selection on belief updates, which also casts doubt on the justification channel.
	However, we note that the null results may come from the small sample size due to filtering.
}

A second potential mechanism is motivated reasoning.
If the rational information acquisition framework holds, 
subjects should expect that, on average, new information will lead to more optimistic beliefs about vaccine effectiveness.
Given this expectation, subjects may react more strongly to information that is optimistic---that is, signals that vaccine effectiveness exceeds their prior beliefs.
This form of motivated reasoning could generate asymmetric updating between subjects who initially under- or overestimate vaccine effectiveness.
To test this, we re-estimate the benchmark regressions from Table \ref{tab:update} separately for subjects who under- and overestimated effectiveness, as shown in Table \ref{tab:update_under} and Table \ref{tab:update_over}.
Among those who underestimated effectiveness (i.e., received signals above their prior), the positive effect of selecting the information persists.
In contrast, for those who overestimated effectiveness (i.e., received signals below their prior), the effect of selection becomes statistically insignificant.
This asymmetric updating pattern provides suggestive evidence of motivated reasoning, consistent with directional sensitivity in the information acquisition model.

\subsubsection{Indirect Update}

In both real-world settings and our experiment, subjects may update their beliefs about one vaccine based on information about another
(e.g., updating the beliefs about Pfizer when only Moderna information is available).
In this part, we discuss the indirect update patterns in our experiment.
To quantify indirect updates, we incorporate the signal strength of other vaccines into the regression models.
Specifically, we define this variable as the mean signal strength across other vaccines for which the subject received information.\footnote{
	For the specifications that include signal strength of other vaccines,
	the observations are dropped if there is no information about ``other vaccines'' received.
	Two possible cases are included: (1) the subject acquires information about zero vaccine and is assigned to the Full Compliance group, and (2) the subject acquires information about one vaccine and is assigned to the Full Compliance group while the observation is of the exact vaccine that she selects and receives the information.
}

We focus on how much subjects respond to the indirect information.
Intuitively, indirect updates should be smaller than direct ones, as indirect information reflects only shared uncertainty across vaccines---not vaccine-specific signals.\footnote{
	This statement is shown formally in Lemma \ref{lemma:multi_vaccine_update} in the appendix.
}
(For example, information about Pfizer may inform general vaccine effectiveness but reveals little about Moderna-specific performance.)

We estimate an alternative version of Equation \ref{eq:reg_persuation} that includes signal disagreement from other vaccines. Results are presented in Table \ref{tab:update_other}.
Models (1) and (3) focus on cases where subjects did not receive direct information about the focal vaccine.
The estimated coefficients for indirect updates (from other-vaccine signal disagreement) are smaller than their direct-update counterparts in Table \ref{tab:update}.
Models (2) and (4) examine cases where subjects receive both direct and indirect information.
In these cases, indirect information has no significant effect on belief updating once direct information is available.
These findings are qualitatively consistent with the theoretical framework.

\begin{result}[Belief Update] 
	\quad
\begin{itemize}
\item Belief updating is larger when the information received is more surprising (i.e., greater disagreement with pre-treatment beliefs).
\item Belief updating is larger when subjects select the information themselves.
\item When subjects receive information about multiple vaccines, they update more based on the direct information than the indirect information (other vaccines).
\end{itemize}
\end{result}

\subsection{Preference Changes}
We next examine whether receiving information alters subjects' vaccine preferences or their willingness to receive specific vaccines.
Table \ref{preference-treatment} summarizes changes in vaccine preferences across the different information-demand groups.
On average, stated vaccine preferences increase by 0.03 points (on a 0-100 scale) after the treatment phase (SD $= 18.12$),
while the average absolute change is 10.12 points (SD $= 15.03$).
We also elicited how many weeks subjects were willing to wait to receive each specific vaccine.
On average, subjects were willing to wait 0.99 additional weeks post-treatment (SD $= 9.40$),
with an average absolute change of 4.74 weeks (SD $= 8.17$).
Nonetheless, changes in willingness to wait are relatively modest.
In 37.5\% of observations, subjects reported the same number of weeks before and after the treatment phase,
and in 50.6\%, the change was less than or equal to one week.

We estimate the following model to examine how vaccine preferences evolve in response to updated beliefs:
\begin{align}
&(\text{Preference}_{i,v}^{Post}-\text{Preference}_{i,v}^{Pre})  \nonumber\\
= & \alpha + \sum_{l \in \mathcal{A}} \beta_l(\text{Effectiveness}_{i,v}^{l,Post}-\text{Effectiveness}_{i,v}^{l,Pre}) \nonumber \\
&+ \sum_{k\in \mathcal{D}}\delta_k \text{DemandCategory}^k_{i,v} 
+ X'_{i}\xi ,  \label{eq:preference_reg}
\end{align}
where the outcome is the change in stated preference for vaccine $v$ for subject $i$ before and after the treatment phase.
The set $\mathcal{A}$ includes the vaccine characteristics: efficacy, hospitalization prevention rate, adverse event rate, and severe adverse event rate.
The set $\mathcal{D}$ indexes the six information demand categories, defined by whether the vaccine was: (i) selected, (ii) not selected but ranked in the top 3, or (iii) not in the top 3, crossed with whether the subject received information or not.\footnote{
	\textit{Not received not top 3} is set as the baseline group.
} 
In short, this model captures how changes in vaccine preferences are associated with belief updates and subjects' demand for vaccine information.
If $\beta_l >0$,  subjects' preferences respond positively to changes in beliefs---consistent with rational updating.
We can further identify treatment effects within each group---Selected, Not Selected Top 3, and Not Top 3---by comparing the coefficients $\delta_k$ for information received versus not received within the same group.

The results are summarized in Table \ref{tab:preference_reg}.
Column (1) reports results for changes in stated preferences.
We find that positive changes in beliefs---such as higher efficacy or hospital prevention rates, or lower adverse event rates---are associated with more favorable vaccine preferences.
Furthermore, we find that, unconditionally, receiving information about a vaccine in the Selected Top 3 group increases stated preference by 2.77 points---equivalent to 0.16 standard deviations of the average change.
However, as shown in Column (2), we do not find significant effects on the number of weeks subjects are willing to wait.
Although the point estimates are directionally consistent with the preference results, the effects are not statistically significant.

To examine the extensive margin, we analyze responses to the \textit{never taking this vaccine} option included in the willingness-to-wait question.
Subjects who are unwilling to receive a vaccine even with zero wait time could select this \textit{never taking this vaccine} option.
Columns (3) and (4) of Table \ref{tab:preference_reg} report extensive-margin results, where the dependent variable is a post-treatment binary indicator for selecting the \textit{never taking} option.
Column (3) restricts the sample to cases where the \textit{never taking} option was selected pre-treatment, allowing us to interpret the outcome as a reversal of prior refusal---i.e., the extensive margin of vaccine uptake.
We find that receiving selected information reduces the share of never-takers by 18.38\%, compared to reductions of 10.53\% for non-selected Top 3 vaccines and 2.87\% for non-Top 3 vaccines.
Column (4) includes only observations where subjects did not select \textit{never taking} pre-treatment. In this sample, we find no notable differences in treatment effects across demand groups.
These findings offer additional support for the rational information acquisition framework: selected information has persuasive power and can shift decisions, even among subjects who initially refused certain vaccines.\footnote{
	Among selected information observations, only 5\% pertain to vaccines not considered prior to the intervention. As such, these results should be interpreted as suggestive---rather than conclusive---evidence of pivotal information choice. }

\begin{result}[Decision with Information]
If a subject receives the vaccine information she selected,
the preference of that vaccine will increase,
and she will be persuaded to take that vaccine if it was not considered before.
Additionally, subjects' preferences of vaccines positively correlates with the upward belief changes in vaccine effectiveness.
\end{result}

%
%

\section{Concluding Discussions}\label{sec:conclusion}

Our experimental design, which isolates each stage of the information acquisition process, reveals a distinct pattern: individuals systematically select information about vaccines they perceive as effective and are already familiar with. 
Consequently, when exposed to self-selected information, they exhibit more substantial belief revisions compared to non-selected information. 
This demonstrates that endogenous information demand is a central determinant of belief updating, even when information exposure itself is randomized.

This insight offers a policy implication for introducing new technologies. Because agents screen out information about unfamiliar options, simply increasing the supply of data is unlikely to increase engagement. Instead, governments may actively lower the barrier to entry. When introducing a new vaccine with comparable effectiveness to existing options, authorities may need to incentivize initial information access to overcome the rational inattention that prevents individuals from considering new alternatives.


There are several limitations to our experimental design.
First, we did not incentivize the elicitation of beliefs and vaccine preferences.
In our effort to maintain a naturalistic environment for subjects, we opted against using incentive-compatible mechanisms commonly used in laboratory settings.
As a result, we must assume that subjects report their beliefs and preferences truthfully and without bias, though the absence of incentives may introduce measurement noise.
Second, while belief variance plays a key role in the rational information acquisition framework, we did not directly elicit subjective variance.
Although we asked the self-reported familiarity with each vaccine as a proxy, this measure may not perfectly capture the uncertainty in subjects' beliefs.
While familiarity may partially capture belief dispersion, it could also reflect cognitive uncertainty or other factors (see \cite{enke2023} for discussion).

Our findings are relevant beyond the context of vaccines.
They may extend to any environment characterized by correlated information sources and endogenous information acquisition---such as financial markets or real estate decisions.
Because our design accommodates settings with correlated signal structures, it can serve as a framework for future studies to identify the mechanisms underlying information selection and belief updating in more complex, real-world environments.

\section*{Acknowledgment}
This project was approved by UCSB Human Subjects Committee (Protocol Number 23-21-0619) and NTU Research Ethics Committee (Protocol Number 202102HS016).
The project was pre-registered on AEA RCT registry (registry number AEARCTR-0008272).
We thank Lawrence Cheng-Hsien Hsu for his insights and knowledge about COVID-19 vaccines,
and we thank MobLab for providing aids during the online experiment sessions.
We thank Erik Eyster, Po-Hsuan Lin, Heather Royer,  Sevgi Yuksel, and the participants of ESA global meeting in Boston 2022, ESA North America Meeting in Charlotte 2023, and Western Economic Association International Annual Conference in San Francisco 2025 for their comments.
The authors utilized AI software for proofreading and minor stylistic improvements. All content and analytical interpretations remain the original work of the authors.
The authors are responsible for all mistakes.

\bibliographystyle{aer_abb}
\bibliography{vaccine}

\newpage
\section*{Tables}

\begin{table}[hp]
\caption{Summary of the Belief Changes}\label{tab:summary_main}
\centering
	\footnotesize
	\begin{tabular}{p{0.41\textwidth} 
		>{\centering\arraybackslash}p{0.12\textwidth} 
		>{\centering\arraybackslash}p{0.12\textwidth} 
		>{\centering\arraybackslash}p{0.12\textwidth} 
		>{\centering\arraybackslash}p{0.12\textwidth}}
	\hline\hline
	                    &    Efficacy&Hospitalization&Adverse Event&   Severe AE\\
\hline
Pre-treatment Beliefs&       69.00&       71.00&       60.76&       32.76\\
                    &     (21.33)&     (24.05)&     (27.26)&     (29.33)\\
[1em]
Post-treatment Beliefs&       72.10&       75.78&       52.60&       29.16\\
                    &     (23.26)&     (25.26)&     (29.30)&     (30.40)\\
[1em]
Adjustment in Beliefs (Post $-$ Pre)&       3.100&       4.780&      -8.161&      -3.601\\
                    &     (16.63)&     (21.77)&     (29.00)&     (24.18)\\
[1em]
$|$ Adjustment in Beliefs $|$&       11.41&       14.63&       21.34&       15.45\\
                    &     (12.49)&     (16.81)&     (21.27)&     (18.94)\\
[1em]
Pre-treatment Belief $-$ Information &       -12.98&       -25.40&       15.46&       32.02\\
                    &     (22.45)&     (25.05)&     (35.85)&     (29.34)\\
[1em]
$|$ Pre-treatment Belief $-$ Information $|$&       18.61&       26.25&       32.27&       32.07\\
                    &     (18.05)&     (24.16)&     (21.98)&     (29.28)\\

	\hline\hline
	\end{tabular}
	\begin{tfootnote}
		Standard deviations in the parentheses. 
		The information variable is the statistic on the page of vaccine information.
		The unit of the beliefs and the information is percentage (0--100).
		The statistics in this table include only data from the three main treatments.
		The information for Adverse Event and Severe Adverse Events is based on 
		the statistics of the treatment group in the report.
	\end{tfootnote}
\end{table}

\clearpage

\begin{table}[htp]
\centering
\caption{Information Preference and Selection \label{tab:info_demand}}
	{\footnotesize
	\begin{tabular}{l*{4}c}\hline\hline
&
\multicolumn{4}{c}{\textit{Dependent
Variables}}\\
&
\multicolumn{2}{c}{Info
Rank}
&
\multicolumn{2}{c}{Selected}\\
                    &\multicolumn{1}{c}{(1)}         &\multicolumn{1}{c}{(2)}         &\multicolumn{1}{c}{(3)}         &\multicolumn{1}{c}{(4)}         \\
\textit{Explanatory Variables}  &Belief Ranking         &Gap from Highest         &Belief Ranking          &Gap from Highest         \\
\hline
Efficacy            &        0.29\sym{***}&       -0.02\sym{***}&        7.20\sym{***}&       -0.63\sym{***}\\
                    &      (0.02)         &      (0.00)         &      (0.75)         &      (0.07)         \\
Hospitalization Prevention&        0.17\sym{***}&       -0.01\sym{***}&        4.47\sym{***}&       -0.15\sym{*}  \\
                    &      (0.02)         &      (0.00)         &      (0.78)         &      (0.07)         \\
Adverse Events      &        0.04\sym{*}  &        0.00\sym{**} &        2.27\sym{***}&        0.14\sym{**} \\
                    &      (0.02)         &      (0.00)         &      (0.64)         &      (0.05)         \\
Severe Adverse Events&        0.01         &        0.00\sym{*}  &       -0.03         &        0.01         \\
                    &      (0.02)         &      (0.00)         &      (0.60)         &      (0.07)         \\
Familiarity         &        0.21\sym{***}&        0.25\sym{***}&        7.29\sym{***}&        8.36\sym{***}\\
                    &      (0.01)         &      (0.01)         &      (0.60)         &      (0.62)         \\
Constants           &        0.57\sym{***}&        2.52\sym{***}&      -32.01\sym{***}&       17.73\sym{***}\\
                    &      (0.09)         &      (0.08)         &      (4.70)         &      (4.62)         \\
\hline
Observations        &        3145         &        3145         &        3145         &        3145         \\
Subjects            &         632         &         632         &         632         &         632         \\
 $ R^2 $            &        0.36         &        0.32         &        0.24         &        0.21         \\
Mean of Dep. Variable&           3         &           3         &        45.2         &        45.2         \\
 \hline\hline \end{tabular}                 \begin{tfootnote}                 Clustered (at subject level) standard errors in parentheses.  The subjects' family income, college majors, and sex are controlled.                  The coefficients and the mean of the dependent variable in (3) and (4) are in percentage.                 \sym{*} \(p<0.05\), \sym{**} \(p<0.01\), \sym{***} \(p<0.001\) . \end{tfootnote}

	}
\end{table}
\clearpage

\begin{table}[hp]
\centering
	\caption{Update in Beliefs (Direct) \label{tab:update}}
	{\footnotesize
	\begin{tabular}{l*{6}c}\hline\hline
&
\multicolumn{6}{c}{\emph{Belief
Update}:
}\\
&
\multicolumn{6}{c}{Post-Treatment
Belief
$-$
Pre-Treatment
Belief
}\\
                    &\multicolumn{1}{c}{(1)}         &\multicolumn{1}{c}{(2)}         &\multicolumn{1}{c}{(3)}         &\multicolumn{1}{c}{(4)}         &\multicolumn{1}{c}{(5)}         &\multicolumn{1}{c}{(6)}         \\
                    &    \multicolumn{3}{c}{Efficacy}         &\multicolumn{3}{c}{Hospitalization}  \\
\hline
Signal Disagreement &        0.30\sym{***}&        0.27\sym{***}&        0.28\sym{***}&        0.52\sym{***}&        0.63\sym{***}&        0.64\sym{***}\\
                    &      (0.03)         &      (0.03)         &      (0.03)         &      (0.04)         &      (0.05)         &      (0.05)         \\
Familiarity         &        0.58\sym{*}  &                     &        0.25         &        1.58\sym{***}&                     &        0.95\sym{*}  \\
                    &      (0.29)         &                     &      (0.31)         &      (0.45)         &                     &      (0.44)         \\
Selected Top 3      &                     &        3.05\sym{**} &        2.86\sym{*}  &                     &        4.60\sym{**} &        3.86\sym{**} \\
                    &                     &      (1.16)         &      (1.13)         &                     &      (1.43)         &      (1.42)         \\
Not Selected Top 3  &                     &        0.21         &        0.30         &                     &        0.87         &        0.59         \\
                    &                     &      (1.44)         &      (1.45)         &                     &      (1.74)         &      (1.75)         \\
Constants           &       -1.73         &       -1.22         &       -2.27         &      -14.58\sym{***}&      -10.64\sym{**} &      -14.10\sym{***}\\
                    &      (2.28)         &      (2.15)         &      (2.60)         &      (3.87)         &      (3.44)         &      (3.90)         \\
\hline
Observations        &        1754         &        1762         &        1754         &        1754         &        1762         &        1754         \\
Subjects            &         611         &         611         &         611         &         611         &         611         &         611         \\
 $ R^2 $            &        0.16         &        0.19         &        0.19         &        0.29         &        0.34         &        0.34         \\
Mean of Dep. Variable&        4.29         &        4.26         &        4.29         &        6.64         &        6.65         &        6.64         \\
Pre-treatment Beliefs Controlled?&          No         &         Yes         &         Yes         &          No         &         Yes         &         Yes         \\
 \hline\hline \end{tabular}\\ \begin{tfootnote}                 Clustered (at subject level) standard errors in parentheses. 
    Observations are at the vaccine level, including only the vaccines that subjects received the information about.
    Due to the missing responses on the familiarity question, the sample size is different in models (2) and (5).
    The subjects' family income, college majors, and sex are controlled in all models.                  Pre-treatment beliefs are controlled in models (2), (3), (5), and (6).                 \sym{*} \(p<0.05\), \sym{**} \(p<0.01\), \sym{***} \(p<0.001\) . \end{tfootnote}

	}
\end{table}

\clearpage

\begin{table}[htbp]
\centering
	\caption{Update in Beliefs (Indirect) \label{tab:update_other}}
	{\footnotesize
	\begin{tabular}{l*{4}c}\hline\hline
&
\multicolumn{4}{c}{\emph{Belief
Update}
}\\
                    &\multicolumn{1}{c}{(1)}         &\multicolumn{1}{c}{(2)}         &\multicolumn{1}{c}{(3)}         &\multicolumn{1}{c}{(4)}         \\
                    &    \multicolumn{2}{c}{Efficacy}         &\multicolumn{2}{c}{Hospitalization } \\
\hline
Signal Disagreement &                     &        0.26\sym{***}&                     &        0.58\sym{***}\\
                    &                     &      (0.03)         &                     &      (0.05)         \\
Signal Disagreement of Other Vaccines&        0.13\sym{**} &        0.05         &        0.37\sym{***}&        0.05         \\
                    &      (0.04)         &      (0.04)         &      (0.04)         &      (0.04)         \\
Selected Top 3      &        5.77\sym{***}&        2.83\sym{*}  &        1.95         &        4.36\sym{**} \\
                    &      (1.09)         &      (1.16)         &      (1.53)         &      (1.43)         \\
Not Selected Top 3  &       -0.85         &       -0.23         &        0.53         &        0.59         \\
                    &      (2.30)         &      (1.44)         &      (2.41)         &      (1.76)         \\
Constants           &       -2.95         &       -1.77         &       -9.59\sym{**} &      -11.14\sym{**} \\
                    &      (2.87)         &      (2.11)         &      (3.69)         &      (3.51)         \\
\hline
Observations        &        1209         &        1741         &        1209         &        1741         \\
Subjects            &         590         &         590         &         590         &         590         \\
 $ R^2 $            &        0.11         &        0.20         &        0.21         &        0.34         \\
Mean of Dep. Variable&        1.15         &        4.24         &        2.35         &        6.63         \\
Info Recevied?      &          No         &         Yes         &          No         &         Yes         \\
 \hline\hline \end{tabular}\\ \begin{tfootnote}[0.8\textwidth]                 Clustered (at subject level) standard errors in parentheses. 
    Observations are at the vaccine level.
    (1) and (3) includes only the vaccines that subjects did not receive the information about,
    and (2) and (4) includes only the observations that the subjects received both the information about the vaccine itself and other vaccines.
    The subjects' family income, college majors, sex, and pre-treatment beliefs  are controlled in all models.                  \sym{*} \(p<0.05\), \sym{**} \(p<0.01\), \sym{***} \(p<0.001\) . \end{tfootnote}

	}
\end{table}

\clearpage

\begin{table}[p]
\begin{center}
\caption{The Preferences of Vaccines: By Categories of Information Demand}
\label{preference-treatment}
\resizebox{\textwidth}{!}{
\begin{tabular}{@{\extracolsep{4pt}}lccccccc}
\hline\hline
& \multicolumn{2}{c}{Selected Top 3}
& \multicolumn{2}{c}{Non-Selected Top 3}
& \multicolumn{2}{c}{Not Top 3}&\\
\cline{2-3}\cline{4-5}\cline{6-7}
            &\multicolumn{1}{c}{Received}&\multicolumn{1}{c}{Not Received}&\multicolumn{1}{c}{Received}&\multicolumn{1}{c}{Not Received}&\multicolumn{1}{c}{Received}&\multicolumn{1}{c}{Not Received}&\multicolumn{1}{c}{Total}\\
\hline
\emph{Preference}&            &            &            &            &            &            &            \\
Pre-treatment&       82.90&       76.68&       75.17&       73.27&       50.99&       43.27&       66.09\\
            &     (21.56)&     (25.41)&     (25.49)&     (28.50)&     (36.36)&     (35.50)&     (33.78)\\
Post-treatment&       84.61&       75.64&       73.32&       73.24&       49.89&       42.52&       66.12\\
            &     (22.18)&     (27.65)&     (27.76)&     (29.48)&     (37.77)&     (36.13)&     (35.09)\\
Post $-$ Pre&        1.71&       -1.04&       -1.85&       -0.03&       -1.10&       -0.75&        0.03\\
            &     (15.06)&     (17.49)&     (20.32)&     (16.89)&     (22.21)&     (19.52)&     (18.12)\\
$|$ Post $-$ Pre $|$&        8.32&        9.68&       11.90&        9.11&       13.40&       10.99&       10.12\\
            &     (12.67)&     (14.59)&     (16.56)&     (14.21)&     (17.73)&     (16.15)&     (15.03)\\
\emph{Weeks Willing to Wait}&            &            &            &            &            &            &            \\
Pre-treatment&       16.56&       17.07&       16.80&       18.17&       17.31&       14.12&       16.37\\
            &     (14.79)&     (15.53)&     (15.49)&     (15.95)&     (15.77)&     (13.72)&     (14.96)\\
Post-treatment&       17.84&       17.39&       18.26&       19.02&       17.09&       14.70&       17.26\\
            &     (15.43)&     (15.88)&     (16.59)&     (15.30)&     (15.69)&     (14.09)&     (15.40)\\
Post $-$ Pre&        1.21&        0.94&        2.24&        0.38&       -0.85&       -0.27&        0.39\\
            &     (11.81)&     (10.20)&     (15.72)&     (13.58)&     (14.12)&     (14.98)&     (13.06)\\
$|$ Post $-$ Pre $|$&        5.50&        4.62&        8.13&        6.93&        6.42&        6.15&        5.59\\
            &     (10.52)&      (9.14)&     (13.64)&     (11.67)&     (12.60)&     (13.67)&     (11.81)\\
Never Take (Pre) (\%)&        5.39&        7.22&       10.27&       16.02&       40.40&       52.02&       24.10\\
            &     (22.59)&     (25.93)&     (30.41)&     (36.77)&     (49.14)&     (49.99)&     (42.77)\\
Never Take (Post) (\%)&        5.39&        9.03&       12.93&       15.05&       38.40&       50.60&       23.48\\
            &     (22.59)&     (28.71)&     (33.61)&     (35.84)&     (48.70)&     (50.02)&     (42.39)\\

\hline\hline
\end{tabular}
}
\end{center}
\begin{minipage}{\linewidth}
\footnotesize
\emph{Notes.} \textit{Preference} has the scale of 0--100.
The \textit{Weeks Willing to Wait} has the scale of 0--52.
The subjects can also choose ``never receiving this vaccine'',
in which case the willingness to wait of that observation is not counted.
\end{minipage}
\end{table}

\clearpage

\begin{table}[htp]
\centering
{\renewcommand{\arraystretch}{0.8}
\caption{Vaccine Preference and Beliefs in Effectiveness}\label{tab:preference_reg}
{\scriptsize
\begin{tabular}{l*{4}c}\hline\hline
&
\multicolumn{2}{c}{\underline{
PostPreference
$-$
PrePreference}
}
&
\\
                    &\multicolumn{1}{c}{(1)}         &\multicolumn{1}{c}{(2)}         &\multicolumn{1}{c}{(3)}         &\multicolumn{1}{c}{(4)}        \\
                    &Preference (0-100)         &  WTW (0-52)         
                    & \multicolumn{2}{c}{Never Take After Treatment (\%)}   \\
\hline
\emph{Belief Differences}&                     &                     &                     &           \\
Efficacy            &        0.08\sym{**} &        0.01         &        0.04         &       -0.09\sym{*}  \\
                    &      (0.03)         &      (0.02)         &      (0.09)         &      (0.04)         \\
Hospitalization     &        0.11\sym{***}&       -0.01         &       -0.18\sym{*}  &       -0.04         \\
                    &      (0.02)         &      (0.02)         &      (0.07)         &      (0.03)         \\
Adverse Events      &       -0.05\sym{**} &        0.00         &        0.10\sym{*}  &        0.02         \\
                    &      (0.02)         &      (0.01)         &      (0.05)         &      (0.02)         \\
Severe Adverse Events&       -0.03         &       -0.01         &       -0.00         &        0.02         \\
                    &      (0.02)         &      (0.02)         &      (0.07)         &      (0.02)         \\
\emph{Information Demand}&                     &                     &                     &                     \\
Selected Top 3 -- Received&        1.41         &        1.50\sym{*}  &      -24.65\sym{***}&       -5.52\sym{***}\\
                    &      (0.80)         &      (0.62)         &      (7.07)         &      (1.33)         \\
-- Not Received     &       -1.35         &        1.37         &       -6.27         &       -4.34\sym{**} \\
                    &      (1.22)         &      (0.81)         &      (7.64)         &      (1.62)         \\
Not Selected Top 3 -- Received&       -1.64         &        2.59\sym{*}  &      -16.90\sym{*}  &       -2.00         \\
                    &      (1.61)         &      (1.27)         &      (8.41)         &      (1.96)         \\
 -- Not Received    &        0.33         &        0.81         &       -6.37         &       -5.50\sym{**} \\
                    &      (1.29)         &      (1.11)         &      (9.46)         &      (1.77)         \\
Not Top 3 -- Received&       -1.16         &       -0.36         &       -2.87         &       -1.72         \\
                    &      (1.34)         &      (0.92)         &      (3.41)         &      (2.05)         \\
Constants           &        1.63         &        1.20         &       78.98\sym{***}&       11.59\sym{***}\\
                    &      (2.07)         &      (1.65)         &      (7.34)         &      (2.43)         \\
\hline
Treatment Effects in \dots&                     &                     &                     &                     \\
Selected Top 3      &        2.77\sym{**} &        0.13         &      -18.38         &       -1.19         \\
                    &      (1.02)         &      (0.76)         &      (9.90)         &      (1.12)         \\
Not Selected Top 3  &       -1.97         &        1.79         &      -10.53         &        3.51         \\
                    &      (1.91)         &      (1.56)         &     (12.44)         &      (2.19)         \\
Not Top 3           &       -1.16         &       -0.36         &       -2.87         &       -1.72         \\
                    &      (1.34)         &      (0.92)         &      (3.41)         &      (2.05)         \\
\hline
Subgroup            &         All         &         All         &  Never Take         &Willing to Take         \\
Observations        &        3160         &        3160         &         759         &        2401         \\
Subjects            &         632         &         632         &         491         &         616         \\
 $ R^2 $            &       0.052         &       0.025         &        0.11         &       0.035         \\
Mean of Dep. Variable&       0.035         &        0.56         &        86.0         &        4.00         \\
Subgroup            &         All         &         All         &  Never Take         &Willing to Take         \\
 \hline\hline \end{tabular} \begin{tabular}{m{0.8\textwidth}}
\scriptsize \textit{Notes.}
 Clustered (at subject level) standard errors in parentheses.                 The subjects' family income, college majors, and sex are controlled.                 The dependent variables of the first two colunms are the differences between the preference variables before and after the treatments.                 The first column is the changes in preference evaluations of the vaccines,                 and the second column is the changes in numbers of weeks that the subjects are willing to wait for the vaccines.                 The dependent variable of the last three columns is the binary variable of whether the subject never takes the vaccine;                 the coefficients are measured in percentage term.                 Column  (4) includes observations reported \textit{never taking that vaccine} prior to the treatment,                 and column (5) includes observations reported \textit{willing to wait for that vaccine} prior to the treatment.                 \sym{*} \(p<0.05\), \sym{**} \(p<0.01\), \sym{***} \(p<0.001\) \end{tabular}

}
}
\end{table}

\clearpage
\section*{Figures}

\begin{figure}[htbp] 
\centering
	\begin{tikzpicture}[scale=2 ,domain=-2.5:2.5,samples=40]
	\def \MeanPrior {-0.4}
	\def \MeanPost {1}
	\def \SignalStar {1.2}
	\draw[draw=black!50, thin, smooth]
        plot (\x,{gauss(\MeanPrior,0.5)}) ;
        \node [black!50, left] at (\MeanPrior,0.5) {\small Prior Belief};
    \draw[draw=red, thin, smooth]
        plot (\x,{gauss(\MeanPost,0.25)});
        \node [red, right] at (\MeanPost,1) {\small Posterior Belief given $s$};
    \draw [->, thick](-2.5,0)--(2.5,0) node [below] {$\theta$ (effectiveness);};
	    \node at (2.5,-0.5) {$s$ (signal received)};
	\fill[fill=purple!20, thin, smooth,domain=-2.5: 2,samples=40]
        plot (\x,{gauss(\MeanPrior,0.65)-2}) ;
	\draw[draw=purple!50, thin, smooth,domain=-2.5 : 2,samples=40]
        plot (\x,{gauss(\MeanPrior,0.65)-2}) ;
        \node [black!50] at (\MeanPrior,-2.2) {\small Distribution of $s$};
	\draw [->,thick] (-2.3,-0.2)--(-2.3,2) node [left] {$f(\theta)$; $f_{s}(\theta)$};   
    \draw[dashed, color=black!50] (\MeanPrior,2)--(\MeanPrior,0) node[below] {$\mu_\theta$};
    \draw[thin,dotted,blue] (\SignalStar,2)--(\SignalStar,0) node[below] {$s^*$};
    \draw[thin,dashed] (1.7,2)--(1.7,0) node[below] {$s$};
    \draw[dashed, orange] (0.5,2)--(0.5,0) node[below]{$\vbar$};
    \draw [-> , thin, red!50] (0.1,0.8)--(0.4,0.8);
    \draw [-> , thin, red!50] (0.1,0.6)--(0.4,0.6); 
    \node[orange] at (0.5,-0.5) {\small (reservation value)};
    \draw [color=Bittersweet] (-2.5,-1.5)--(\SignalStar,-1.5)--(2.5,-0.8)
    	node [right] {$v(s)$};
    \draw[thin,dotted,blue] (\SignalStar,-0.7)--(\SignalStar,-2) node[below] {$s^*$};
    \draw [->,thick] (-2.5,-2)--(2.5,-2) node [below] {$s$ (signal received)};
    \draw [->,thick] (-2.3,-2.2)--(-2.3,-0.5) node [left] {$v(s)$};
    \node [left] at (-2.5,-1.5) {$\vbar$};
    \node [black!80] at (-0.55, -1.7) {\small (not taking vaccine)};
    \node [black!80] at (2, -1.7) {\small (taking vaccine)};
	\end{tikzpicture}
\caption{A Demonstration of the Belief Update in Vaccine Effectiveness \label{fig:model}}
	\begin{tfootnote}
	The top panel demonstrates the probability density of the vaccine effectiveness, $\theta$.
	The gray distribution represents the prior distribution, 
	where $\mu_\theta$ is its mean.
	After receiving a signal of $s$, the belief is adjusted to the posterior distribution (the red curve).
	$\vbar$ is the reservation value;
	when the posterior mean exceeds $\vbar$, the DM will take the vaccine,
	and the expected value from taking it is $\E{\theta|s}$.
	$s^*$ is the critical signal that switches the DM's decision of taking the vaccine or not.
	The bottom figure demonstrates the value function after receiving the signal $s$.
	On the left of the critical signal $s^*$, 
	the value is the reservation value $\vbar$, as the DM will not take the vaccine.
	On the right of $s^*$, the DM will take the vaccine,
	and the value becomes the expected efficacy, $\E{\theta|s}$.
	\end{tfootnote}
\end{figure}

\clearpage

\clearpage

\newpage
\appendix
\section*{Online Appendix}
\renewcommand{\thesection}{\Alph{section}.}
\renewcommand{\thesubsection}{\Alph{section}.\arabic{subsection}}

\section{Proofs and Supplementary Theory}

\subsection{Proof of Proposition \ref{theory:comp_stat}} \label{sec:proof_comp_stat}

Define the relative accuracy of the prior belief $\gamma\equiv\frac{\sigs^2}{\sigt^2}$,
then $\E{\theta|s} = \frac{\gamma \mu_\theta+s }{1+\gamma}$ and $s^* = \overline{v}+\gamma(\overline{v}-\mu_\theta)$.
Denote $z^* = \frac{s^* - \mu_\theta}{\sqrt{\sigs^2 + \sigt^2}}$.
Also denote $\Phi(\cdot)$ and $\phi(\cdot)$ as the cumulative distribution function (cdf) and the probability density function (pdf) of standard normal distribution.

Given the setting, we can find the value of the information as follows:
\begin{align}
&\Ex{s}{v(s)}-v_0 \nonumber \\
=& \overline{v}\Pr(s \leq s^*)  + \E{\frac{\gamma\mu_\theta + s}{1+\gamma} \bigg| s > s^* } \Pr(s > s^*)-v_0 \nonumber \\
=&\overline{v}\cdot \Phi\left(\frac{s^* - \mu_\theta}{\sqrt{\sigs^2 + \sigt^2}}\right) 
+ \left(\frac{\gamma \mu_\theta+\E{s|s>s^*}}{1+\gamma}\right) \left( 1- \Phi\left(\frac{s^* - \mu_\theta}{\sqrt{\sigs^2 + \sigt^2}}\right) \right)-v_0 \nonumber \\
=&-\overline{v}(1 - \Phi\left(z^*\right)) + \left(\frac{\gamma \mu_\theta+\mu_\theta + \sqrt{\sigs^2 + \sigt^2}\phi(z^*)/\left( 1- \Phi\left(z^*\right) \right) }{1+\gamma} \right)\left( 1- \Phi\left(z^*\right) \right) - (v_0-\overline{v})\nonumber  \\
=& (\mu_\theta - \overline{v})(1 - \Phi\left(z^*\right)) +
\frac{\sqrt{\sigs^2 + \sigt^2}\phi(z^*) }{1+\gamma} - (v_0-\overline{v}). \label{eq:proof_v0}
\end{align}
When $\mu_\theta \geq \vbar$, $v_0=\mu_\theta$, and (\ref{eq:proof_v0}) becomes
$$ -(\mu_\theta - \overline{v}) \Phi\left(z^*\right) +
\frac{\sqrt{\sigs^2 + \sigt^2}\phi(z^*) }{1+\gamma} ,$$
which decreases in $\mu_\theta-\vbar$. 
When  $\mu_\theta < \vbar$, $v_0=\vbar$, and (\ref{eq:proof_v0}) becomes
$$ (\mu_\theta - \overline{v})(1 - \Phi\left(z^*\right)) +
\frac{\sqrt{\sigs^2 + \sigt^2}\phi(z^*) }{1+\gamma},$$
which increases in $\mu_\theta-\vbar$. 

We then show (\ref{eq:proof_v0}) increases in $\sigma_\theta^2$.
Note that 
$$ \frac{\dd \gamma}{\dd \left(\sigma_\theta^2\right)} = -\frac{\sigma_s^2}{(\sigma_\gamma^2)^2} = -\frac{\gamma}{\sigma_\theta^2}$$
and
\begin{align*}
\frac{\dd z^*}{\dd \left(\sigma_\theta^2\right)} =& 
\frac{-\frac{\gamma}{\sigma_\theta^2}(\vbar-\mu_\theta)\sqrt{\sigma_s^2+\sigma_\theta^2}
	-\frac{(1+\gamma)(\vbar-\mu_\theta)}{2\sqrt{\sigma_s^2+\sigma_\theta^2}}}{\sigma_s^2+\sigma_\theta^2}
= \frac{-\frac{\gamma}{\sigma_\theta^2}(\vbar-\mu_\theta)\sqrt{\sigma_s^2+\sigma_\theta^2}
	-\frac{z^*}{2}}{\sigma_\theta^2 (1+\gamma)} .
\end{align*}
Then
\begin{align*}
&\frac{\dd }{\dd \left(\sigma_\theta^2\right)}\left\{ (\mu_\theta - \overline{v})(1 - \Phi\left(z^*\right)) +
\frac{\sqrt{\sigs^2 + \sigt^2}\phi(z^*) }{1+\gamma} - (v_0-\overline{v})\right\} \\
=& -(\mu_\theta - \overline{v})\phi\left(z^*\right)\frac{\dd z^*}{\dd \sigma_\theta^2}
+\frac{\frac{1+\gamma}{2\sqrt{\sigs^2 + \sigt^2}} - \frac{-\gamma\sqrt{\sigs^2 + \sigt^2}}{\sigma_\theta^2}}{(1+\gamma)^2}\phi(z^*) 
+ \frac{\sqrt{\sigs^2 + \sigt^2}}{1+\gamma}\cdot (-z^*)\phi(z^*) \frac{\dd z^*}{\dd \sigma_\theta^2}\\
=& (\vbar - \mu_\theta )\phi\left(z^*\right)\frac{\dd z^*}{\dd \sigma_\theta^2}
+\frac{\frac{\sigma_\theta^2(1+\gamma)}{2\sqrt{\sigs^2 + \sigt^2}} + \gamma\sqrt{\sigs^2 + \sigt^2}}{\sigma_\theta^2(1+\gamma)^2}\phi(z^*) 
 -(\vbar - \mu_\theta)\phi(z^*) \frac{\dd z^*}{\dd \sigma_\theta^2} \\
=& \frac{\sigma_\theta^2(1+\gamma)+ 2\gamma(\sigs^2 + \sigt^2)}{2
\sqrt{\sigs^2 + \sigt^2}\sigma_\theta^2(1+\gamma)^2}\phi(z^*) \\
=&\frac{\sigma_\theta^2(1+\gamma)+ 2\gamma\sigma_\theta^2(1+\gamma)}{2
\sqrt{\sigs^2 + \sigt^2}\sigma_\theta^2(1+\gamma)^2}\phi(z^*) 
= \frac{1+ 2\gamma}{2\sqrt{\sigs^2 + \sigt^2}(1+\gamma)}\phi(z^*) >0.
\end{align*}
Lastly, we show (\ref{eq:proof_v0}) decreases in $\sigma_s^2$.
$$ \frac{\dd \gamma}{\dd \left(\sigma_s^2\right)} = \frac{1}{\sigma_\gamma^2} $$
and
\begin{align*}
\frac{\dd z^*}{\dd \left(\sigma_s^2\right)} =& 
\frac{\frac{1}{\sigma_\theta^2}(\vbar-\mu_\theta)\sqrt{\sigma_s^2+\sigma_\theta^2}
	-\frac{(1+\gamma)(\vbar-\mu_\theta)}{2\sqrt{\sigma_s^2+\sigma_\theta^2}}}{\sigma_s^2+\sigma_\theta^2}
= \frac{\frac{1}{\sigma_\theta^2}(\vbar-\mu_\theta)\sqrt{\sigma_s^2+\sigma_\theta^2}
	-\frac{z^*}{2}}{\sigma_\theta^2 (1+\gamma)} .
\end{align*}
Then
\begin{align*}
&\frac{\dd }{\dd \left(\sigma_s^2\right)}\left\{ (\mu_\theta - \overline{v})(1 - \Phi\left(z^*\right)) +
\frac{\sqrt{\sigs^2 + \sigt^2}\phi(z^*) }{1+\gamma} - (v_0-\overline{v})\right\} \\
=& -(\mu_\theta - \overline{v})\phi\left(z^*\right)\frac{\dd z^*}{\dd \sigma_s^2}
+\frac{\frac{1+\gamma}{2\sqrt{\sigs^2 + \sigt^2}} - \frac{\sqrt{\sigs^2 + \sigt^2}}{\sigma_\theta^2}}{(1+\gamma)^2}\phi(z^*) 
+ \frac{\sqrt{\sigs^2 + \sigt^2}}{1+\gamma}\cdot (-z^*)\phi(z^*) \frac{\dd z^*}{\dd \sigma_s^2}\\
=& (\vbar - \mu_\theta )\phi\left(z^*\right)\frac{\dd z^*}{\dd \sigma_\theta^2}
+\frac{\frac{\sigma_\theta^2(1+\gamma)}{2\sqrt{\sigs^2 + \sigt^2}} -\sqrt{\sigs^2 + \sigt^2}}{\sigma_\theta^2(1+\gamma)^2}\phi(z^*) 
 -(\vbar - \mu_\theta)\phi(z^*) \frac{\dd z^*}{\dd \sigma_\theta^2} \\
=& \frac{\sigma_\theta^2(1+\gamma)-2(\sigs^2 + \sigt^2)}{2
\sqrt{\sigs^2 + \sigt^2}\sigma_\theta^2(1+\gamma)^2}\phi(z^*) \\
=&\frac{\sigma_\theta^2(1+\gamma)-2\gamma\sigma_\theta^2(1+\gamma)}{2
\sqrt{\sigs^2 + \sigt^2}\sigma_\theta^2(1+\gamma)^2}\phi(z^*) 
= \frac{-1}{2\sqrt{\sigs^2 + \sigt^2}(1+\gamma)}\phi(z^*) <0.
\end{align*}
Thus (\ref{eq:proof_v0}) decreases in $\gamma\left(\equiv\frac{\sigs^2}{\sigt^2}\right).$

\subsection{The Case of Multiple Vaccines}
\label{sec:multi_vaccine}

\setcounter{prop}{0}
\renewcommand{\theassumption}{\thesection\arabic{assumption}}
\renewcommand{\thelemma}{\thesection\arabic{lemma}}
\renewcommand{\theprop}{\thesection\arabic{prop}}

The basic decision structure is the same as the one-vaccine case,
despite that the beliefs are extended to the case of five vaccines.
We divide the vaccine effectiveness into two parts: 
the common factor among different vaccines, and the vaccine specific factor.
Specifically, let $\theta_j = \overline{\theta} + \tilde{\theta}_j$,
where $\overline{\theta}\sim N(\overline{\mu},\overline{\sigma}_\theta^2)$ is the \textit{common} belief of the vaccine effectiveness across brands,
and $\tilde{\theta}_j \sim N(\tilde{\mu}_j, \tilde{\sigma}_j^2)$ is the \textit{vaccine $j$ specific} belief of the effectiveness.
Then we write
$$\theta_j \sim N(\mu_j,\sigma^2_j),$$
where $\mu_j = \overline{\mu}+\tilde{\mu}_j$.
We formally state two more assumptions to capture the division.
\begin{assumption} 
	\begin{enumerate}[(1)]
	\item The common factor is independent with the vaccine specific factor 
		(or formally, $\tilde{\theta}_j \perp \overline{\theta}$).
	\item The vaccine specific factors are independent between any two distinct vaccines $j$ and $k$
		(or formally, $\tilde{\theta}_j \perp \tilde{\theta}_k$ for any $j\neq k$.)
	\end{enumerate}
\end{assumption}
Notice that 
$$\Cov{\theta_j,\theta_k} 
= \Cov{\overline{\theta} + \tilde{\theta}_j,\overline{\theta} + \tilde{\theta}_k}
=\Cov{\overline{\theta},\overline{\theta}} = \Var{\overline{\theta}} = \overline{\sigma}_\theta^2.$$
Intuitively, as the only factor that any two vaccines share is the common factor,
the covariance between vaccines beliefs is the variance of the common factor.

Let $\bm{\theta}=(\theta_1,\dots,\theta_J)^T$. Then we can find $\theta$ follows multivariate normal distribution:
$$\bm{\theta}\sim N\left(\bm{\mu}, \bm{\Sigma}_{\bm{\theta}} \right),$$
where $\bm{\mu}=(\mu_1,\dots,\mu_J)^T$, ${\bm{\Sigma}_{\bm{\theta}}}_{jj}= \sigma^2_j$, 
and ${\bm{\Sigma}_{\bm{\theta}}}_{jk}= \overline{\sigma}_\theta^2$ for every $j\neq k$.

The information of the vaccine $j$, $s_j$, is centered at $\theta_j$ with a normally distributed error term, $\varepsilon_j \sim N(0,\sigma^2_s)$,
$$s_j = \theta_j + \varepsilon_j.$$
We add two more assumptions on the information structure.
\begin{assumption} 
	\begin{enumerate}[(1)]
	\item The disturbance in the information about two vaccines are independent
		(or formally, $\varepsilon_j \perp\varepsilon_k$ for every $j\neq k$.)
	\item The disturbance in the information is independent with the effectiveness of the vaccines 
		(or formally, $\varepsilon_j \perp\bm{\theta}$.)
	\end{enumerate}
\end{assumption}
Intuitively, we assume that the disturbances from the signal 
(for example, the errors from the laboratory trials)
do not relate to the effectiveness of the vaccine itself,
and they are independent among different vaccine brands.
The two assumption imply that $\Var{s_j|\bm{\theta}}=\Var{\varepsilon_j}=\sigma^2_s$ for every $j$.
Then we can determine the conditional distribution of $\bm{s}=(s_1,\dots,s_J)^T$,
$$\bm{s} | \bm{\theta} \sim N(\bm{\theta}, \bm{\Sigma}_{\bm{s}}),$$
where $\bm{\Sigma}_{\bm{s}} = \sigma^2_s\bm{I}_J.$

Let $D_j\in\{0,1\}$ be the decision of whether to acquire the information of vaccine $j$ 
($D_j=1$ means the vaccine information is received, and $D_j=0$ means the opposite),
and $\bm{D}=(D_1,\dots,D_J)^T$.
We define $\bm{s}^*=(D_1s_1,\dots,D_Js_J)^T$, 
and $\bm{\Omega}^*_{\bm{s}}=\frac{1}{\sigma^2_s}(\bm{D}\bm{D}^T)$.
Intuitively, we assign the deterministic value 0 to the vaccines that the agent does \textit{not} receive.
The corresponding distribution, $\bm{s}^*$, follows a \textit{degenerate multivariate normal distribution},
$$\bm{s}^*|\bm{\theta} \sim N(\theta,\bm{\Omega}).$$
Then the Bayesian posterior of $\bm{\theta}$ given $\bm{s}^*$ obeys the following distribution,
$$\bm{\theta}|\bm{s}^*\sim 
N\left(\left(\bm{\Omega}+\bm{\Sigma}_{\bm{\theta}}^{-1}\right)^{-1}
	\left(\bm{s}^T\bm{\Omega}+\bm{\mu}^T\bm{\Sigma}_{\bm{\theta}}^{-1}\right)^T,
\left(\bm{\Omega}+\bm{\Sigma}_{\bm{\theta}}^{-1}\right)^{-1}\right).$$

The decision environment is the similar to the single vaccine case.
Let $\vbar$ be the reservation value without any vaccine.
Given the information, the agent compares the vaccine with the highest posterior mean with the reservation value.
Denote $(\hat{\mu}_1|\textbf{s}^*,\dots,\hat{\mu}_J|\textbf{s}^* ) \equiv \E{\bm{\theta}|\bm{s}^*} $,
and $\hat{\mu}^*|\textbf{s}^* = \max_{j\in\{1,\dots,J\}} \hat{\mu}_j|\textbf{s}^*$. 
That is, $\hat{\mu}^*|\textbf{s}^*$ is the effectiveness of the best vaccine given the information received.
Call $j^*$ the best vaccine.
Then the agent chooses vaccine $j^*$ if $\hat{\mu}^*|\textbf{s}^* \geq \overline{v}$ and stay at the reservation value otherwise,
and the value function becomes
\begin{equation} \label{eq:value_s_mul}
v(\textbf{s}^*) = 
\begin{cases}
\hat{\mu}^*|\textbf{s}^*& \text{ if } \hat{\mu}^*|\textbf{s}^* \geq \overline{v}\\
\overline{v}& \text{ if } \hat{\mu}^*|\textbf{s}^* <\overline{v} \end{cases}.
\end{equation} 

Hence, we can follow the same setting in Section \ref{sec:theory_single} and calculate the value of information.
The DM acquire the information if 
$$ \Ex{\bm{s}}{v(\bm{s}^*)} - C(\bm{D}) \geq \overline{v}, $$
where $C(\cdot)$ is the cost function of acquiring information.
For simplicity, we assume the cost of acquiring the information about each brand of vaccine to be a constant $c$,
so the total cost of acquiring information about $n$ vaccines becomes $nc$.

In the following subsection, we provide an example of two vaccines.

\subsubsection{An Example of Two Vaccines}
Following the settings, we list the environment of the two-vaccine case as the following assumptions.

\begin{assumption}\label{ass:2vaccine_eff}
Let the effectiveness of vaccine 1 be $\theta_1 = \overline{\theta} + \tilde{\theta}_1$ and the effectiveness of vaccine 2 be $\theta_2 = \overline{\theta} + \tilde{\theta}_2$,
where $\overline{\theta} \sim N(\overline{\mu}, \overline{\sigma}_\theta^2)$,
$\tilde{\theta}_j \sim N(\tilde{\mu}_j , \tilde{\sigma}^2_{j})$, $\tilde{\theta}_j \perp \overline{\theta}$, 
and $\tilde{\theta}_1 \perp \tilde{\theta}_2$.
Denote $\mu_j = \overline{\mu} + \tilde{\mu}_j$ and $\sigma^2_j = \overline{\sigma}^2_\theta + \tilde{\sigma}^2_j$.
Thus $\bm{\theta} = (\theta_1,\theta_2)$ follows a bivariate normal distribution,
$$ \bm{\theta} \sim N\left(\left[\begin{array}{c}\mu_1 \\ \mu_2 \end{array}  \right],
\left[\begin{array}{cc}\sigma^2_1 & \overline{\sigma}_\theta^2 \\ 
\overline{\sigma}_\theta^2 & \sigma^2_2 \end{array}  \right]\right).$$
\end{assumption}
\begin{assumption} \label{ass:2vaccine_signal}
The information that the DM receives for the vaccines are $s_1 = \theta_1 + \varepsilon_1$ and $s_2= \theta_2 + \varepsilon_2$,
where $\varepsilon_j \sim N(0,\sigma_s^2)$, $\varepsilon_1 \perp \varepsilon_2$, and $\varepsilon_j \perp \bm{\theta}$.
\end{assumption}
Given the cases that the DM receives only $s_1$, only $s_2$, or both $s_1$ and $s_2$,
the means of the Bayesian posteriors are jointly normally distributed.
\begin{lemma}\label{lemma:multi_vaccine_update}
Given Assumptions \ref{ass:2vaccine_eff} and \ref{ass:2vaccine_signal},
the mean of the posterior beliefs given receiving the signal sets 
$\{s_1\}$, $\{s_2\}$, and $\{s_1, s_2\}$ are respectively
\begin{align*}
\E{\bm{\theta} | s_1} &=	\left[
	\begin{array}{c}
	\mu_1 +\frac{\sigma_1^2  }{\sigma_s^2+ \sigma_1^2 }(s_1-\mu_1) \\
	\mu_2 +\frac{\overline{\sigma}^2_\theta  }{\sigma_s^2+ \sigma_1^2 } (s_1-\mu_1)
	\end{array}
	\right]. \\	
\E{\bm{\theta} | s_2} &=	\left[
	\begin{array}{c}
	\mu_1 +\frac{\overline{\sigma}^2_\theta   }{\sigma_s^2+ \sigma_2^2 }(s_2-\mu_2) \\
	\mu_2 +\frac{\sigma_2^2  }{\sigma_s^2+ \sigma_2^2 } (s_2-\mu_2)
	\end{array}
	\right]. \\	
\E{\bm{\theta} | s_1,s_2} &=	\left[
	\begin{array}{c}
	\mu_1 +\frac{\sigma_1^2 }{\sigma_s^2+ \sigma_1^2 + \sigma^2_2}(s_1-\mu_1) 
			+\frac{\overline{\sigma}^2_\theta}{\sigma_s^2+ \sigma_1^2 + \sigma^2_2}(s_2-\mu_2) 
			+\frac{\sigma^2_1\sigma^2_2 - (\overline{\sigma}^2_\theta)^2}{\sigma^2_s(\sigma_s^2+ \sigma_1^2 + \sigma^2_2)}s_1\\
	\mu_2 +\frac{\overline{\sigma}^2_\theta }{\sigma_s^2+ \sigma_1^2 + \sigma^2_2}(s_1-\mu_1) 
			+\frac{\sigma^2_2}{\sigma_s^2+ \sigma_1^2 + \sigma^2_2}(s_2-\mu_2) 
			+\frac{\sigma^2_1\sigma^2_2 - (\overline{\sigma}^2_\theta)^2}{\sigma^2_s(\sigma_s^2+ \sigma_1^2 + \sigma^2_2)}s_2
	\end{array}
	\right]. 
\end{align*}
\end{lemma}
When the information about the vaccine 1 only is received, 
the update in the mean of the belief of vaccine 1 effectiveness is determined by the weighted difference between the information ($s_1$) and the mean of vaccine 1's prior belief ($\mu_1$), where
when the signal is relatively more accurate ($\sigma_s^2$ higher or $\sigma_1$ lower),
the magnitude of the belief update is larger.
This prediction is identical with the single vaccine version.
Furthermore,
the update in the mean of the belief of vaccine 2 effectiveness is also determined by the weighted difference between the information and the mean of vaccine 1's prior belief,
while the weight is lower than the effect of the information on vaccine 1 ($\overline{\sigma}^2_\theta \leq \sigma_1^2$).
Intuitively, the information about vaccine 1 only helps the inference of the common factor between vaccine 1 and vaccine 2,
so the magnitude of the update is.

When the DM receives the information about both vaccines, 
the posterior is determined by the weighted difference between the prior means of the vaccines and the information.
For each of the vaccines, there is an additional (positive) term on the signal,
representing the adjustment from iterated updating process.

With the posterior belief given the signal realization, 
the DM can decide whether to take one of the two vaccines or not taking anyone by comparing the mean of the posterior beliefs with the reservation value $\vbar$.
If the mean of either of the vaccines exceeds $\vbar$, 
then the DM takes the one has the higher mean;
if the means of both of the vaccines do not exceed $\vbar$,
then the DM does not take any vaccine and take $\vbar$.

For each of the three possible information combinations,
there are thresholds for the realized signals.
\begin{lemma}\label{lemma:multi_vaccine_cons}
Suppose that $\vbar\geq \mu_1 \geq \mu_2$.
	\begin{enumerate}[(a)]
	\item When only $s_i$ is received: DM chooses vaccine $i$ if $s_i \geq s_i^*$
		and receives the value of $\E{\theta_i|s_i}$, where 
		$$ s_i^* = \mu_i + \frac{\sigma_i^2 + \sigma_s^2}{\sigma_i^2} \left(\vbar - \mu_i\right).$$
		Otherwise, the DM rejects both of the vaccines and receives the value of $\vbar$.
	\item When $s_1,s_2$ are both received: DM chooses vaccine $i$ against vaccine $j$ if 
		\begin{enumerate}[(i)]
		\item $(\mu_i-\mu_j) +\frac{\sigma_i^2-\overline{\sigma}^2_\theta }{\sigma_s^2+ \sigma_1^2 + \sigma^2_2}(s_i-\mu_i) 
			-\frac{\sigma^2_j - \overline{\sigma}^2_\theta}{\sigma_s^2+ \sigma_1^2 + \sigma^2_2}(s_j-\mu_j) 
			+\frac{\sigma^2_1\sigma^2_2 - (\overline{\sigma}^2_\theta)^2}{\sigma^2_s(\sigma_s^2+ \sigma_1^2 + \sigma^2_2)}(s_i-s_j)\geq 0 $
		\item $\mu_i +\frac{\sigma_i^2 }{\sigma_s^2+ \sigma_1^2 + \sigma^2_2}(s_i-\mu_i) 
			+\frac{\overline{\sigma}^2_\theta}{\sigma_s^2+ \sigma_1^2 + \sigma^2_2}(s_j-\mu_j) 
			+\frac{\sigma^2_1\sigma^2_2 - (\overline{\sigma}^2_\theta)^2}{\sigma^2_s(\sigma_s^2+ \sigma_1^2 + \sigma^2_2)}s_i \geq \vbar$,	
		\end{enumerate}
		and the DM the value of $\E{\theta_i|s_1,s_2}$.
		Otherwise, the DM rejects both vaccines and receives the value of $\vbar$.
	\end{enumerate}
\end{lemma}
When the DM receives only the information about one vaccines,
the decision problem degenerates to the case where only the information about one vaccine is available.
When the DM receives the information about both vaccines, 
the decision depends on two criteria: 
(1) which of the two vaccines has higher posterior mean, and
(2) whether the mean of this better vaccine excess the default value $\vbar$.
%

The \textit{ex-post} utility level given the information can then be derived from Lemma \ref{lemma:multi_vaccine_cons},
which we denote as $v_1(s_1)$, $v_2(s_2)$, or $v_{1,2}(s_1,s_2)$ given different bundles of information acquired.
The DM can then choose either to acquire only the information about vaccine 1, only about vaccine 2,
or acquire the information about both vaccines.
The DM first finds the expected value given the signal combinations, 
and the she chooses the optimal information bundle.

We denote the following value functions:
\begin{align*}
V_1(\mu_1,\mu_2,\overline{v},\sigma_1^2,\sigma_2^2,\overline{\sigma}_\theta^2,\sigma_s^2,c)
	&= \Ex{s_1}{\E{v_1(s_1)|s_1}}-c \\
V_2(\mu_1,\mu_2,\overline{v},\sigma_1^2,\sigma_2^2,\overline{\sigma}_\theta^2,\sigma_s^2,c)
	&= \Ex{s_2}{\E{v_2(s_2)|s_2}}-c \\
V_{1,2}(\mu_1,\mu_2,\overline{v},\sigma_1^2,\sigma_2^2,\overline{\sigma}_\theta^2,\sigma_s^2,c)
	&= \Ex{s_1,s_2}{\E{v_{1,2}(s_1,s_2)|s_1}}-2c \\
\end{align*}
Note that the cost of the information bundle of both vaccines is ${2c}$.

The following table displays the decision rule given the parameters.
\begin{center}
\hspace{-1cm}
	\small
	\begin{tabular}{rcccc}
	\hline\hline 
	&\multicolumn{4}{c}{Which of the following is the largest?}\\
	\hline
	the largest:&
		$V_1(\cdot)$&$V_2(\cdot)$
		&$V_{1,2}(\cdot)$& $\vbar$\\
	then acquire:&
		$s_1$&$s_2$&$s_1$ and $s_2$& no info\\
		\hline\hline
	\end{tabular}
\end{center}

Then we can give the following similar prediction as in Section \ref{sec:theory_single}.
\begin{prop}
Suppose $\max\{\mu_1,\mu_2\} \leq \vbar$.
Then $V_1$, $V_2$, and $V_{1,2}$
\begin{enumerate}[(i)]
\item increase in $\mu_1$ and $\mu_2$,
\item decrease in $\sigma_1^2$, $\sigma_2^2$, and $\sigma_\theta^2$,
\item increase in $\sigma_s^2$.
\end{enumerate}

\end{prop}


\newpage
\section{Experiment Details}\label{sec:experiment_details}
\setcounter{table}{0}
\renewcommand{\thetable}{\Alph{section}\arabic{table}}
\setcounter{figure}{0}
\renewcommand{\thefigure}{\Alph{section}\arabic{figure}}

\begin{figure}[htbp]
\centering
\resizebox{\textwidth}{!}{
\begin{tikzpicture}[every node/.style={scale=0.8},
		squarenode/.style={rectangle, draw =blue!60, fill=blue!5, very thick},
		sheetnode/.style={rectangle, draw =black!60, fill=black!5,minimum width=5cm,thin},
		rsquarenode/.style={rectangle, draw=red!60, fill=red!5, very thick},
		treatnode/.style={rectangle, draw={rgb:red,119;green,66;blue,141}, fill={purple!5}, 
			very thick,rounded corners,minimum width=3cm,minimum height=0.6cm}]
	\node[rsquarenode] (prebeliefs) {
		\begin{tabular}{c}
		Pre-treatment Beliefs\\
		Effectiveness of Vaccines/Preference to Vaccines
		\end{tabular}
		};
	\node[squarenode] (order) [below =0.3cm of prebeliefs] {
		\begin{tabular}{c}
		\textbf{Information Demand Q1: Ranking}\\
		``Please rank the vaccines according to your williness to read the information about it.''
		\end{tabular}
		};
	\draw[->] (prebeliefs)--(order);

	\node[squarenode] (numbers) [below =0.3cm  of order] {
		\begin{tabular}{c}
		\textbf{Information Demand Q2: Selection}\\
		``I'd like to read my top 3/2/1 or none vaccine info.''
		\end{tabular}
		};
	\draw[->] (order)--(numbers);
	
	\node[rectangle, rounded corners, dashed, draw=black,below=0.5cm of numbers] (treatments){ 
		\begin{tabular}{c}
		\textbf{Treatment---Information Exposure}\\
		\begin{tikzpicture}[every node/.style={scale=1},
		squarenode/.style={rectangle, draw =blue!60, fill=blue!5, very thick},
		sheetnode/.style={rectangle, draw =black!60, fill=black!5,minimum width=5cm,thin},
		rsquarenode/.style={rectangle, draw=red!60, fill=red!5, very thick},
		treatnode/.style={rectangle, draw={rgb:red,119;green,66;blue,141}, fill={purple!5}, 
			very thick,rounded corners,minimum width=3cm,minimum height=0.6cm}]
	\node[treatnode] (t3) {
		\begin{tabular}{c}
		\textbf{Random Assign}\\
		{\scriptsize Receive 3 vaccine info}\\
		{\scriptsize regardless answers in Q1 and Q2}
		\end{tabular}
		};
	\node[treatnode] (t2) [left=0.2 cm of t3]{
		\begin{tabular}{c}
		\textbf{Top 3}\\
		{\scriptsize Receive top 3 vaccine info}\\
		{\scriptsize based on answers in Q1}
		\end{tabular}
		};
	\node[treatnode] (t1) [left=0.2 cmof t2]{
		\begin{tabular}{c}
		\textbf{Full Compliance}\\
		{\scriptsize Receive top n vaccine info}\\
		{\scriptsize based on answers in Q1 and Q2}
		\end{tabular}
		};
	\node[treatnode,dashed] (t4) [right=0.2 cmof t3]{
		\begin{tabular}{c}
		\textbf{Top $3^*$}\\
		{\scriptsize Same as \textbf{Top 3}}\\
		{\scriptsize while Q2 not asked}
		\end{tabular}
		};
	\node[treatnode,dashed] (t5) [right=0.2 cm of t4]{
		\begin{tabular}{c}
		\textbf{Random Assign$^*$}\\
		{\scriptsize Same as \textbf{Random Assign}}\\
		{\scriptsize while Q1 and Q2 not asked}
		\end{tabular}
		};
		\end{tikzpicture}
		\end{tabular}
	};

	\draw[->] (numbers)--(treatments);

	\node[rsquarenode] (postbeliefs)[below=0.5cm of treatments] {
		\begin{tabular}{c}
		Post-treatment Beliefs\\
		Effectiveness of Vaccines/Preference to Vaccines
		\end{tabular}
		};
	\draw[->] (treatments)--(postbeliefs);

	\end{tikzpicture}
}
	\caption{The Experiment Survey Flow}\label{schedule}
\end{figure}

\begin{figure}[htbp]
\centering
\begin{subfigure}{0.48\textwidth}
	\includegraphics[width=\textwidth]{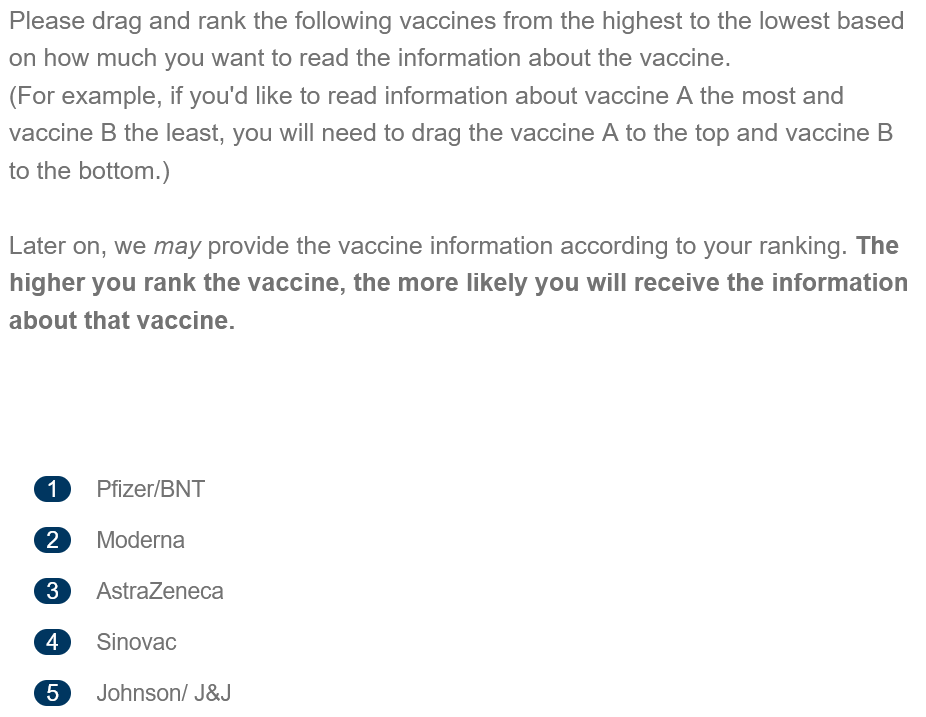}
	\caption{Ranking Question}\label{fig:ranking_question}
\end{subfigure}
\begin{subfigure}{0.48\textwidth}
	\includegraphics[width=\textwidth]{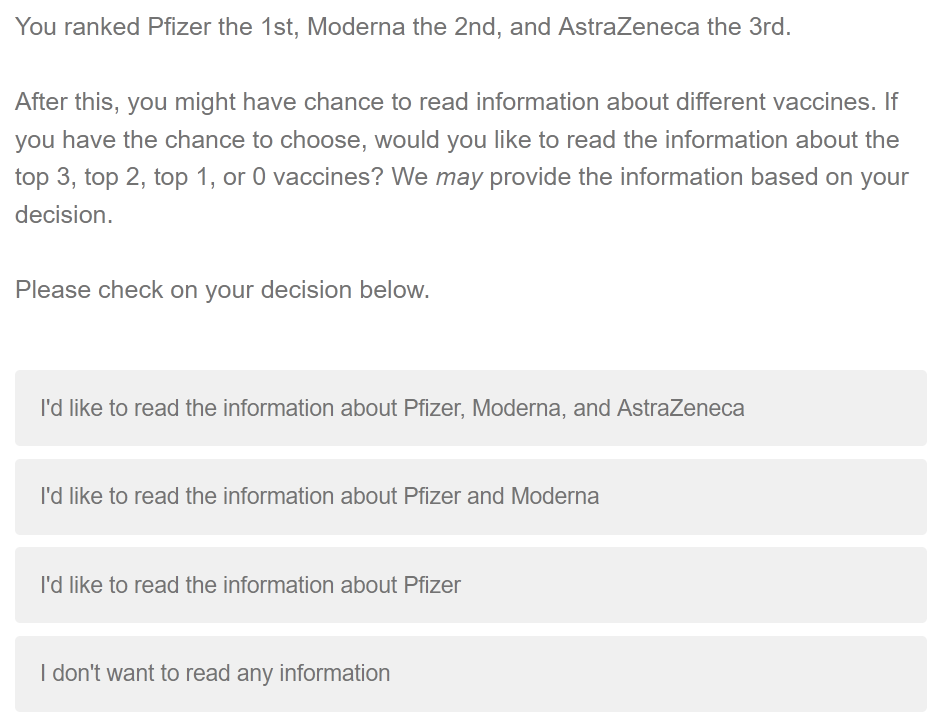}
	\caption{Selection Question}\label{fig:selection_question}
\end{subfigure}
\medskip
\begin{subfigure}{0.48\textwidth}
	\includegraphics[width=\textwidth]{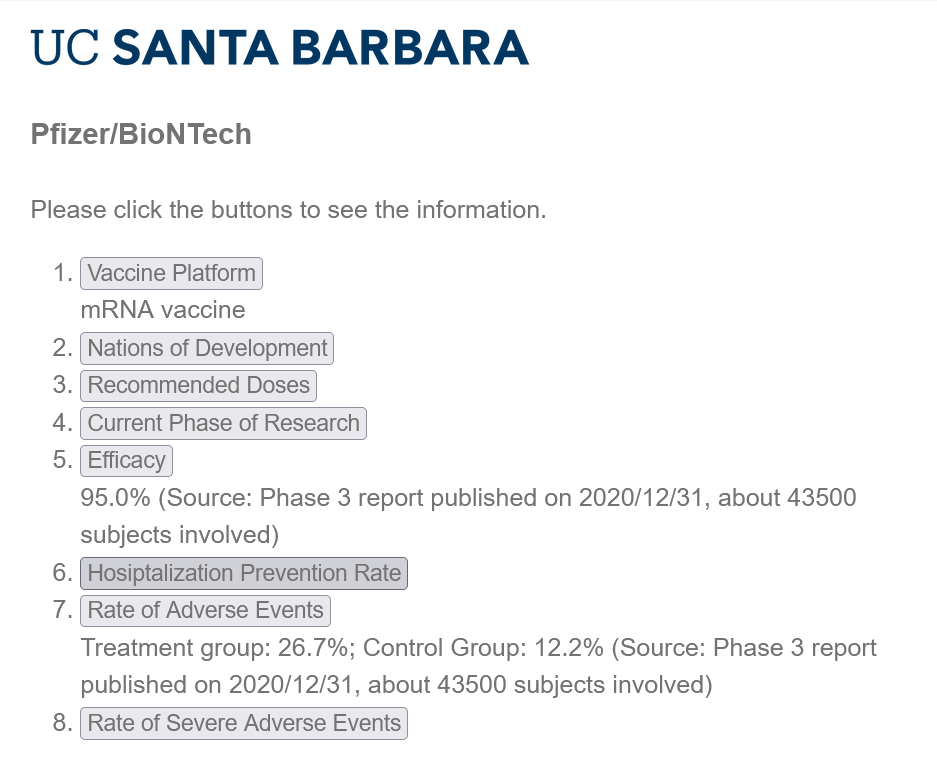}
	\caption{Sample Vaccine Information Sheet}\label{fig:info_sheet}
\end{subfigure}

\caption{Experimental Interface}\label{fig:info_demand_elicitation}
{\begin{tfootnote} 
	The original interface is in Chinese.
	We translated the interface into English for demonstration purposes.
	Panel (a) is the ``Ranking'' question, and Panel (b) is the ``Selection'' question.
	Panel (c) is a sample of the information sheet that subjects observed.
	\end{tfootnote}}
\end{figure}

\clearpage

\label{sec:subject_detail}
\begin{table}[H]
\caption{Summary Statistics}\label{tab:subject_detail}
\begin{center} \small
\begin{tabular}{lcccccc}
\hline\hline
	                &       FC&       T3&       RA&   T3$^*$&   RA$^*$&    Total\\
\hline
\# of Subjects  &   207.00&   217.00&   208.00&   233.00&   201.00&   213.78\\
Female (\%)     &    61.84&    64.06&    66.83&    61.37&    61.19&    63.04\\
Age             &    29.59&    28.58&    29.67&    28.57&    28.65&    29.00\\
Household Yearly Income $\geq$ 40k (\%)&    10.63&     8.29&     8.65&    11.16&     9.45&     9.66\\
Received At Least 1 Vaccine (\%)&    91.30&    93.55&    90.38&    94.42&    89.55&    91.93\\

 \hline\hline
\end{tabular}
\begin{tfootnote}[0.8\textwidth]
 FC represents the treatment Free Choice, T3 the Top 3 Choices, and RA the Random Assignment.
The treatments with asteroids represents the treatments with no unused information preference
questions.
\end{tfootnote}
\end{center}

\end{table}

\label{vaccine_detail}
\begin{table}[H]
\caption{Vaccine Performance Information}\label{tab:vaccine_info_detail}
\begin{center}
\resizebox{0.95\textwidth}{!}{
\begin{tabular}{lccccccccc}
\hline\hline\\
Vaccine &  
\begin{minipage}[c]{3cm}
\centering
Platform
\end{minipage} &
\begin{minipage}[c]{4cm}
\centering
Development \\ Countries
\end{minipage} & 
\begin{minipage}[c]{4cm}
\centering
Recommended \\ Doses
\end{minipage}& 
\begin{minipage}[c]{4cm}
\centering
Current \\ Phase
\end{minipage}\\[1em]  \hline
AstraZeneca 
	& virus vector 	&UK/Sweden	& 2; Day 0 $+$ 28 & 4\\
Johnson \& Johnson  
	& virus vector	&Netherland/Belgium/US & 1 & 4\\
Moderna 
	& mRNA			&US 		&2; Day 0 $+$ 28 & 4\\
Pfizer 
	& mRNA			&US/Germany &2; Day 0 $+$ 21 &4\\
Sinovac 
	& inactive virus&China 		&2; Day 0 $+$ 14 & 4\\
\hline\hline\\
Vaccine &  
\begin{minipage}[c]{3cm}
\centering
Efficacy
\end{minipage} &
\begin{minipage}[c]{4cm}
\centering
Hospitalization \\ Prevention Rate
\end{minipage} & 
\begin{minipage}[c]{4cm}
\centering
Adverse  \\ Event Rates
\end{minipage}& 
\begin{minipage}[c]{4cm}
\centering
Severe Adverse\\ Event Rates
\end{minipage}\\[1em]  \hline
AstraZeneca 
	& 70.4\%	& 100\%	
	& \begin{minipage}[c]{4cm}\centering Vaccinated: 27.03\%\\ Placebo: 16.33\% \end{minipage}
	& \begin{minipage}[c]{4cm}\centering Vaccinated: 0.7\%\\ Placebo: 0.8\% \end{minipage}\\[1em]
Johnson \& Johnson  
	& 66.9\%	& 93.1\%	
	& \begin{minipage}[c]{4cm}\centering Vaccinated: 68.1\%\\ Placebo: 29.4\% \end{minipage}
	& \begin{minipage}[c]{4cm}\centering Vaccinated: 0.1\%\\ Placebo: 0.1\% \end{minipage}\\[1em]
Moderna 
	& 94.1\%	& 100\%	
	& \begin{minipage}[c]{4cm}\centering Vaccinated: 79.4\%\\ Placebo: 36.5\% \end{minipage}
	& \begin{minipage}[c]{4cm}\centering Vaccinated: 1.5\%\\ Placebo: 1.3\% \end{minipage}\\[1em]
Pfizer 
	& 95.0\%	& 88.9\%	
	& \begin{minipage}[c]{4cm}\centering Vaccinated: 26.7\%\\ Placebo: 12.2\% \end{minipage}
	& \begin{minipage}[c]{4cm}\centering Vaccinated: 1.1\%\\ Placebo: 0.6\% \end{minipage}\\[1em]
Sinovac 
	& 83.5\%	& 100\%	
	& \begin{minipage}[c]{4cm}\centering Vaccinated: 18.9\%\\ Placebo: 16.9\% \end{minipage}
	& \begin{minipage}[c]{4cm}\centering Vaccinated: 0.3\%\\ Placebo: 0.2\% \end{minipage}\\[1em]
\hline\hline \\
\end{tabular}
}
\begin{minipage}{0.8\textwidth}
\footnotesize {\it Note:} See \cite{voysey2021}, \cite{sadoff2021}, \cite{baden2021}, \cite{polack2020}, \cite{tanriover2021}.
\end{minipage}
\end{center}
\end{table}

\begin{table}[H]
\caption{Interaction Data With the Vaccine Information\label{tab:info_interaction}}
\centering \small
	\begin{tabular}{lcccc}
	\hline\hline
	& Top-Ranked & Requested  & All \\ \hline
	\textit{Bottom Clicked (\%)} \\
	Vaccine Platform &	83.4	&	82.5	&	79.7\\
	Countries		&	82.3	&	82.2	&	79.3 \\
	Doses			&	82.3	&	82.1	&	79.1\\
	Research Phase	&	82.7	&	82.3	&	79.1\\
	Efficacy		&	82.7	&	82.9	&	80.0\\
	Hospital		&	82.1	&	81.9	&	79.1\\
	Adverse Event	&	82.3	&	81.9	&	78.7\\
	Severe AE		&	80.7	&	79.8	&	76.7\\
	\textit{Time on page (sec)}\\
	Mean			& 	52.90	&	50.00 	& 	47.06 \\
	Median			& 	45.44	&	39.55	&	37.44 \\ \hline
	$N$ 			&	481		&	1,150	&	1,762 \\ \hline\hline
	\end{tabular}
	\begin{tfootnote}[0.6\textwidth]
	The table includes all subjects from the three main treatment groups.
	The first column contains observations of the highest-ranked information (according to \textbf{Ranking} question).
	The second column contains observations of the requested information.
	\end{tfootnote}
\end{table}

\newpage
\section{Appendix Figures}
\setcounter{figure}{0}
\renewcommand{\thefigure}{\Alph{section}\arabic{figure}}

\begin{figure}[htb]
\includegraphics[width=\textwidth]{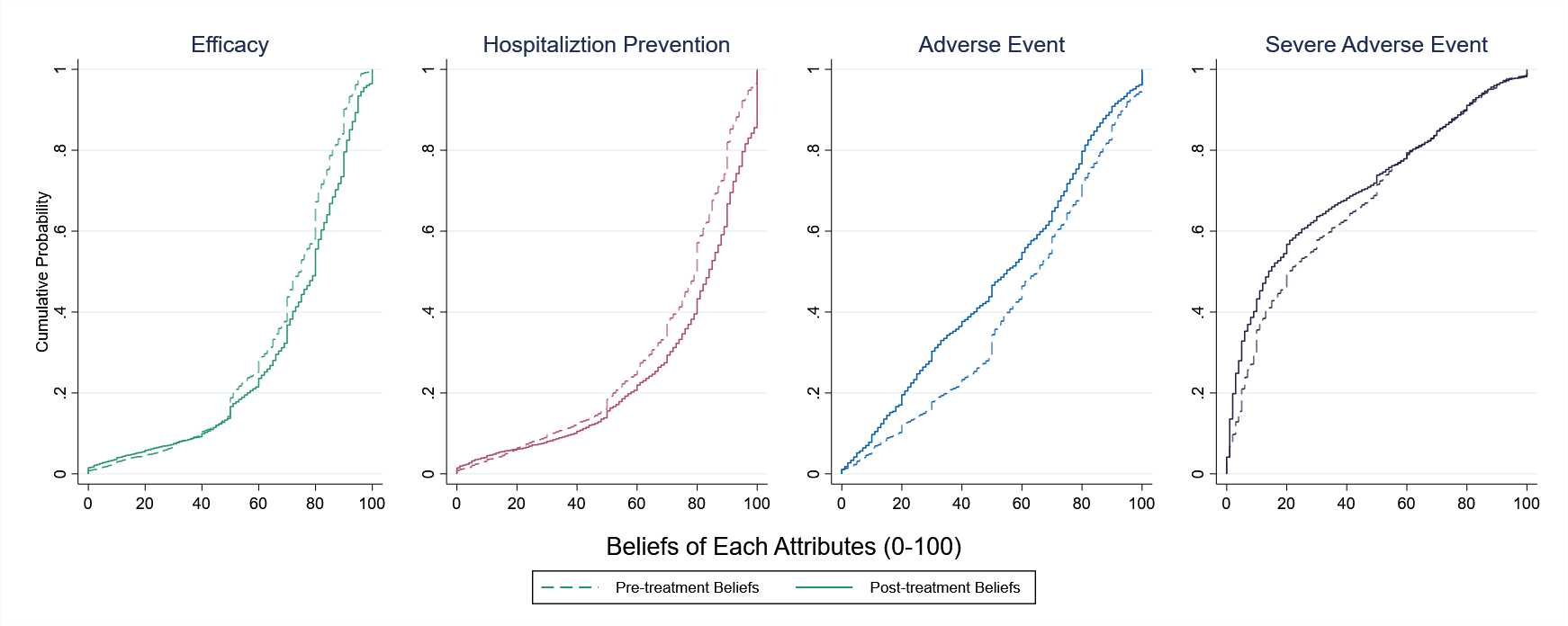}
\caption{Distribution of Pre-treatment and Post-treatment Beliefs\label{fig:cdf_belief}}
\begin{tfootnote}
The dashed lines represent the distribution of beliefs before the treatment.
The solid lines represent the distribution of beliefs after the treatment.
\end{tfootnote}
\end{figure}

\begin{figure}[htb]
{
\centering
	\begin{subfigure}[b]{0.48\textwidth}
	\centering
	\includegraphics[width=\textwidth]{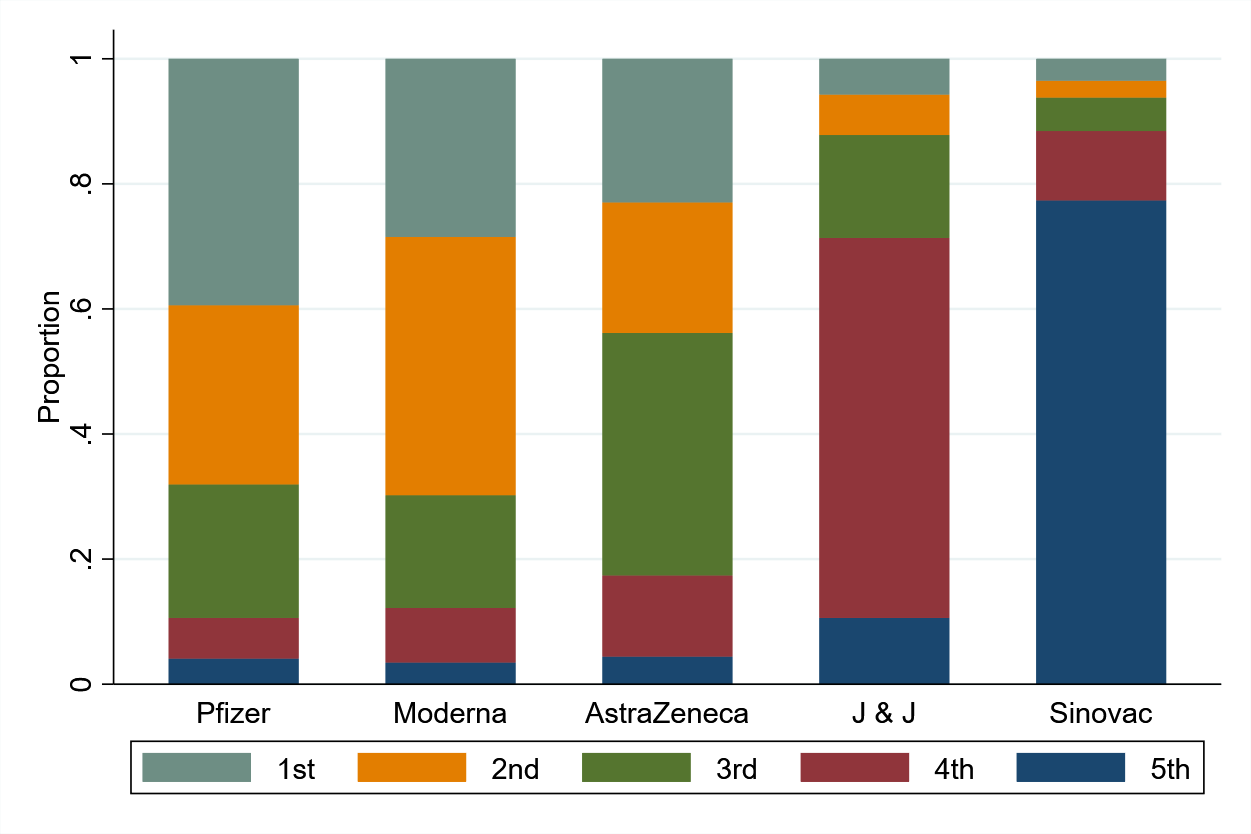}	
	\caption{Vaccine Information Ranking}
	\end{subfigure}
	\hfill
	\begin{subfigure}[b]{0.48\textwidth}
	\centering
	\includegraphics[width=\textwidth]{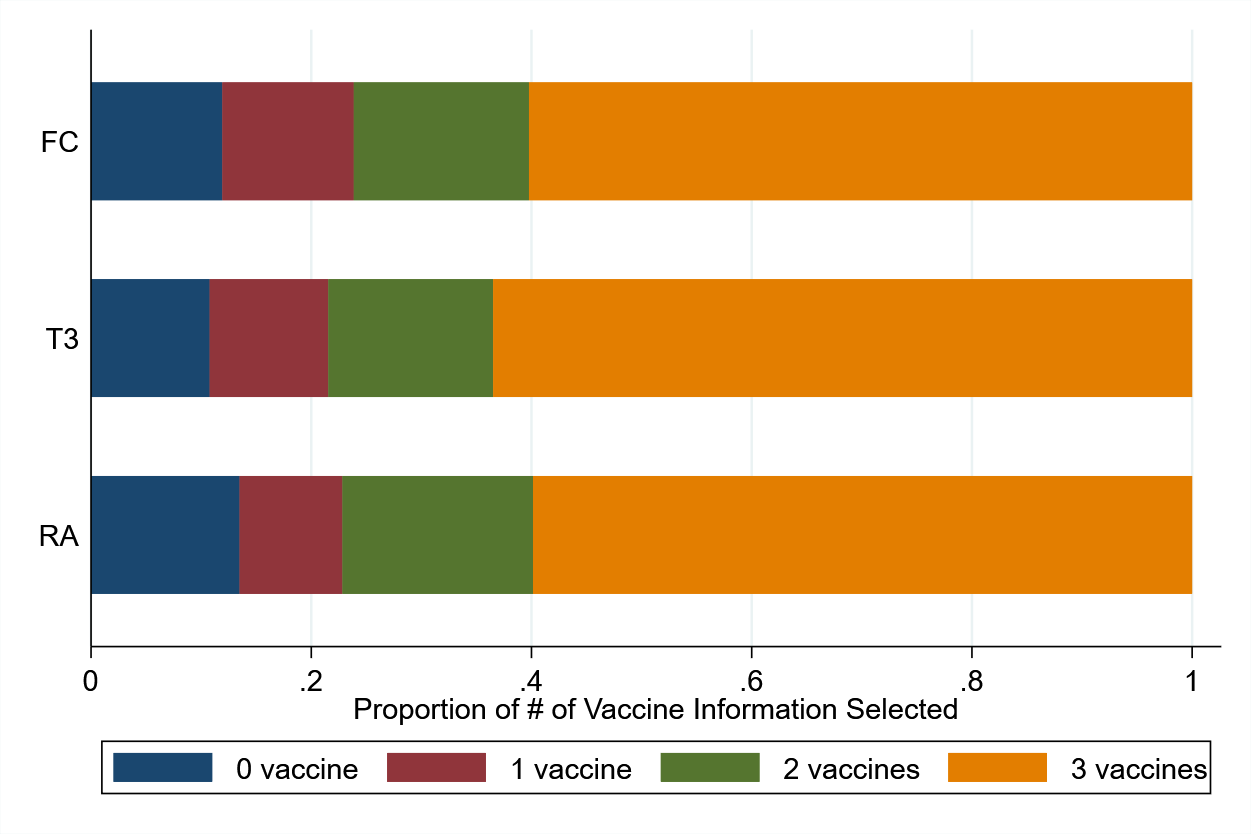}
	\caption{Information Selected in Each Treatment}
	\end{subfigure}
\caption{Vaccine Information Demands}\label{fig:info_consumption}
}
{\begin{tfootnote} Panel (a) shows the ranking distribution for each vaccine.
Panel (b) shows the number of vaccines selected in the ``Number'' question in each treatment arm.
\end{tfootnote}}
\end{figure}

\begin{figure}[htb]
{
\centering
	\includegraphics[width=0.45\textwidth]{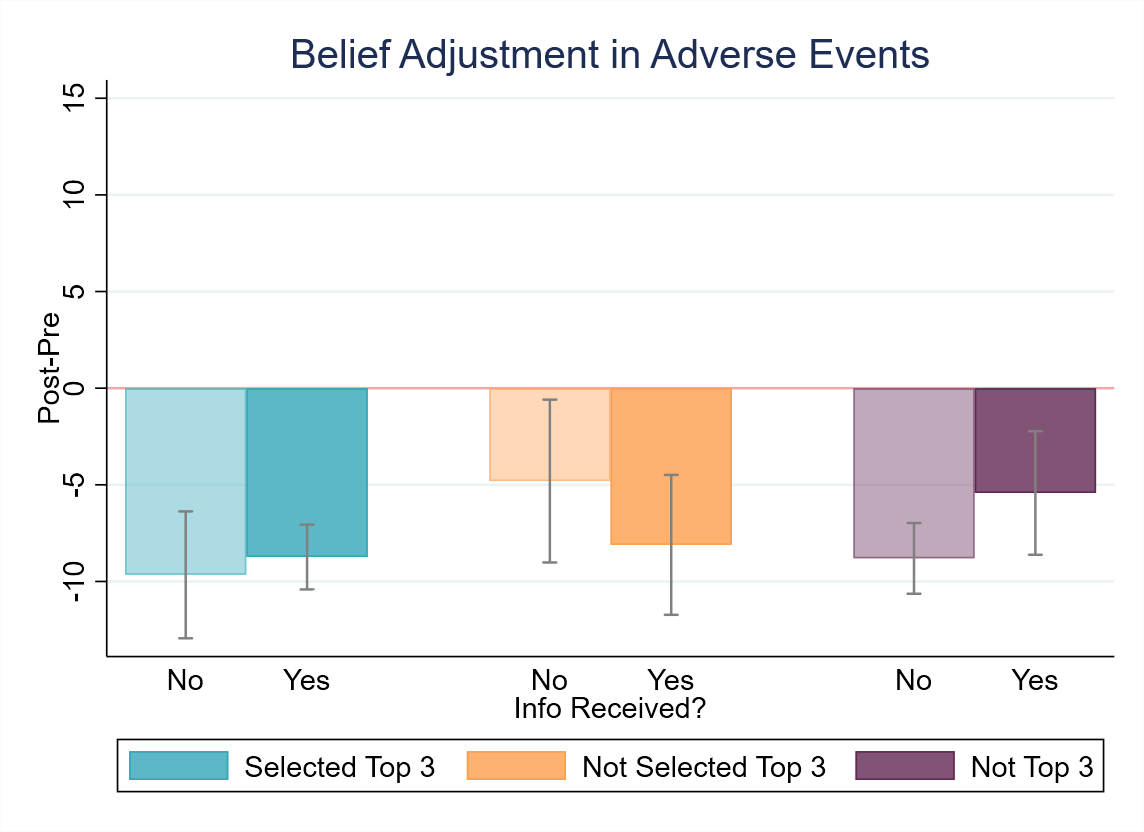}
	\includegraphics[width=0.45\textwidth]{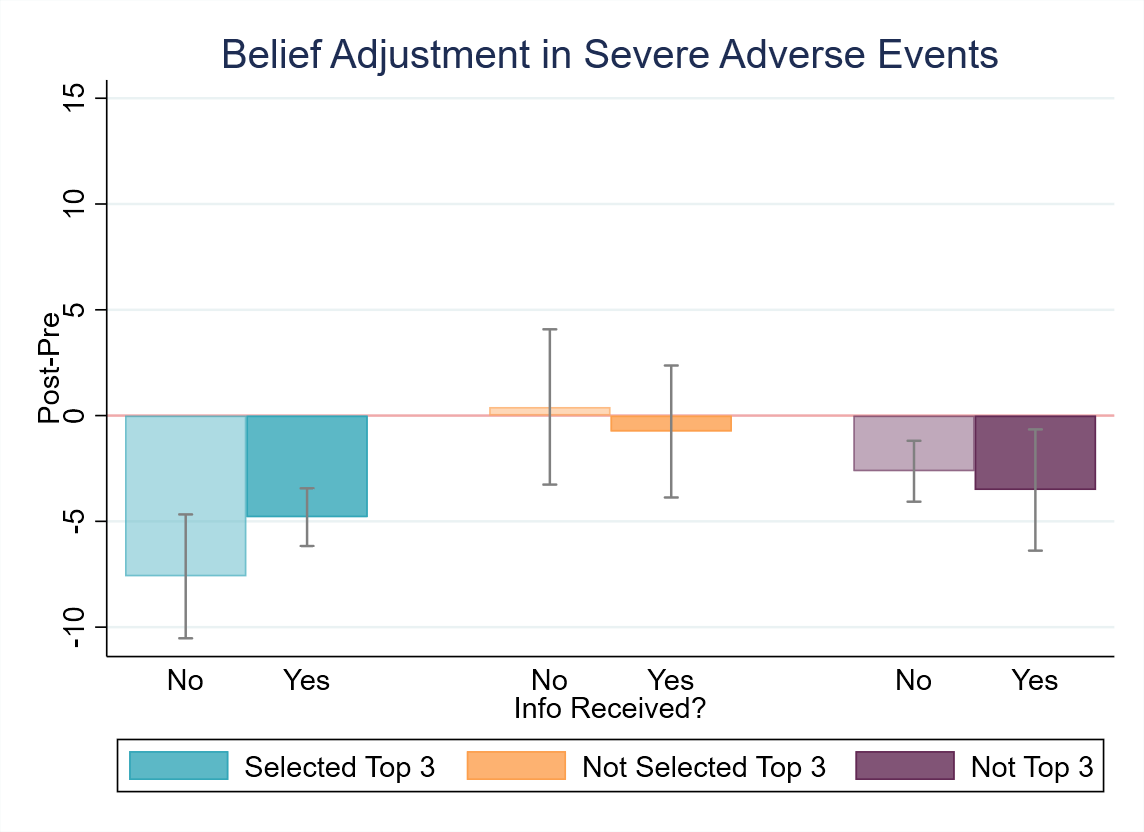}
	\caption{Changes in Beliefs}
	\label{fig:change_in_beliefs_adverse}
	}
\begin{tfootnote}
The mean of the changes in beliefs before and after the information exposure stage are depicted.
The change is defined by Post-treatment belief $-$ Pre-treatment belief.
Within each category, the left/lighter bar is for the observations that the subjects do not receive the vaccine information,
and the right/deeper bar is for the observations that the subjects receive the information.
95\% confidence intervals of the means are plotted for each category.
\end{tfootnote}
\end{figure}

\clearpage

\section{Appendix Tables}
\setcounter{table}{0}
\renewcommand{\thetable}{\Alph{section}\arabic{table}}
\begin{table}[H]
\caption{Summary by Vaccine}
\centering
\resizebox{0.45\textwidth}{!}{
\begin{tabular}{lccccc}
	\hline\hline
	&\multicolumn{5}{c}{Efficacy}\\
	\cmidrule(lr){2-6}
	                    &      Pfizer&     Moderna& AstraZeneca&        J\&J&     Sinovac\\
\hline
Pre-treatment Beliefs&       78.03&       79.72&       72.51&       66.27&       48.49\\
                    &     (16.18)&     (15.15)&     (15.09)&     (19.16)&     (23.51)\\
[1em]
Post-treatment Beliefs&       82.69&       83.72&       76.95&       68.46&       48.70\\
                    &     (15.27)&     (14.90)&     (16.01)&     (21.02)&     (26.87)\\
[1em]
Adjustment in Beliefs (Post $-$ Pre)&       4.652&       4.003&       4.435&       2.193&       0.217\\
                    &     (13.81)&     (13.40)&     (14.30)&     (18.26)&     (21.52)\\
[1em]
$| \text{Adjustment in Beliefs} |$&       9.896&       9.566&       10.47&       12.44&       14.68\\
                    &     (10.70)&     (10.19)&     (10.69)&     (13.53)&     (15.72)\\
[1em]
$\frac{| \text{Adjustment in Beliefs} |}{\text{Pre-treatment Beliefs}}$   &       0.186&       0.209&       0.220&       0.326&       0.527\\
                    &     (0.622)&     (1.442)&     (0.930)&     (1.057)&     (1.560)\\
[1em]
$|$ Pre-treatment Error $|$&       17.19&       14.59&       11.29&       14.29&       35.71\\
                    &     (15.94)&     (14.95)&     (10.22)&     (12.77)&     (22.44)\\
[1em]
$|$ Post-treatment Error $|$&       13.01&       11.43&       13.34&       15.83&       35.97\\
                    &     (14.68)&     (14.11)&     (11.00)&     (13.91)&     (25.27)\\
[1em]
Learning            &       4.177&       3.157&      -2.052&      -1.538&      -0.263\\
                    &     (13.36)&     (12.82)&     (11.32)&     (14.27)&     (19.95)\\

\end{tabular}
}
\resizebox{0.45\textwidth}{!}{
\begin{tabular}{lccccc}
	\hline\hline
	&\multicolumn{5}{c}{Hospital Prevention Rate}\\
	\cmidrule(lr){2-6}
	                    &      Pfizer&     Moderna& AstraZeneca&        J\&J&     Sinovac\\
\hline
Pre-treatment Beliefs&       78.06&       79.00&       75.33&       69.86&       52.76\\
                    &     (20.50)&     (20.01)&     (19.98)&     (22.47)&     (26.51)\\
[1em]
Post-treatment Beliefs&       84.02&       85.53&       82.47&       73.34&       53.56\\
                    &     (17.82)&     (18.20)&     (19.64)&     (23.19)&     (30.24)\\
[1em]
Adjustment in Beliefs (Post $-$ Pre)&       5.962&       6.530&       7.133&       3.476&       0.801\\
                    &     (20.47)&     (20.83)&     (20.27)&     (22.40)&     (24.05)\\
[1em]
$| \text{Adjustment in Beliefs} |$&       13.25&       13.56&       14.39&       15.31&       16.63\\
                    &     (16.70)&     (17.10)&     (15.96)&     (16.71)&     (17.38)\\
[1em]
$\frac{| \text{Adjustment in Beliefs} |}{\text{Pre-treatment Beliefs}}$ &       0.498&       0.429&       0.358&       0.426&       0.708\\
                    &     (3.540)&     (2.429)&     (1.404)&     (1.263)&     (3.476)\\
[1em]
$|$ Pre-treatment Error $|$&       14.33&       21.00&       24.67&       23.99&       47.24\\
                    &     (18.23)&     (20.01)&     (19.98)&     (21.66)&     (26.51)\\
[1em]
$|$ Post-treatment Error $|$&       11.59&       14.47&       17.53&       21.14&       46.44\\
                    &     (14.38)&     (18.20)&     (19.64)&     (21.94)&     (30.24)\\
[1em]
Learning            &       2.738&       6.530&       7.133&       2.853&       0.801\\
                    &     (18.89)&     (20.83)&     (20.27)&     (21.30)&     (24.05)\\

\end{tabular}
}
\resizebox{0.45\textwidth}{!}{
\begin{tabular}{lccccc}
	\hline\hline
	&\multicolumn{5}{c}{Adverse Event Rate}\\
	\cmidrule(lr){2-6}
	                    &      Pfizer&     Moderna& AstraZeneca&        J\&J&     Sinovac\\
\hline
Pre-treatment Beliefs&       58.50&       62.94&       72.73&       56.52&       53.13\\
                    &     (27.73)&     (26.33)&     (24.26)&     (25.91)&     (27.74)\\
[1em]
Post-treatment Beliefs&       51.85&       54.96&       59.84&       50.20&       46.16\\
                    &     (29.52)&     (29.48)&     (28.76)&     (27.15)&     (29.80)\\
[1em]
Adjustment in Beliefs (Post $-$ Pre)&      -6.649&      -7.981&      -12.89&      -6.320&      -6.972\\
                    &     (29.75)&     (28.78)&     (28.27)&     (28.19)&     (29.58)\\
[1em]
$| \text{Adjustment in Beliefs} |$&       21.46&       21.05&       21.44&       21.00&       21.73\\
                    &     (21.64)&     (21.18)&     (22.47)&     (19.82)&     (21.22)\\
[1em]
$\frac{| \text{Adjustment in Beliefs} |}{\text{Pre-treatment Beliefs}}$ &       0.656&       0.587&       0.528&       0.601&       0.652\\
                    &     (1.616)&     (1.641)&     (1.872)&     (1.185)&     (1.394)\\
[1em]
$|$ Pre-treatment Error $|$&       36.03&       23.51&       42.28&       22.02&       37.49\\
                    &     (21.95)&     (20.27)&     (18.63)&     (17.88)&     (23.14)\\
[1em]
$|$ Post-treatment Error $|$&       32.13&       29.41&       33.36&       25.95&       32.25\\
                    &     (21.69)&     (24.53)&     (20.32)&     (19.58)&     (24.29)\\
[1em]
Learning            &       3.891&      -5.893&       8.922&      -3.932&       5.234\\
                    &     (22.54)&     (24.84)&     (19.96)&     (21.60)&     (24.59)\\

	\hline\hline
\end{tabular}
}
\resizebox{0.45\textwidth}{!}{
\begin{tabular}{lccccc}
	\hline\hline
	&\multicolumn{5}{c}{Severe Adverse Event Rate}\\
	\cmidrule(lr){2-6}
	                    &      Pfizer&     Moderna& AstraZeneca&        J\&J&     Sinovac\\
\hline
Pre-treatment Beliefs&       30.36&       32.10&       38.20&       30.44&       32.71\\
                    &     (28.17)&     (29.43)&     (31.52)&     (27.52)&     (29.24)\\
[1em]
Post-treatment Beliefs&       27.28&       28.58&       32.23&       27.16&       30.56\\
                    &     (29.89)&     (30.69)&     (32.38)&     (28.44)&     (30.27)\\
[1em]
Adjustment in Beliefs (Post $-$ Pre)&      -3.084&      -3.521&      -5.967&      -3.280&      -2.152\\
                    &     (24.71)&     (24.09)&     (24.94)&     (23.17)&     (23.87)\\
[1em]
$| \text{Adjustment in Beliefs} |$&       15.44&       15.14&       16.50&       14.86&       15.33\\
                    &     (19.53)&     (19.06)&     (19.62)&     (18.07)&     (18.41)\\
[1em]
$\frac{| \text{Adjustment in Beliefs} |}{\text{Pre-treatment Beliefs}}$ &       0.942&       0.990&       0.834&       1.021&       1.048\\
                    &     (2.905)&     (4.686)&     (3.847)&     (4.773)&     (3.840)\\
[1em]
$|$ Pre-treatment Error $|$&       29.32&       30.71&       37.52&       30.35&       32.44\\
                    &     (28.11)&     (29.31)&     (31.50)&     (27.51)&     (29.21)\\
[1em]
$|$ Post-treatment Error $|$&       26.28&       27.28&       31.58&       27.07&       30.30\\
                    &     (29.80)&     (30.51)&     (32.33)&     (28.43)&     (30.23)\\
[1em]
Learning            &       3.050&       3.434&       5.942&       3.278&       2.145\\
                    &     (24.69)&     (24.00)&     (24.91)&     (23.16)&     (23.84)\\

	\hline\hline
\end{tabular}
}
\end{table}

\begin{table}[htp]
\caption{Reported Ranking on Each Vaccine}
\begin{center}
	\begin{tabular}{lccccc}
	\hline\hline
	Rank&AstraZeneca&J \& J&Moderna&Pfizer&Sinovac\\ \hline
	1st	&	212	&	50	&	240	&	342	&	26	\\
		&(24.37\%)&(5.75\%)&(27.59\%)&(39.31\%)&(2.99\%)\\
	2nd	&	174	&	55	& 	368	&	249	&	24	\\
		&(20.00\%)&(6.32\%)&(42.30\%)&(28.62\%)&(2.76\%)\\
	3rd	&	335	&	138	&	159	&	185	&	53	\\
		&(38.51\%)&(15.86\%)&(18.28\%)&(21.26\%)&(6.09\%)\\
	4th	&	110	&	528	&	73	&	58	&	101\\
		&(12.64\%)&(60.69\%)&(8.39\%)&(6.67\%)&(11.61\%)\\
	5th	&	39	&	99	&	30	&	36	&	666\\
		&(4.48\%)&(11.38\%)&(3.45\%)&(4.14\%)&(76.55\%)\\ \hline\hline
	\end{tabular}
\end{center}
\end{table}

\begin{table}[htp]
\caption{The Number of Information Sheets Requested}
\begin{center}
	\begin{tabular}{lcccc}
	\hline\hline
	Treatments	& Full Compliance	&	Top 3	& Assigned	& Total\\ \hline
	None	&	21	&	28	&	29	&	78\\
		&(10\%)&(13\%)&(14\%)&(12\%)\\
	Top 1	&	22	&	23	&	22	&	67\\
		&(10\%)&(11\%)&(10\%)&(11\%)\\
	Top 2	&	31	&	35	&	39	&	105	\\
		&(15\%)&(16\%)&(19\%)&(17\%)\\
	Top	3	&	136	&	132	&	118	&	286\\
		&(65\%)&(61\%)&(57\%)&(61\%)\\ \hline\hline
	\end{tabular}
\end{center}
\end{table}

\begin{table}
\caption{Vaccine Background Knowledge and Receptions}\label{vaccine_bg}
\begin{center}
	\begin{adjustbox}{max width=\textwidth}
	\begin{tabular}{lcccccc}
	\hline\hline
	                &   Pfizer&  Moderna&AstraZeneca&   J \& J&  Sinovac&  Medigen\\
\hline
Correct Recommended Doses (\%)&    81.33&    87.90&    87.71&    41.18&    33.30&    74.77\\
                &  (38.98)&  (32.63)&  (32.85)&  (49.24)&  (47.15)&  (43.46)\\
[1em]
Correct Platform (\%)&    62.48&    64.07&    39.59&    18.11&    23.92&    49.44\\
                &  (48.44)&  (48.00)&  (48.93)&  (38.52)&  (42.68)&  (50.02)\\
[1em]
Familiarity (1-7)&    4.624&    4.662&    4.933&    3.127&    2.712&    4.216\\
                &  (1.239)&  (1.216)&  (1.208)&  (1.325)&  (1.395)&  (1.484)\\
[1em]
Registered (\%) &    47.94&    49.25&    67.45&    3.752&    1.876&    12.85\\
                &  (49.98)&  (50.02)&  (46.88)&  (19.01)&  (13.57)&  (33.48)\\
[1em]
Received (\%)   &    27.39&    9.287&    52.35&    0.657&    0.563&    7.129\\
                &  (44.62)&  (29.04)&  (49.97)&  (8.081)&  (7.485)&  (25.74)\\

	\hline\hline \\
	\multicolumn{7}{l}{\emph{Note:} Standard deviations are in the parentheses.}
	\end{tabular}
	\end{adjustbox}
\end{center}
\end{table}

\begin{table}[htb]
\centering
\caption{Information Preference and Selection---Only Available Vaccines \label{tab:info_demand_available}}
	{\footnotesize
	\begin{tabular}{l*{4}c}\hline\hline
&
\multicolumn{4}{c}{\textit{Dependent
Variables}}\\
&
\multicolumn{2}{c}{Info
Rank}
&
\multicolumn{2}{c}{Selected}\\
                    &\multicolumn{1}{c}{(1)}         &\multicolumn{1}{c}{(2)}         &\multicolumn{1}{c}{(3)}         &\multicolumn{1}{c}{(4)}         \\
                    &Belief Ranking         &Gap from Highest         &Belief Ranking          &Gap from Highest         \\
\hline
Efficacy            &        0.12\sym{***}&       -0.01\sym{**} &        2.83\sym{**} &       -0.34\sym{*}  \\
                    &      (0.03)         &      (0.00)         &      (0.97)         &      (0.15)         \\
Hospitalization Prevention&        0.08\sym{**} &       -0.01\sym{**} &        2.88\sym{**} &       -0.30         \\
                    &      (0.03)         &      (0.00)         &      (1.11)         &      (0.18)         \\
Adverse Events      &       -0.03         &       -0.00         &        0.70         &        0.08         \\
                    &      (0.02)         &      (0.00)         &      (0.85)         &      (0.06)         \\
Severe Adverse Events&       -0.03         &       -0.00         &       -1.17         &       -0.15         \\
                    &      (0.02)         &      (0.00)         &      (0.85)         &      (0.10)         \\
Familiarity         &        0.04\sym{*}  &        0.05\sym{**} &        3.43\sym{**} &        3.60\sym{***}\\
                    &      (0.02)         &      (0.02)         &      (1.05)         &      (1.06)         \\
Constants           &        2.89\sym{***}&        3.65\sym{***}&       19.20\sym{*}  &       43.85\sym{***}\\
                    &      (0.18)         &      (0.11)         &      (9.25)         &      (7.55)         \\
\hline
Observations        &        1891         &        1891         &        1891         &        1891         \\
Subjects            &         632         &         632         &         632         &         632         \\
 $ R^2 $            &       0.046         &       0.038         &       0.039         &       0.038         \\
Mean of Dep. Variable&        3.73         &        3.73         &        66.5         &        66.5         \\
 \hline\hline \end{tabular}                 \begin{tfootnote}                 Clustered (on subject level) standard errors in parentheses.  The subjects' family income, college majors, and sex are controlled.                  The coefficients and the mean of the dependent variable in (3) and (4) are in percentage.                 \sym{*} \(p<0.05\), \sym{**} \(p<0.01\), \sym{***} \(p<0.001\) . \end{tfootnote}

	}
\end{table}

\begin{table}[htb]
\centering
\caption{Information Preference and Selection---Adding Vaccine Reception History \label{tab:info_demand_received}}
	{\footnotesize
	\begin{tabular}{l*{4}c}\hline\hline
&
\multicolumn{4}{c}{\textit{Dependent
Variables}}\\
&
\multicolumn{2}{c}{Info
Rank}
&
\multicolumn{2}{c}{Selected}\\
                    &\multicolumn{1}{c}{(1)}         &\multicolumn{1}{c}{(2)}         &\multicolumn{1}{c}{(3)}         &\multicolumn{1}{c}{(4)}         \\
                    &Belief Ranking         &Gap from Highest         &Belief Ranking          &Gap from Highest         \\
\hline
Efficacy            &        0.25\sym{***}&       -0.02\sym{***}&        5.28\sym{***}&       -0.49\sym{***}\\
                    &      (0.02)         &      (0.00)         &      (0.74)         &      (0.07)         \\
Hospitalization Prevention&        0.14\sym{***}&       -0.01\sym{***}&        2.87\sym{***}&       -0.05         \\
                    &      (0.02)         &      (0.00)         &      (0.75)         &      (0.07)         \\
Adverse Events      &        0.03         &        0.00\sym{*}  &        1.77\sym{**} &        0.11\sym{*}  \\
                    &      (0.02)         &      (0.00)         &      (0.62)         &      (0.05)         \\
Severe Adverse Events&        0.00         &        0.00         &       -0.24         &       -0.02         \\
                    &      (0.02)         &      (0.00)         &      (0.57)         &      (0.06)         \\
Familiarity         &        0.19\sym{***}&        0.21\sym{***}&        5.96\sym{***}&        6.51\sym{***}\\
                    &      (0.01)         &      (0.01)         &      (0.57)         &      (0.57)         \\
Received Vaccine Before&        0.63\sym{***}&        0.72\sym{***}&       31.81\sym{***}&       34.00\sym{***}\\
                    &      (0.06)         &      (0.06)         &      (2.29)         &      (2.24)         \\
Constants           &        0.59\sym{***}&        2.22\sym{***}&      -30.76\sym{***}&        3.71         \\
                    &      (0.08)         &      (0.08)         &      (4.41)         &      (4.19)         \\
\hline
Observations        &        3145         &        3145         &        3145         &        3145         \\
Subjects            &         632         &         632         &         632         &         632         \\
 $ R^2 $            &        0.40         &        0.38         &        0.32         &        0.31         \\
Mean of Dep. Variable&           3         &           3         &        45.2         &        45.2         \\
 \hline\hline \end{tabular}                 \begin{tfootnote}                 Clustered (at subject level) standard errors in parentheses.  The subjects' family income, college majors, and sex are controlled.                  The coefficients and the mean of the dependent variable in (3) and (4) are in percentage.                 \sym{*} \(p<0.05\), \sym{**} \(p<0.01\), \sym{***} \(p<0.001\) . \end{tfootnote}

	}
\end{table}

\begin{table}[htbp]
\centering
	\caption{Update in Beliefs---Adding Vaccine Reception History \label{tab:update_vaccined}}
	{\footnotesize
	\begin{tabular}{l*{6}c}\hline\hline
&
\multicolumn{6}{c}{\emph{Belief
Update}:
}\\
&
\multicolumn{6}{c}{Post-Treatment
Belief
$-$
Pre-Treatment
Belief
}\\
                    &\multicolumn{1}{c}{(1)}         &\multicolumn{1}{c}{(2)}         &\multicolumn{1}{c}{(3)}         &\multicolumn{1}{c}{(4)}         &\multicolumn{1}{c}{(5)}         &\multicolumn{1}{c}{(6)}         \\
                    &    \multicolumn{3}{c}{Efficacy}         &\multicolumn{3}{c}{Hospitalization}  \\
\hline
Signal Disagreement &        0.31\sym{***}&        0.28\sym{***}&        0.28\sym{***}&        0.52\sym{***}&        0.63\sym{***}&        0.64\sym{***}\\
                    &      (0.03)         &      (0.04)         &      (0.04)         &      (0.04)         &      (0.05)         &      (0.05)         \\
Familiarity         &        0.45         &                     &        0.13         &        1.43\sym{**} &                     &        0.86         \\
                    &      (0.31)         &                     &      (0.33)         &      (0.48)         &                     &      (0.47)         \\
Selected Top 3      &                     &        2.82\sym{*}  &        2.74\sym{*}  &                     &        4.40\sym{**} &        3.80\sym{**} \\
                    &                     &      (1.15)         &      (1.12)         &                     &      (1.42)         &      (1.41)         \\
Not Selected Top 3  &                     &        0.06         &        0.21         &                     &        0.75         &        0.54         \\
                    &                     &      (1.43)         &      (1.44)         &                     &      (1.73)         &      (1.74)         \\
Received the Vaccine Before&        1.86\sym{*}  &        1.95\sym{**} &        1.85\sym{*}  &        1.77         &        1.89\sym{*}  &        1.20         \\
                    &      (0.74)         &      (0.66)         &      (0.73)         &      (0.92)         &      (0.77)         &      (0.89)         \\
Constants           &       -1.69         &       -1.55         &       -2.14         &      -14.39\sym{***}&      -10.91\sym{**} &      -13.95\sym{***}\\
                    &      (2.28)         &      (2.13)         &      (2.59)         &      (3.90)         &      (3.44)         &      (3.92)         \\
\hline
Observations        &        1754         &        1762         &        1754         &        1754         &        1762         &        1754         \\
Subjects            &         611         &         611         &         611         &         611         &         611         &         611         \\
 $ R^2 $            &        0.16         &        0.20         &        0.20         &        0.29         &        0.34         &        0.34         \\
Mean of Dep. Variable&        4.29         &        4.26         &        4.29         &        6.64         &        6.65         &        6.64         \\
Pre-treatment Beliefs Controlled?&          No         &         Yes         &         Yes         &          No         &         Yes         &         Yes         \\
 \hline\hline \end{tabular}\\ \begin{tfootnote}                 Clustered (at subject level) standard errors in parentheses. The subjects' family income, college majors, and sex are controlled in all models.                  Pre-Treatment beliefs are controlled in models (2), (3), (5), and (6).                 Only the observations that subjects have received that vaccine before are included.                 \sym{*} \(p<0.05\), \sym{**} \(p<0.01\), \sym{***} \(p<0.001\) . \end{tfootnote}

	}
\end{table}

\begin{table}[htbp]
\centering
	\caption{Update in Beliefs---Only Underestimate (Information $\geq$ Belief) \label{tab:update_under}}
	{\footnotesize
	\begin{tabular}{l*{6}c}\hline\hline
&
\multicolumn{6}{c}{\emph{Belief
Update}:
}\\
&
\multicolumn{6}{c}{Post-Treatment
Belief
$-$
Pre-Treatment
Belief
}\\
                    &\multicolumn{1}{c}{(1)}         &\multicolumn{1}{c}{(2)}         &\multicolumn{1}{c}{(3)}         &\multicolumn{1}{c}{(4)}         &\multicolumn{1}{c}{(5)}         &\multicolumn{1}{c}{(6)}         \\
                    &    \multicolumn{3}{c}{Efficacy}         &\multicolumn{3}{c}{Hospitalization}  \\
\hline
Signal Strength     &        0.34\sym{***}&        0.31\sym{***}&        0.31\sym{***}&        0.53\sym{***}&        0.66\sym{***}&        0.67\sym{***}\\
                    &      (0.05)         &      (0.06)         &      (0.06)         &      (0.04)         &      (0.05)         &      (0.05)         \\
Familiarity         &        0.56         &                     &        0.10         &        1.51\sym{**} &                     &        0.84         \\
                    &      (0.41)         &                     &      (0.42)         &      (0.52)         &                     &      (0.52)         \\
Selected Top 3      &                     &        3.47\sym{*}  &        3.38\sym{*}  &                     &        5.07\sym{***}&        4.44\sym{**} \\
                    &                     &      (1.44)         &      (1.40)         &                     &      (1.53)         &      (1.52)         \\
Not Selected Top 3  &                     &        0.41         &        0.57         &                     &        0.97         &        0.73         \\
                    &                     &      (1.77)         &      (1.79)         &                     &      (1.87)         &      (1.89)         \\
Constants           &       -2.82         &       -2.42         &       -2.97         &      -15.84\sym{***}&      -12.82\sym{***}&      -15.73\sym{***}\\
                    &      (2.94)         &      (2.57)         &      (3.18)         &      (4.30)         &      (3.78)         &      (4.30)         \\
\hline
Observations        &        1273         &        1278         &        1273         &        1531         &        1539         &        1531         \\
Subjects            &         598         &         598         &         598         &         605         &         605         &         605         \\
 $ R^2 $            &        0.14         &        0.19         &        0.19         &        0.28         &        0.34         &        0.34         \\
Mean of Dep. Variable&        6.24         &        6.22         &        6.24         &        8.07         &        8.07         &        8.07         \\
Pre-treatment Beliefs Controlled?&          No         &         Yes         &         Yes         &          No         &         Yes         &         Yes         \\
 \hline\hline \end{tabular}\\ \begin{tfootnote}                 Clustered (at subject level) standard errors in parentheses. The subjects' family income, college majors, and sex are controlled in all models.                  Pre-Treatment beliefs are controlled in models (2), (3), (5), and (6).                 Only the observations with underestimated pre-treatment beliefs (relative to the information) are included.                 \sym{*} \(p<0.05\), \sym{**} \(p<0.01\), \sym{***} \(p<0.001\) . \end{tfootnote}

	}
\end{table}
\begin{table}[htbp]
\centering
	\caption{Update in Beliefs---Only Overestimate (Information $<$ Belief) \label{tab:update_over}}
	{\footnotesize
	\begin{tabular}{l*{6}c}\hline\hline
&
\multicolumn{6}{c}{\emph{Belief
Update}:
}\\
&
\multicolumn{6}{c}{Post-Treatment
Belief
$-$
Pre-Treatment
Belief
}\\
                    &\multicolumn{1}{c}{(1)}         &\multicolumn{1}{c}{(2)}         &\multicolumn{1}{c}{(3)}         &\multicolumn{1}{c}{(4)}         &\multicolumn{1}{c}{(5)}         &\multicolumn{1}{c}{(6)}         \\
                   &    \multicolumn{3}{c}{Efficacy}         &\multicolumn{3}{c}{Hospitalization}  \\
\hline
Signal Strength     &        0.33\sym{***}&        0.34\sym{***}&        0.34\sym{***}&        0.84\sym{***}&        0.79\sym{**} &        0.85\sym{***}\\
                    &      (0.08)         &      (0.08)         &      (0.08)         &      (0.23)         &      (0.24)         &      (0.25)         \\
Familiarity         &        0.62         &                     &        0.50         &        0.91         &                     &        0.88         \\
                    &      (0.44)         &                     &      (0.45)         &      (0.74)         &                     &      (0.71)         \\
Selected Top 3      &                     &        1.56         &        1.39         &                     &       -1.44         &       -1.68         \\
                    &                     &      (1.74)         &      (1.72)         &                     &      (2.31)         &      (2.27)         \\
Not Selected Top 3  &                     &       -0.13         &       -0.04         &                     &        0.96         &        0.84         \\
                    &                     &      (2.06)         &      (2.07)         &                     &      (2.52)         &      (2.42)         \\
Constants           &       -1.69         &        0.17         &       -1.72         &       -3.92         &        1.12         &       -2.28         \\
                    &      (3.17)         &      (3.12)         &      (3.79)         &      (5.07)         &      (3.62)         &      (4.96)         \\
\hline
Observations        &         481         &         484         &         481         &         223         &         223         &         223         \\
Subjects            &         395         &         397         &         395         &         203         &         203         &         203         \\
 $ R^2 $            &       0.082         &        0.10         &        0.10         &       0.065         &       0.072         &       0.078         \\
Mean of Dep. Variable&       -0.89         &       -0.92         &       -0.89         &       -3.16         &       -3.16         &       -3.16         \\
Pre-treatment Beliefs Controlled?&          No         &         Yes         &         Yes         &          No         &         Yes         &         Yes         \\
 \hline\hline \end{tabular}\\ \begin{tfootnote}                 Clustered (at subject level) standard errors in parentheses. The subjects' family income, college majors, and sex are controlled in all models.                  Pre-Treatment beliefs are controlled in models (2), (3), (5), and (6).                 Only the observations with overestimated pre-treatment beliefs (relative to the information) are included.                 \sym{*} \(p<0.05\), \sym{**} \(p<0.01\), \sym{***} \(p<0.001\) . \end{tfootnote}

	}
\end{table}

\begin{table}[htp]
\caption{Changes in Beliefs of Different Demands}\label{tab:error_group}
{\renewcommand{\arraystretch}{0.8}
\begin{adjustbox}{max width=\textwidth}
	\begin{tabular}{l*{4}c}\hline\hline
&
\multicolumn{4}{c}{$|$
Post-error
$|$}\\
                    &\multicolumn{1}{c}{(1)}&\multicolumn{1}{c}{(2)}&\multicolumn{1}{c}{(3)}&\multicolumn{1}{c}{(4)}\\
                    &\multicolumn{1}{c}{Efficacy}&\multicolumn{1}{c}{Hospitalization}&\multicolumn{1}{c}{Efficacy}&\multicolumn{1}{c}{Hospitalization}\\
\hline
$|$ Pre-error $|$   &       0.664\sym{***}&       0.578\sym{***}&       0.781\sym{***}&       0.807\sym{***}\\
                    &    (0.0257)         &    (0.0323)         &    (0.0306)         &    (0.0324)         \\
Selected Top 3 -- Received&      -8.086\sym{***}&      -12.98\sym{***}&      -3.756\sym{***}&       0.113         \\
                    &     (0.641)         &     (0.959)         &     (0.995)         &     (1.137)         \\
\qquad\qquad\qquad -- Not Received&      -6.595\sym{***}&      -10.72\sym{***}&      -1.638         &       2.710         \\
                    &     (0.784)         &     (1.308)         &     (1.228)         &     (1.722)         \\
Not Selected Top 3 -- Received&      -5.724\sym{***}&      -9.249\sym{***}&      -1.394         &       1.664         \\
                    &     (1.117)         &     (1.671)         &     (1.484)         &     (2.107)         \\
\qquad\qquad\qquad -- Not Received&      -3.736\sym{**} &      -7.652\sym{***}&       2.317         &       2.813         \\
                    &     (1.156)         &     (1.662)         &     (1.421)         &     (2.528)         \\
Not Top 3 -- Received&      -5.220\sym{***}&      -7.452\sym{***}&      -1.215         &      -0.487         \\
                    &     (0.971)         &     (1.401)         &     (1.290)         &     (1.990)         \\
\qquad\qquad\qquad -- Not Received&           0         &           0         &           0         &           0         \\
                    &         (.)         &         (.)         &         (.)         &         (.)         \\
$|$ Pre-error $|$ $\times $ Received Selected Top 3&                     &                     &      -0.216\sym{***}&      -0.498\sym{***}\\
                    &                     &                     &    (0.0654)         &    (0.0406)         \\
$|$ Pre-error $|$ $\times $ Not Received Selected Top 3&                     &                     &      -0.258\sym{**} &      -0.464\sym{***}\\
                    &                     &                     &    (0.0819)         &    (0.0715)         \\
$|$ Pre-error $|$ $\times $ Received Non-selected Top 3&                     &                     &      -0.206\sym{*}  &      -0.350\sym{***}\\
                    &                     &                     &     (0.104)         &    (0.0992)         \\
$|$ Pre-error $|$ $\times $ Not Received Non-selected Top 3&                     &                     &      -0.290\sym{***}&      -0.313\sym{**} \\
                    &                     &                     &    (0.0768)         &     (0.102)         \\
$|$ Pre-error $|$ $\times $ Received Non-Top 3&                     &                     &      -0.164\sym{**} &      -0.191\sym{**} \\
                    &                     &                     &    (0.0582)         &    (0.0584)         \\
Constants           &       8.952\sym{***}&       19.15\sym{***}&       6.184\sym{***}&       11.16\sym{***}\\
                    &     (1.417)         &     (2.818)         &     (1.470)         &     (2.773)         \\
\hline
Observations        &        3160         &        3160         &        3160         &        3160         \\
Subjects            &         632         &         632         &         632         &         632         \\
 \hline\hline \multicolumn{3}{l}{Standard errors in parentheses.  Family income, college majors, and sex are controlled.} \\                 \multicolumn{3}{l}{\sym{*} \(p<0.05\), \sym{**} \(p<0.01\), \sym{***} \(p<0.001\)} \end{tabular}

\end{adjustbox}
}
\end{table}


\end{document}